\documentclass[twocolumn,floats,prx,showpacs]{revtex4-1}
\usepackage{amsfonts}
\usepackage{amsmath}
\usepackage{pspicture}
\usepackage{epsfig}
\usepackage{calc}
\usepackage{times}
\usepackage{color}
\usepackage{subfigure}
\usepackage{graphicx}
\usepackage{enumerate}
\usepackage{amssymb}
\usepackage{amsthm}
\usepackage{adjustbox}
\usepackage{appendix}
\newcommand{\appref}[1]{\hyperref[#1]{{Appendix~\ref*{#1}}}}
\newcommand{\cancel}[1]{}
\usepackage{comment}

\usepackage[table,usenames,dvipsnames]{xcolor}
\usepackage[colorlinks=true,
			hypertexnames=false
			]{hyperref}
\PassOptionsToPackage{linktocpage}{hyperref} %link numbers in TOC for arxiv
\newcommand{\bc}{\begin{center}}
\newcommand{\ec}{\end{center}}

\newcommand{\id}{\mathbb{I}}

\usepackage{amsfonts}

\def\id{\mathbb{I}}

\def\01{\{0,1\}}

\newcommand{\ket}[1]{|#1\rangle}
\newcommand{\bra}[1]{\langle#1|}

\definecolor{jens}{rgb}{0,0,0}
\newcommand{\je}[1]{{\color{jens} #1}}

\definecolor{caro}{rgb}{0,0,0}
\newcommand{\cw}[1]{{\color{caro} #1}}

\definecolor{alex}{rgb}{0,0,0}
\newcommand{\alex}[1]{{\color{alex} #1}}

\definecolor{rein}{rgb}{0,0,0}

\begin{document}

\title{Simulating topological tensor networks with Majorana qubits} 

\author{C. Wille,$^1$ R. Egger,$^2$ J. Eisert,$^{1,3}$ and A. Altland$^4$}
\affiliation{${}^1$~Dahlem Center for Complex Quantum Systems, Physics Department, Freie Universit\"at Berlin, D-14195 Berlin, Germany,\\
${}^2$~Institut f\"ur Theoretische Physik, Heinrich-Heine-Universit\"at, D-40225 D\"usseldorf, Germany\\
${}^3$~Department of Mathematics and Computer Science, Freie Universit\"at Berlin, D-14195 Berlin, Germany,\\
${}^4$~Institut f\"ur Theoretische Physik, Universit\"at zu K\"oln, Z\"ulpicher Str.~77, D-50937 K\"oln, Germany}

\begin{abstract}
The realization of topological quantum phases of matter remains a key challenge to condensed matter physics and quantum information science. In this work, we  demonstrate that progress in this direction can be made by combining concepts of tensor network theory with Majorana device technology. Considering the topological double semion string-net phase as an example, we exploit the fact that the representation of topological phases by tensor networks can be significantly simpler than their description by lattice Hamiltonians. The building blocks defining the tensor network are tailored to realization via simple units of capacitively coupled Majorana bound states. In the case under consideration,  this yields a remarkably simple blueprint of a synthetic double semion string-net,
and one may be optimistic that the required device technology will be available soon.  Our results indicate that the implementation of  tensor network structures via mesoscopic quantum devices opens up a powerful novel avenue toward the realization and quantum simulation of synthetic topological quantum matter.
\end{abstract}
\maketitle

\section{Introduction}

How can complex, or even topological, states of matter be realized in a physical system under precisely controlled conditions? Both from a conceptual and an applied perspective this is one of the most pressing questions of contemporary physics. Much of the dramatic recent progress in the  theory of quantum matter is owed to the identification of a vast spectrum of candidate topological phases of matter. At the same time, only few of these phases have been seen in experiment. Except for the quantum Hall states, and perhaps a few systems promising to harbor $\mathbb{Z}_2$-gauge symmetries, long range entangled topological phases of matter remain elusive at this point \cite{qc3,Wen2017}. This is a
\je{highly} unfortunate situation, not only from a condensed matter point of view: Conventional approaches to universal quantum computation based on surface codes \cite{TopologicalQuantumMemory} require  enormous overheads in magic state distillation \cite{PhysRevA.71.022316}, which may well turn out to be impractical. By contrast, the excitations supported by topological phases of matter have much stronger computational potential \cite{Kitaev-AnnPhys-2003,qc3,RevModPhys.87.307} and would arguably define a faster pathway, if only they could be realized.   

The apparent gap between theoretical understanding and the reality of materials is best illustrated by the concept of string-nets. Proposed more than a decade ago as a theoretical paradigm perhaps large enough to include all non-chiral gapped bosonic phases of matter with topological  order \cite{Levin2005,LevinWen,Wen2017}, they generically contain carefully balanced 12-spin interactions. No real life material will deliver such type of effective interactions out of the box. The standard way of reading the situation is to consider the string-net Hamiltonians as effective fixed points inside their topological phases. However, from a pragmatic perspective this only sidesteps the problem, and the identification of the respective basins of attraction in relation to existing quantum matter remains a daunting challenge. 
There are two principal approaches to make progress in this situation. The first aims to identify quantum matter in a prospected topological phase without direct reference to the representing string-net fixed point. As indicated above, progress in this direction has been rather limited so far.
The second aims to \textit{engineer} synthetic string-nets from well controlled device building blocks. Adopting the principles of 
quantum simulation~\cite{CiracZollerSimulation,Roadmap} this involves versatile design strategies to construct complex models from basic building blocks. Following this approach,
we here suggest a class of mesoscopic systems as a platform for  string-net engineering. A key advantage of this setup is that, to a very good approximation, interactions of low order can be systematically avoided. In this way, we suggest an  
architecture whereby, e.g., a 12-spin Hamiltonian can be engineered in a controlled manner. It stands to reason that this represents a qualitative design advantage over established platforms for quantum simulations (such as 
ultracold atom gases in optical lattices
\cite{BlochSimulation}) which operate on the basis of restricted classes of  2-local interactions. \alex{However, there remains the question \textit{which} of the many conceivable string net architectures to target in a goal oriented effort of this type:}

\je{From the perspective of topological quantum computing, the material class of string nets spans a vast spectrum of platforms, from  simple surface codes to  networks harboring non-abelian excitations capable of universal quantum computation. At this point, much of the experimental effort focuses on the realization of abelian surface codes for their relatively simple layout principles. Implementing additional design elements such as magic states and possibly lattice surgery, even such architectures may become capable of universal computation. However, this comes at the expense of formidable hardware overhead~\cite{Vijay2015,PhysRevB.94.174514,PhysRevLett.116.050501,PhysRevLett.108.260504,Hoffman2016} and  practicability is far from granted. On the other extreme of the spectrum, much theoretical work has been invested in exploring the computational power of non-abelian string net anyons (e.g., of Fibonacci type) where recent work demonstrates their capacity to support fault tolerant quantum computation, including under the averse conditions of finite temperature and
noise~\cite{NonAbelianDecoders,PhysRevA.95.022309}. However, given that no practical realization schemes have been suggested as yet, these approaches remain academic at this  stage. 

Rather than focusing on these extremes, we here reason that string net architectures of intermediate complexity may eventually provide optimal compromises between computational performance and realistic design principle. On a theoretical level, this pragmatic approach is only now beginning to be appreciated in the literature where recent work shows that, e.g., an abelian semion network --- the system next in complexity to the elementary toric code --- can be upgraded to become a fully error correcting code. More specifically, a variant of the double semion Hamiltonian has recently been shown \cite{SemionErrorCorrection} to provide a topological quantum error correcting code breaking with the paradigm of  
Calderbank-Shor-Steane (CSS) constructed from pairs of classical error-correcting codes. While the concrete perspectives of such approaches, too, cannot yet be foreseen with reliability, we see compelling motivation to explore hardware designs of intermediate 
complexity.}

\je{In this work, we move towards this goal via the combination of two  principles: the first is a novel application of tensor network theory \cite{Orus-AnnPhys-2014,AreaReview,VerstraeteBig,SchuchReview}. 
Tensor networks are known to describe string-nets in ways drastically reduced in complexity compared to the native Hamiltonian. This insight enters our design, which essentially implements a tensor network, \je{rather} than a Hamiltonian. Second, we propose the realization of these two-dimensional (2D) structures in a hardware platform featuring Majorana bound states \cite{Alicea2012,Leijnse2012,Beenakker2013,Sarma2015,Aguado2017,Lutchyn2018,Mourik2012,Albrecht2016,Deng2016,Nichele2017,Gazi2017,Zhang2018}.

Both directions are novel.  We emphasize  that the Majorana qubit introduced in Refs.~\cite{Beri2012,Beri2013,Altland2013,Plugge2017,Karzig2017} is essential for its unique capability in describing electronic and Pauli spin correlations. However, the hardware overhead required on top of the Majorana states is comparatively modest and does not go beyond that otherwise required to realize small-scale qubit networks. Our design is thus formulated in terms of device technology that is not yet available  but with a bit of optimism will be very soon \cite{Lutchyn2018}.

Tensor networks provide a very natural environment for the description of topological matter, and in particular for string\nobreakdash -nets. In fact, the fusion rules of topological quasi-particles find a mathematical abstraction in tensor categories,  whose representation in  the language of string-nets represents a tensor network almost by design. The formulation of a tensor network involves a   'physical' and a 'virtual' space, where the former represents physical  degrees of freedom (spins on a lattice, etc.) and the second serves to generate entanglement and capture the correlation structure of the state.  While the representation of string-net ground states as tensor networks is relatively straightforward~\cite{Gu2009,Buerschaper2009}, a drastic simplification emerges from the idea~\cite{Schuch_MPS,1409.2150,FermionicMPO,PhysRevB.95.245127} to inject, or encode, the physical space into the larger space of virtual states. In exchange for the seeming redundancy of this encoding, one gains a much simpler local description in terms of an effective Hamiltonian \cite{Brell2014PEPS} based on Hamiltonian gadget ideas \cite{Kempe-SIAM-2006, PhysRevA.77.062329}. At least, it is simple enough to afford a direct realization in quantum devices.}

\alex{We will illustrate the principles of the approach  on the example of the double semion string-net \cite{Levin2005}. As mentioned above, this system comes next in complexity to the toric code but is already rich enough to define a quantum error correcting code \cite{SemionErrorCorrection}. 
% make novel directions in error correction possible.  \cw{Although it is not yet 
% %fully 
% computationally universal}, its realization 
% would be a key step forward in the context of quantum information processing. In fact}, \je{concomitant with the program laid out above, a variant of the double semion Hamiltonian has recently been shown to provide a topological quantum error correcting code \cite{SemionErrorCorrection},
% that breaks with the paradigm of 
% Calderbank-Shor-Steane (CSS) codes which are constructed from pairs of classical error-correcting codes.}
Within the traditional Hamiltonian approach, the description of the double semion model already requires the full complexity of 12-spin correlations \footnote{\alex{For the double semion model restricted to the ground state space of all vertex operators, there is a mapping from a Hamiltonian with 12-local to 6-local plaquette operators \cite{SemionErrorCorrection}. However, the resulting 6-local operators have a more complicated structure and, in particular, they are no longer given by product operators. As a consequence, there is no straightforward way to implement the corresponding Hamiltonian.}} and may well be unrealistic. However, we here exploit that in tensor network virtual space the problem is  reduced down to a more manageable six-spin ring exchange, similar to that defining the physical qubits of a toric code on a hexagonal lattice \cite{Vijay2015}.  To implement the passage to tensor network space, various standard elements of tensor network constructions, including  Bell state projections, one-dimensional (1D) matrix product operators (MPOs), or repetition code qubits, need to be implemented on the device level. A principal message of this work is that networks of tunnel-coupled low capacitance quantum dots harboring Majorana bound states  --- Majorana Cooper boxes (MCBs) in the jargon of the field \cite{Beri2012,Beri2013,Altland2013,Plugge2017,Karzig2017} --- are optimally suited to this task. }

While this may be true also for more general tensor networks, it certainly is the case for the projected entangled pair states (PEPS) required in our present application.  For a preliminary impression of the hardware setup required to realize a double semion string-net, see Fig.~\ref{fig:DS} below, where the boxes represent MCBs, red and blue lines are tunneling bridges (short quantum wires), and the triangles are mere guides to the eye. (Not shown in the figure are side gates next to the tunneling links required to run a one-time calibration procedure, likewise required in the realization of, e.g., a Majorana surface code 
\cite{PhysRevLett.116.050501,PhysRevB.94.174514,PhysRevLett.108.260504,PhysRevX.6.031016,Hoffman2016,Roy2017}.) We emphasize that the structure does not include elements truly adverse to present-date experimental realization. Specifically, their operation does not require magnetic field interferometry or time varying external voltage sources. The structures rather are static in that the ground state of the tunnel- and capacitively coupled network is the double semion state. AC operations may be required to create anyonic excitations in the system or for the measurement of local state characteristics. However, these are operations beyond the scope of the current text. 

In essence, we propose the double semion system as a case study  for a conceptually novel avenue to the realization and quantum simulation of topological quantum matter. The approach involves bridging between Hamiltonian and tensor network descriptions of topological matter
in a fresh mindset, and from there on to hardware implementations. Here tensor networks are thus not considered as a numerical 
framework to study interacting quantum matter \cite{Orus-AnnPhys-2014,AreaReview,VerstraeteBig,Orus-AnnPhys-2014}, 
for which tensor network states and, in particular, PEPS \cite{PEPSOld, iPEPS}
are well known and for which they have originally been suggested \cite{DMRGWhite92}. They are also not used in an
a-posteriori approach of capturing complex phases of matter in a mathematically precise and conceptually clear
fashion \cite{TopologicalOrderInPEPS,PEPSTopology,ClassificationPhases,PhysRevB.81.064439,PhysRevB.95.245127,FermionicMPO,1409.2150}. 
Instead, they here serve as immediate candidates for a solid-state device realization. 
The double semion system is ideally suited to illustrate the principle because in spite of its relative simplicity, a 'direct' realization in quantum matter and/or hardware constructions seems illusive in view of the required 12-spin correlations \cite{Note1}. \je{The} essential insight which brings the system into reach is that its encoding in the virtual space of a tensor network dramatically reduces the complexity required from the hardware. While the generalization to branching string configurations on a lattice is straightforward, the realization of a tensor structure reliably describing a universal phase (such as the double Fibonacci phase) remains a nontrivial task. However, we hope that the concepts and device building blocks introduced below will be instrumental to progress in this direction as well. 
 
The remainder of this work is structured as follows. In Sec.~\ref{sec2}, we discuss the low-energy theory for arbitrary MCB networks.  In particular, we introduce novel design principles based on destructive interference mechanisms, see Sec.~\ref{sec2d}, which are key to the constructions put forward below.  In Sec.~\ref{sec3}, we show that high-order qubit interactions can be engineered in simple MCB networks based on these design principles.  In particular, for 1D structures of coupled MCBs, we establish a link to MPOs and other key elements of tensor network states.  Next, after a general discussion of string-nets (in particular, of the double semion system) and their representation in terms of tensor networks in~Secs.~\ref{sec4a} and \ref{sec4b}, we present the MCB network realization for the ground state of the double semion model in Sec.~\ref{sec4c}.  This work concludes with an outlook in Sec.~\ref{sec5}. Technical details 
related to Sec.~\ref{sec4c} can be found in the Appendix.

\section{Majorana Cooper Box Networks}\label{sec2}

In this section we discuss how networks of tunnel coupled MCBs give rise to an effective  theory containing engineered many-body interactions. We begin with a review of the smallest unit, the MCB  \cite{Beri2012,Beri2013,Altland2013,Plugge2017,Karzig2017}, and then apply perturbation theory to demonstrate how several MCBs connected by tunnel bridges define an effective low energy effective theory. We conclude the section with a discussion of design principles controlling and constraining the correlations present in this network.

\subsection{Majorana Cooper Box}\label{sec2a}

\begin{figure}[t]
\includegraphics[width=\columnwidth]{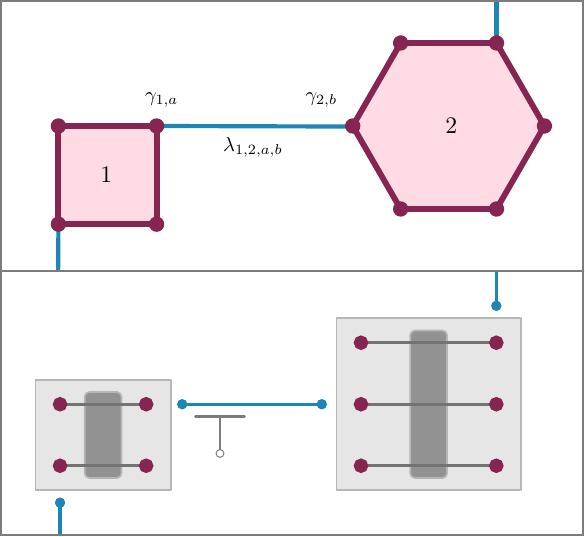}
\caption{Upper panel: Part of a network containing one tetron (MCB $1$) and one hexon (MCB $2$). In the ground state, each MCB abides in a definite eigenstate of the parity operator~\eqref{eq:ParityOperator}. Individual MBSs of different MCBs are coupled via complex tunneling amplitudes indicated in blue. Bottom panel: A somewhat less schematic representation of the same architecture, emphasizing that each MCB is a mesoscopic island comprising $n$ parallel quantum wires proximitized by the same superconductor (shaded vertical structure). Tunneling links between neighboring MCBs are tunable via local gate voltage changes.} 
\label{fig:MCBHardware}
\end{figure}

A single MCB can be conceptualized as a system of mesoscopic size, $i$ labelling spatial sites, 
harboring an even number $2n$ of spatially localized Majorana bound states (MBSs). The latter correspond to 
Majorana operators $\gamma_{i,a}=\gamma^\dagger_{i,a}$ with $a=1,\dots,2n$ and  the anticommutator algebra 
\begin{equation}
\{ \gamma_{i,a},\gamma_{j,b}\}=2\delta_{i,j}\delta_{a,b}\; . 
\end{equation}
For recent reviews of MBS physics, see Refs.~\cite{Alicea2012,Leijnse2012,Beenakker2013,Sarma2015,Aguado2017,Lutchyn2018}.  
The upper panel in Fig.~\ref{fig:MCBHardware} shows examples with $n=2$ and $n=3$, called \emph{tetron} and \emph{hexon} \cite{Karzig2017}, respectively.   The $2^n$-dimensional Hilbert space representing each box is split into two $2^{n-1}$-dimensional subspaces identified by the eigenvalues of the fermion number parity operator,
\begin{equation}
\label{eq:ParityOperator}
\mathcal P =  \mathrm i ^n\gamma_1 \ldots \gamma_{2n}=\pm 1\;.
\end{equation}
The even- and odd-parity sectors are separated by a large energy gap due to the charging energy $E_C$ of the box (see below).  Without loss of generality, we assume that the sector $\mathcal{P}=-1$ ($\mathcal P=+1$) is energetically favored for every tetron (hexon). The condition \eqref{eq:ParityOperator} implies strong entanglement between the elementary MBSs in each parity sector and will be of crucial importance to the definition of higher levels of entanglement between the base units. Next, Majorana operators $\gamma_{i,a}$ and $\gamma_{j,b}$ belonging to neighboring MCBs $i,j$ may be coupled by tunneling links characterized by tunneling amplitudes $\lambda_{ij,ab}$. 
Before discussing how the units introduced above can be combined to functional networks   \cite{PhysRevB.94.174514,Xu2010,PhysRevLett.108.260504,Roy2017,Nussinov2012,Vijay2015,Vijay2016,Litinski2017}, 
let us briefly review how they could be realized in practice.

 Referring for a detailed discussion to Refs.~\cite{PhysRevLett.116.050501,Plugge2017,Karzig2017}, we note that a $2n$ MCB is a collection of $n$ parallel (semiconductor or topological insulator) quantum wires, all connected to a conventional mesoscopic superconductor. The quantum wires are characterized by strong intrinsic spin-orbit coupling (e.g., InAs or InSb nanowires \cite{Lutchyn2018}) and can be fabricated with their own superconducting shell (e.g., an epitaxial Al surface layer \cite{Albrecht2016}).   All wires are isolated against ground but electrically connected by the superconductor island to form a floating (not grounded) MCB. A schematic of the setup is shown in the bottom panel of Fig.~\ref{fig:MCBHardware}, where the vertical dark shaded bars indicate the superconductor connected to the horizontal wires. The competition of superconductivity, magnetic field, and spin-orbit coupling in the wires drives the system into a topological state whose prime signature is the formation of MBSs at the wire ends \cite{Alicea2012,Leijnse2012,Beenakker2013,Sarma2015,Aguado2017,Lutchyn2018}, see for instance Refs.~\cite{Mourik2012,Albrecht2016,Deng2016,Nichele2017,Gazi2017,Zhang2018} for recent experimental reports of MBS signatures in related setups.  
Since the wires are parallel, a uniform magnetic field along the wire direction will induce the topological transition simultaneously in all wires.

In practice, these MBSs are quantum states of finite extension, and wave function overlap between them should be avoided in order to realize true zero-energy states. However, experiment indicates that for device proportions in the $\mathcal{O}(1\mu\mathrm{m})$ range, this hybridization can be made negligibly small \cite{Albrecht2016,Lutchyn2018}.
When isolated against ground, systems of this size have small electrostatic capacitance, $C$, and as a consequence a large electrostatic charging energy, $E_C(N-n_g)^2$, where $E_C:= e^2/2C\approx 1$~meV \cite{Albrecht2016},  $N$ is the particle number on the MCB and $n_g$ an effective background parameter tunable via side-gate electrodes. Importantly, charging effects are sensitive to the number of fermions occupying the Majorana sector of the MCB, $n_\gamma=\sum_{\alpha=1}^n n_\alpha$ with $n_\alpha:= c^\dagger_\alpha c_\alpha$ and  conventional fermion operators $c_\alpha=(\gamma_{2\alpha-1}+ i \gamma_{2\alpha})/2$ \cite{Fu2010}. In the generic Coulomb valley case, $n_g$ is tuned to a value such that the closest integer, $N$, is electrostatically favored. In fact, the MCB networks described below are operated at integer values of $n_g$ for all MCBs, where the low-energy theory enjoys particle-hole (anti-)symmetry. In any case, the resulting effective constraint to remain at a specific value of $N$ fixes the parity of $n_\gamma$ and thus yields Eq.~\eqref{eq:ParityOperator} \cite{Beri2012,Altland2013,Hyart2013}. (Only the parity is fixed  because a number of, say, $n_\gamma=3$ can be converted to $n_\gamma=1$ by creating a Cooper pair in the superconductor at no energy cost.)  Finally, we also assume that the proximity-induced superconducting gap in the wires is  sufficiently large to justify the neglect of above-gap quasiparticles.

MCB structures similar to the ones sketched above in design and proportions are currently becoming experimental reality. The formation of MBSs, controllable hybridization between MBSs as well as electrostatic charging effects have been evidenced in a number of reports \cite{Mourik2012,Albrecht2016,Deng2016,Nichele2017,Gazi2017,Zhang2018}. While we have not yet witnessed controlled Majorana qubit or MBS braiding experiments, these are the next conceptual steps on the agenda. Once these steps have been achieved, and sources of decoherence are under effective
control, the construction of networks will come into focus.  

\subsection{Tunneling Hamiltonian}\label{sec2b}

Individual MCBs can be connected by placing tunneling bridges between MBSs on different islands, as indicated in blue in Fig.~\ref{fig:MCBHardware}. These phase-coherent connectors (e.g., normal-conducting short nanowires) define the effective couplings $\lambda_{ij,ab}$. Their bare values respond sensitively to variations in the fabrication process and may largely be out of control. However, their values can subsequently be tuned via voltage changes on local side gates, as indicated in the bottom panel of Fig.~\ref{fig:MCBHardware}. This freedom may be applied to adjust individual tunneling amplitudes to premeditated values in a one-time interferometric  calibration procedure prior to the operation of the system,
see Secs.~\ref{sec2e} and \ref{sec4c} below and Refs.~\cite{Fu2010,PhysRevLett.116.050501,Plugge2017,Karzig2017}.  
While current fabrication technology does not exclude crossing links in a quasi-2D architecture \cite{Lutchyn2018}, they are difficult to implement in practice. The networks described below are constructed such that the number of crossings is kept at a minimum.

Our approach will be based on perturbation theory in the parameters $|\lambda_{ij,ab}|/E_C \ll 1$. Physically, this means that state changes of the system are induced by virtual excitations out of the definite parity ground state sector. The formulation of this expansion is facilitated by turning to a charge fractionalized picture wherein all Majorana fermions are considered electrically neutral and the charge balance is described by a number-phase conjugate pair $[\hat\varphi_i,\hat N_j] ={\mathrm i}
\delta_{i,j}$ \cite{Fu2010,Altland2013,Beri2013}. Technically, the passage to this formulation amounts to a gauge transformation, $c_{i,\alpha} \mapsto c_{i,\alpha} e^{\mathrm i\hat\varphi_i}$, applied to the fermion operators of the respective MCB. When represented in this way, the Hamiltonian of the system assumes the form    
$\hat H=\hat  H_0 +  \hat H_t$,
where $\hat H_0=E_C \sum_i (\hat N_i-n_{g,i})^2$ includes the charging Hamiltonians of all MCBs and $\hat H_t$ contains the tunnel couplings connecting different MCBs,
\begin{equation}\label{eq:V}
 \hat H_t=\sum_{i,j,a,b} \lambda_{ij,ab} \gamma_{i,a} \gamma_{j,b} e^{\mathrm i(\hat \varphi_i - \hat \varphi_j)} +\mathrm{h.c.}\;.
\end{equation}
These operators describe the inter-MCB correlation of Majorana bilinears $\gamma_{i,a} \gamma_{j,b}$ via the tunneling amplitudes $\lambda_{ij,ab}$, where  $e^{\mathrm i(\hat \varphi_i - \hat \varphi_j)}$ accounts for the fact that charge is raised/lowered by one unit on MCB $i/j$. Note that the action of $\hat H_t$ on the 
$\hat H_0$ charge ground state generates an excited state. The task of the perturbative program outlined above is the identification of relevant virtual ring-exchange processes wherein multiple tunneling leads back to the ground state.
 
\subsection{MCB Pauli operators}\label{sec2c}

\begin{figure}[t]
\includegraphics[width=\columnwidth]{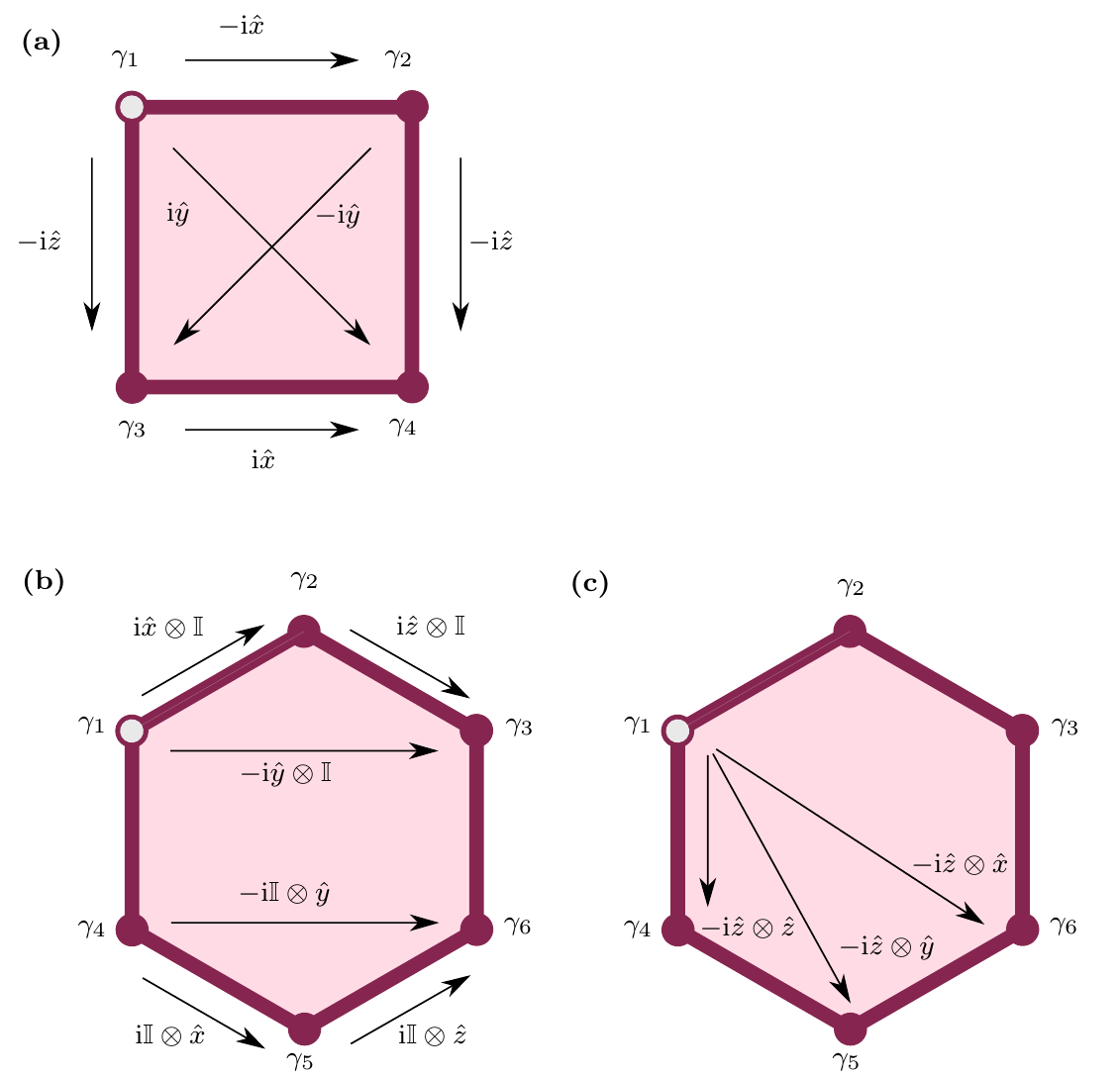}
\caption{Pauli operator representation of Majorana bilinears on tetrons and hexons.  Arrows between vertices (i.e., MBSs) identify Majorana bilinears and their associated single- or two-qubit operators. The marking of $\gamma_1$ (open circle) indicates our Majorana ordering choice. (a) Single-qubit operators for a tetron. (b) Single-qubit operators for a hexon. (c) Several two-qubit operators for a hexon. } 
\label{fig:MCB_legend}
\end{figure} 

For our purposes below, it will be convenient to represent bilinears of Majorana operators acting within a sector of definite parity through 
qubit Pauli operators, that is, $\hat x, \hat y, \hat z$. For example, the ground-state space of a tetron, i.e., the space of two conventional fermions with even parity ($n_\gamma=0,2$), is equivalent to the Hilbert space of a single qubit \cite{Beri2012}. Indeed, the representation
  \begin{equation}\label{eq:tetronqubit}
\gamma_1 \gamma_2=  -{\mathrm i}\hat x \;, \qquad \gamma_2 \gamma_3= -{\mathrm i} \hat y \;, \qquad  \gamma_1 \gamma_3= -{\mathrm i} \hat z \;,
 \end{equation} 
faithfully represents the operator algebra of Majorana bilinears. Combinations involving $\gamma_4$ are fixed by the parity constraint in Eq.~\eqref{eq:ParityOperator}, $\gamma_1\gamma_2\gamma_3\gamma_4=1$, as indicated in Fig.~\ref{fig:MCB_legend}(a). 
Since the definite assignment of an Pauli operator equivalence makes reference to a specific ordering of Majorana operators, we occasionally indicate the position of $\gamma_1$ by an open circle in our figurative representations. 

For hexons, the ground state space is four-fold degenerate and can be interpreted as the Hilbert space of two qubits \cite{Karzig2017}. We partition the six MBSs into two groups of three, with the understanding that Majorana bilinears formed from only one group correspond to single-qubit Pauli operators,
see Fig.~\ref{fig:MCB_legend}(b).  In contrast, bilinears involving MBSs from different groups  
yield two-qubit operators. For instance, $\gamma_1 \gamma_4 = - \mathrm i \hat z \otimes \hat z$ 
follows by using the single-qubit operators in Fig.~\ref{fig:MCB_legend}(b) together with the 
parity constraint in Eq.~\eqref{eq:ParityOperator}, which for hexons is equivalent to $\gamma_1 \gamma_4 = \mathrm i \gamma_2 \gamma_3 \gamma_5 \gamma_6$. Several other two-qubit operators are specified in Fig.~\ref{fig:MCB_legend}(c). 

\subsection{Effective low-energy theory}\label{sec2d}

For a general MCB network Hamiltonian, 
\begin{equation}
\hat H=\hat H_0+\hat H_t,
\end{equation}
each tunneling process involves decharging (charging) the emitting (receiving) MCB by an elementary charge. Since the charging energy $E_C$ is assumed large compared to the tunneling amplitudes, and open tunneling paths inevitably leave the system in an excited state, only  
\emph{closed paths} are physically relevant to the low-energy physics. In the following we show 
%how Schrieffer-Wolff perturbation theory (taken in the standard self-energy approximation) \cite{Bravyi2011} may be applied to
\cw{how a self-energy expansion as performed in Ref. \cite{Kitaev06} (cf. \cite{PhysRevLett.108.260504,1367-2630-13-5-053039}) }can be applied to represent the relevant tunneling processes as products of Pauli operators acting on the low-energy Hilbert space of the system. 
\cw{The starting point of the analysis is the series expansion}
\begin{equation}\label{eq:series} 
\hat H_\text{eff}=\sum_{k=1}^\infty\hat H^{(k)}\;,\quad
\hat H^{(k)}=\hat P_0 \left(\hat H_t \frac{1}{-\hat H_0} \right)^{k-1} \hat H_t \hat P_0 \;,
\end{equation}
where $\hat P_0$ is the ground-state projector of $\hat H_0$.

\begin{figure}%
\includegraphics[width=\columnwidth]{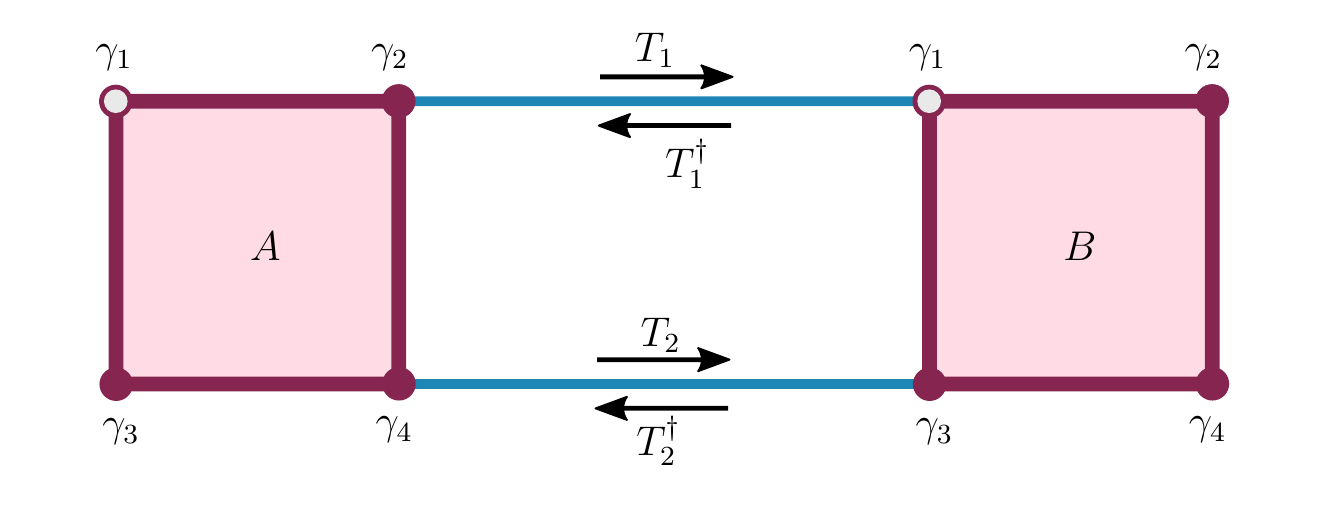}%
\caption{Example of a minimal MCB network, see~Eq.~\eqref{eq:T12}.} %
\label{fig:example}%
\end{figure}

Let us first discuss the structure of Eq.~\eqref{eq:series} on the example of two connected tetrons as in Fig.~\ref{fig:example}. With the notation in Fig.~\ref{fig:example}, the tunneling Hamiltonian is given by  
$\hat H_t= T_1+ T_2 + T_1^\dagger +T_2^\dagger$ with the hopping operators
\begin{equation}
T_1=\lambda \gamma_{2,A} \gamma_{1,B} e^{\mathrm i (\hat\varphi_A-\hat\varphi_B)}\;,  \quad \label{eq:T12}
T_2=\kappa \gamma_{4,A} \gamma_{3,B} e^{\mathrm i (\hat\varphi_A-\hat\varphi_B)}\;.
\end{equation} 
It is useful to represent individual $T_i$ operators as directed arrows from $\gamma_{i,a}$ to $\gamma_{j,b}$ (for different MCBs $a\ne b$), where Hermitian conjugates $T^\dagger$ correspond to reversed arrows, see Fig.~\ref{fig:example}.  
Charge neutrality now requires that each MCB carries an equal number of incoming and outgoing arrows. In this way, terms contributing to Eq.~\eqref{eq:series} are oriented closed paths and the effective Hamiltonian is given by a sum over all closed path contributions.  We first consider only paths without self-intersections, i.e., oriented loops of length $k$, where the orientation sense, $d=\pm$, is determined by the arrow direction and briefly remark on self-intersecting closed paths at the end of this section. Since $\hat H_t$ is Hermitian, every oriented loop comes along with its Hermitian conjugate counterpart, i.e., a loop  with opposite orientation. Moreover, loop contributions to $\hat H^{(k)}$ are distinguished by the ordering sequence, $s$, of individual $T_i$ operators which in general neither commute with each other nor with $1/\hat H_0$.

For our two-tetron example in order $k=2$, we obtain 
\begin{eqnarray} 
\hat H^{(2)}&=& \hat O^{(2)}_{1,2,+}+\hat O^{(2)}_{2,1,+}+\hat O^{(2)}_{1,2,-}+\hat O^{(2)}_{2,1,-}\;, 
\label{eq:ex_H2split} \\ \nonumber
\hat O_{i,j,+}^{(2)}&=&\hat P_0 T_i \frac{1}{-\hat H_0} T_j^\dagger \hat P_0\;,
\end{eqnarray}
where the index specifies $(s,d)$ and reversing the loop orientation equals Hermitian conjugation. 
We note that in Eq.~\eqref{eq:ex_H2split} `diagonal' contributions, $\hat O_{j,j,\pm}^{(2)}$, have been dropped since they only cause an irrelevant overall energy shift.  We now perform the projection to the charge ground state sector in Eq.~\eqref{eq:ex_H2split}, where the first term takes the form
\begin{equation}\label{eq:ex_eval} 
\hat O^{(2)}_{1,2,+}= -\frac{\lambda \kappa^\ast}{2 E_C} \gamma_{2,A} \gamma_{1,B} \gamma_{3,B}\gamma_{4,A} = \frac{\lambda \kappa^\ast}{2 E_C} \hat z_A \hat z_B\;.
\end{equation}
Here we have assumed that both MCBs are operated at the
electron-hole symmetric point (integer $n_{g,A/B}$). In that case, the intermediate virtual state involves a single-electron charging (decharging) of box $A$ ($B$) with excitation energy $2E_C$. For the qubit operator representation, i.e., the second equality in Eq.~\eqref{eq:ex_eval}, we have used the Pauli operators \eqref{eq:tetronqubit} for each tetron.

We are now ready to specify the general operator structure for an ordered oriented loop of length $k$, 
\begin{equation}\label{eq:decompose}
\hat O^{(k)}_{s,d}= \mathrm i^k  A(k,s,d) \hat Q^{(k)} \;,
\quad \hat Q^{(k)}=\hat q_1 \hat q_2 \cdots \hat q_k\;,
\end{equation}
where $\hat Q^{(k)}$ is composed of Pauli operators $\hat q_i$ acting on MCB $i$. Note that $\hat Q^{(k)}$ 
neither depends on the orientation $d$ nor on the sequence $s$. The 
prefactor $A(k,s,d)$ collects all tunneling amplitudes and the (inverse) excitation energies of virtual intermediate states, with $A(k,s,d)\sim  1/E_C^{k-1}$.  Finally, by summing over all possible sequences of $T_i$ operators, we obtain the operator for \emph{unordered} oriented loops,
\begin{equation}
\hat O^{(k)}_d = \sum_{s} \hat O^{(k)}_{s,d} \;. \label{eq:unordered}
\end{equation}

\subsection{Destructive interference mechanisms and Hamiltonian design} \label{sec2e}

Since the prefactor $A(k,s,d)$ in Eq.~\eqref{eq:decompose} decays exponentially with the loop length $k$, dominant contributions to $\hat H_{\rm eff}$ 
originate from short loops involving just a few qubit operators.
This fact poses a serious problem for engineering complex quantum systems. In particular, fixed-point Hamiltonians exhibiting topological order, e.g., string-net models, generically rely on the presence of high-order many-qubit interactions \cite{Levin2005,Fidkowski2009,Wen2017}, see Sec.~\ref{sec4a}. A key point of our work is to design MCB network structures where unwanted low-order contributions due to short loops are automatically eliminated by destructive interference.  
Based on these design ideas, one can largely tune the effective Hamiltonians in a desired fashion. We expect this to be of general interest for quantum simulations, beyond the specific application of generating topological models.
We next present two different mechanisms effecting such type of loop cancelation.

\begin{figure}
\includegraphics[width=\columnwidth]{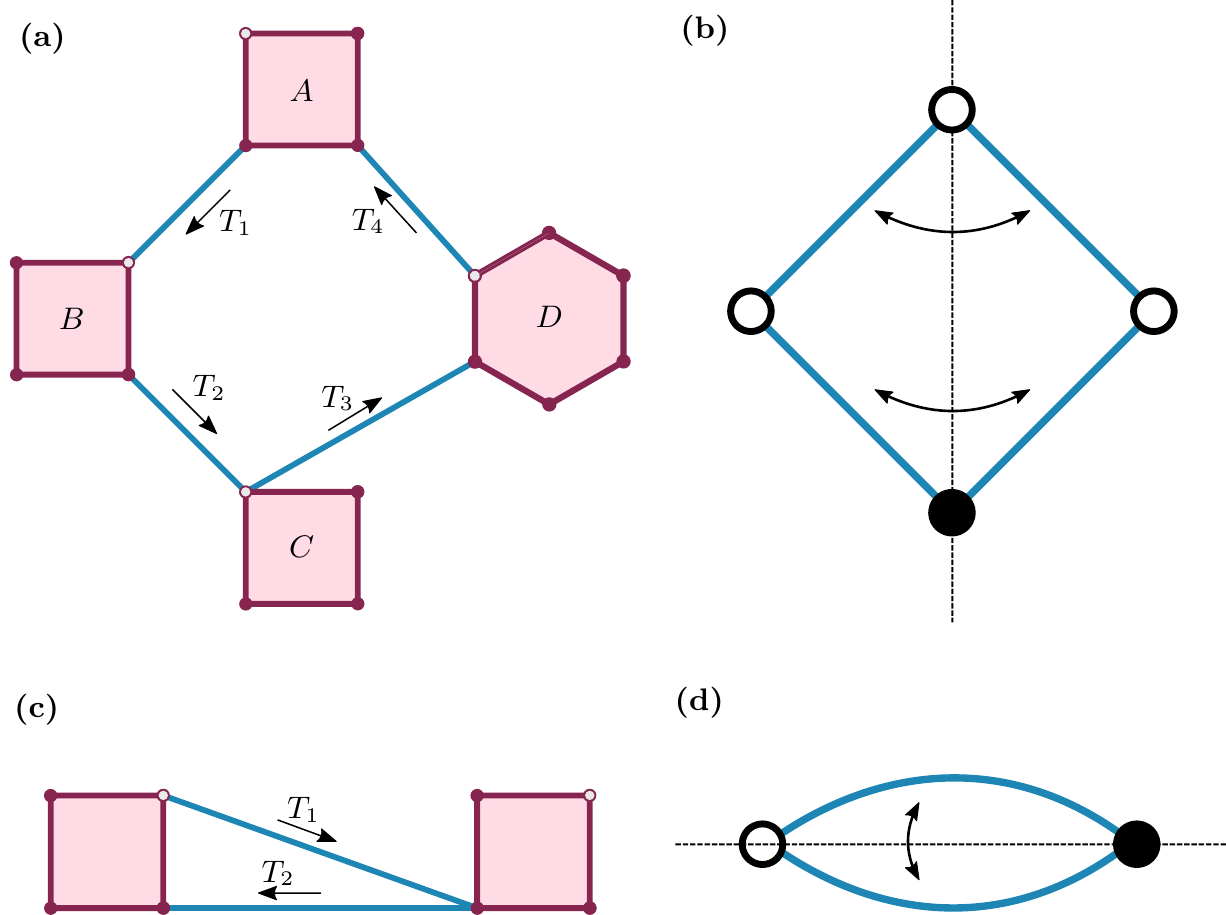}
\caption{Two examples for loop cancellation by symmetries and anticommuting $T_i$ terms. (a) MCB network with overlapping hopping terms $T_2$ and $T_3$, and (b) the corresponding reduced graph (see main text for its definition and the coloring of vertices). The reflection symmetry defining the loop-canceling permutation $\Pi[s]$ is also indicated.  (c) Another example for a network with anticommuting $T_i$ operators and (d) the associated reduced graph. }
\label{fig:ex2}
\end{figure}

\emph{Loop cancellation by symmetry ---}
Unordered oriented loops will automatically vanish if, for each sequence $s$ of tunneling operators, there exists a \emph{permuted sequence} $\Pi[s]$ with opposite prefactor, 
\begin{equation}\label{eq:aop}
A(k, \Pi[s],d) = -A(k,s,d).
\end{equation}
Indeed, in such  cases, Eq.~\eqref{eq:unordered} readily gives  
\begin{equation}
\hat O^{(k)}_d=\sum_{s} \hat O^{(k)}_{s,d}  = 
\sum_{s} \hat O^{(k)}_{\Pi[s],d} = - \sum_{s} \hat O^{(k)}_{s,d} = 0.
\end{equation} 
The next, and more difficult, step is to identify MCB network structures where such loop-canceling permutations exist. To that end, we first observe that for structures containing a pair of anticommuting tunneling operators, $\{ T_i, T_j\}=0$, the prefactor $A(k,s,d)$ will change sign when interchanging $T_i$ and $T_j$ in the sequence $s$.   
Anticommuting $T_i$ and $T_j$ operators share a common Majorana operator and thus represent \emph{overlapping} hopping terms. Two examples for MCB structures with overlapping $T_i$ operators are depicted in Fig.~\ref{fig:ex2}. 

However, an odd number of overlapping $T_i$ operator pairs --- and thus a sign change in $A(k,s,d)$ for the 
permuted sequence --- is only a necessary (but not sufficient) condition for loop cancellation.  
In order to fulfill Eq.~\eqref{eq:aop}, we also have to guarantee that the product of intermediate excitation
energies is identical.  The latter are determined by the charging contribution $\hat H_0$
and therefore only depend on the charge transferred between MCBs, irrespective of precisely which 
MBSs are involved in the tunneling path or whether we have tetrons or hexons. 
Loop structures with an invariant energy product can then be identified 
from a simpler \emph{reduced graph} obtained by collapsing each MCB to a single vertex, 
see Fig.~\ref{fig:ex2}(b,d) for examples.
For convenience, vertices with overlapping $T_i$ operators are now marked by a different color (black in Fig.~\ref{fig:ex2}). If the reduced graph has a reflection symmetry, 
mapping edges to edges and vertices to vertices of the same color, 
every sequence $s$ is uniquely mapped onto a mirror sequence with the same energy product. \alex{We note that this invariance condition does not require equal values of the charging energies on the individual islands. In fact, the charging energies do not even appear in our definition of a reduced graph.  }
If the reflection symmetry also interchanges an odd number of overlapping $T_i$ pairs,
this map defines a loop-canceling permutation $\Pi[s]$.

Let us illustrate this  mechanism for the MCB network in Fig.~\ref{fig:ex2}(a), which has a reflection symmetry of its reduced graph [see Fig.~\ref{fig:ex2}(b)] along the vertical axis.  This symmetry corresponds to the permutation $T_1 \leftrightarrow T_4$ and $T_2 \leftrightarrow T_3$ of tunneling terms, with the anticommutator $\{T_2,T_3\}=0$. Without loss of generality, we set all coupling amplitudes to $\lambda=1$ and evaluate the $k=4$ loop contributions for a specific sequence $s$ and the associated permuted sequence $\Pi[s]$,
\begin{eqnarray} \nonumber
\hat O^{(4)}_{s,+} &=&\hat P_0 T_1 \frac{1}{-\hat H_0} T_3 \frac{1}{-\hat H_0} T_2 \frac{1}{-\hat H_0} T_4 \hat P_0 \\
&=& \frac{\mathrm i}{16E_C^3} \hat x_A \hat x_B \id_C (\hat z \otimes \hat z)_D \;,\\ \nonumber
\hat O^{(4)}_{\Pi[s],+}&=&\hat P_0 T_4 \frac{1}{-\hat H_0} T_2 \frac{1}{-\hat H_0} T_3 \frac{1}{-\hat H_0} T_1 \hat P_0\\\nonumber &=&-\frac{\mathrm i}{16E_C^3} \hat x_A \hat x_B \id_C (\hat z \otimes \hat z)_D \;.
\end{eqnarray}
Clearly, both contributions cancel each other. Such cancellation due to overlapping  $T_i$ operators in combination with geometric symmetries already applies at the level of oriented loops. The vanishing of an oriented loop then implies,  by Hermitian conjugation, the vanishing of its reversed partner. 

\emph{Loop cancellation by phase tuning ---} Alternatively, the cancellation of loops can be achieved via the engineered tuning of complex tunneling couplings, notably their phases. In this way, oriented loops can be converted into anti-Hermitian operators, sign opposite to their reversed partners. The  summation over both orientations in Eq.~\eqref{eq:unordered} then implies the 
exact cancellation of the corresponding \emph{unordered unoriented} loop
according to
\begin{equation}
\hat O^{(k)}=\sum_{d=\pm } \hat O^{(k)}_d = \hat O^{(k)}_+ + \hat O^{(k)\dagger}_+ = 0\;. \label{eq:undirected}
\end{equation}
How the tunneling amplitudes in a given MCB network have to be chosen for this to happen 
depends on the loop length $k$. Excluding loop structures containing
anticommuting $T_i$ terms, we find that for odd (even) $k$, the product of all tunneling amplitudes along the loop path 
has to be purely real (imaginary). In fact, one only needs to tune an overall \emph{loop phase} defined below.

Let us briefly reconsider the two-tetron example in Fig.~\ref{fig:example} to illustrate this mechanism. An unordered oriented loop with $k=2$, winding once around the structure, is described by
\begin{equation}
\hat O^{(2)}_+=\frac{\lambda  \kappa^\ast}{E_C} \hat z_A  \hat z_B, 
\end{equation}
cf.~Eq.~\eqref{eq:ex_eval}.
The unoriented loop contribution $\hat O^{(2)}$ thus exactly vanishes for
$\text{Re}(\lambda  \kappa^*)=0$.  Writing $\lambda=|\lambda|e^{\mathrm i \phi_\lambda}$ and $\kappa=|\kappa|e^{i\phi_\kappa}$, this condition is equivalently 
formulated as a condition on the loop phase, 
$\phi_{\rm loop}:= \phi_\lambda-\phi_{\kappa}=\pm \pi/2$.
This type of phase tuning offers a powerful tool for suppressing few-qubit interactions. 

We now return to the general case of a length-$k$ loop without anticommuting $T_i$ 
terms.   With the tunneling amplitudes $|\lambda_1|e^{\mathrm i \phi_1},\ldots,|\lambda_k|e^{\mathrm i \phi_k}$ along the loop path, the loop phase is 
$\phi_{\rm loop}=\phi_1+\cdots+\phi_k$.  We then observe that for
\begin{equation}\label{eq:cases}
\phi_{\rm loop}= \begin{cases} \pm \pi/2\;,  &k \textrm{ even}\;, \\ 0,\pi\;,  &k \textrm{ odd}\;, \end{cases}
\end{equation}
the unoriented loop vanishes, giving rise to $\hat O^{(k)}=0$.
In practice, the loop phase may be calibrated in an initial step by adjusting the voltage on just one local gate near a tunneling link within the loop. By means of interferometric (e.g., conductance or capacitance spectroscopy) measurements \cite{Plugge2017,Karzig2017}, the value of $\phi_{\rm loop}$ can be determined experimentally.  The local gate voltage is then subsequently readjusted until the desired 
value of $\phi_{\rm loop}$ in Eq.~\eqref{eq:cases} has been realized. We expect this calibration technique to be important when 
building complex structures from elementary building blocks.

\alex{In practice, the above calibration procedure will adjust the loop phases only up to a certain accuracy level, and the  suppression of short-loop contributions will not be perfect.
One may account for this fact by regarding the corresponding qubit operators, $\hat V_i$, as small perturbations to the unperturbed effective Hamiltonian, 
$\hat H_{\rm eff}$,  of  a topologically ordered phase and consider the generalized 
$\hat H_{\rm eff}'= \hat H_{\rm eff} + \sum_i \hat V_i$.
If all $ \|\hat V_{i} \|$ are sufficiently small, the perturbed Hamiltonian
$H'_{\rm eff}$  will satisfy the conditions of Ref.~\cite{1001.0344} where the stability of topological order under small local
perturbations has been demonstrated.  This implies that a finite window for tolerable loop phase calibration errors must exist. Within this window, our construction inherits the 
robustness of topological order as established in Ref.~\cite{1001.0344}.  In Sec.~\ref{sec4c}, we return to this 
point for the specific case of the double semion model. }

\begin{figure}
\includegraphics[width=0.72 \columnwidth]{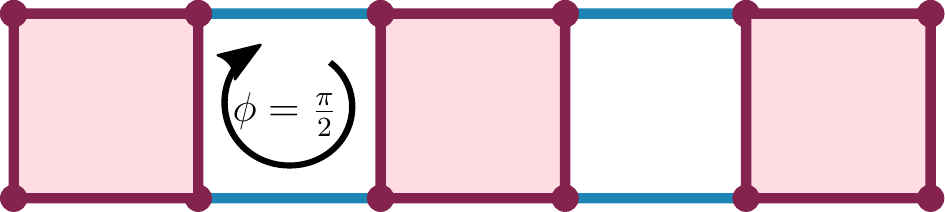}
\caption{An example for cancellation of composed loop paths.}
\label{fig:compose}
\end{figure}

\emph{Composite loop patterns ---}
Finally, we investigate how the design mechanisms can be transferred to general closed paths. It is straightforward to show that the contribution associated to a disjoint union of loops factorizes into the product of the single loop operators up to a real positive proportionality constant (a combinatorial prefactor) and thus such contributions vanish if any of their loops vanish. 
\alex{Loop-cancelling permutations $\Pi$ can also be applied  to graphs with self-intersections which implies that our symmetry cancellation arguments extend to this case. This follows from inspection of the symmetries of the reduced graph, and the sign structure of the permutations acting on it.
% is not restricted to act on a simple closed loop, our
% symmetry cancelation arguments --- examining the symmetries of the reduced graph and the sign change of the induced permutation --- continue to hold.

Alternatively, one may  eliminate  self-intersecting paths by phase tuning. To this end, consider a general oriented closed path with intersections partitioned into non-intersecting closed sub-paths. It is then specified by a collection $\{(s_i,d_i)\}$ of $n$ sub-paths, $s_i$, along with their orientations, $d_i$.  Due to charge neutrality, this is equivalent to an assembly of $n$ closed oriented non-intersecting loops. 
% We next note that a non-oriented parent path defines all different orientation patterns of its sub-paths. 
%From all possible $2^n$ orientation patterns, only 
Only orientation patterns for which the number of paths beginning and ending at all vertices are identical satisfy charge neutrality and need to be taken into account. If all those patterns are such that they contain at least one oriented loop with a loop canceling phase, see Eq.~\eqref{eq:cases}, the parent path will be subject to cancellation as well. For example, consider the (unoriented) loops shown in Fig.~\ref{fig:compose}, where the left length-2 loop, $\hat L= \hat L_+ + \hat L_-=0$, vanishes due to phase cancellation while the right length-2 loop, $\hat R=\hat R_+ + \hat R_-$, has an arbitrary loop phase and does not vanish by itself. The closed length-4 path $\hat O^{(4)}$ obtained by concatenation of $\hat L$ and $\hat R$ can then be decomposed into two oriented segments, namely the left and the right loop with clockwise or anticlockwise orientations. It thus fulfills the aforementioned criterion and $\hat O^{(4)}=O^{(4)}_{++}+O^{(4)}_{+-}+O^{(4)}_{-+}+O^{(4)}_{--} \propto (\hat L_+ + \hat L_-)(\hat R_+ + \hat R_-)=0$ implies that the length-4 path is subject to phase cancellation.
}

\section{Engineering multi-qubit operators}\label{sec3}

We next outline  how the  design principles of Sec.~\ref{sec2e} may be applied to engineer complex quantum systems with \emph{high-order qubit interactions} in MCB networks. 
In order to design 2D topological models, we have to formulate new
design principles to make sure that we arrive at models that are in the anticipated phase. \je{We do so by building upon the 
framework of Hamiltonian gadgets 
\cite{Kempe-SIAM-2006,PhysRevA.77.062329,Brell2014PEPS,Bartlett06}.} \alex{Gadgets 
are tools for generating Hamiltonian terms with high locality (high operator order in the language of many body physics) in perturbation theory. Introduced
in Ref.~\cite{Kempe-SIAM-2006} as a method to generate three-body terms, the idea has been generalized in 
Ref.~\cite{PhysRevA.77.062329} to arbitrary orders. In our work, a variant of Hamiltonian gadgets tailored to the   generation of PEPS from basic building blocks
\cite{Brell2014PEPS,Bartlett06} will be applied. This design element will be key to the direct realization of
topological fixed point models in mesoscopic Majorana platforms.}

\alex{
We start our construction from the Hamiltonians of  $M$ coupled qubits (see Fig.~\ref{fig:tetron_prod}) connected to effectively implement  matrix product operators (MPOs)
\cite{Mixed,raey,MPO_Representations,UndecidableMPO,Bultinck2017}. In  Sec.~\ref{sec4}, these 1D units will be  building blocks for the implementation of 2D tensor networks \cite{Orus-AnnPhys-2014,AreaReview,VerstraeteBig,SchuchReview} supporting topological order. This bottom up construction will make heavy use of the gadget principle which serves to generate effective interactions beyond the nearest neighbor level (sixth order correlations in our case) while terms of lower order are excluded to a high level of accuracy. }

\subsection{Product operators}\label{sec3a}

\begin{figure}
\includegraphics[width=0.35\textwidth]{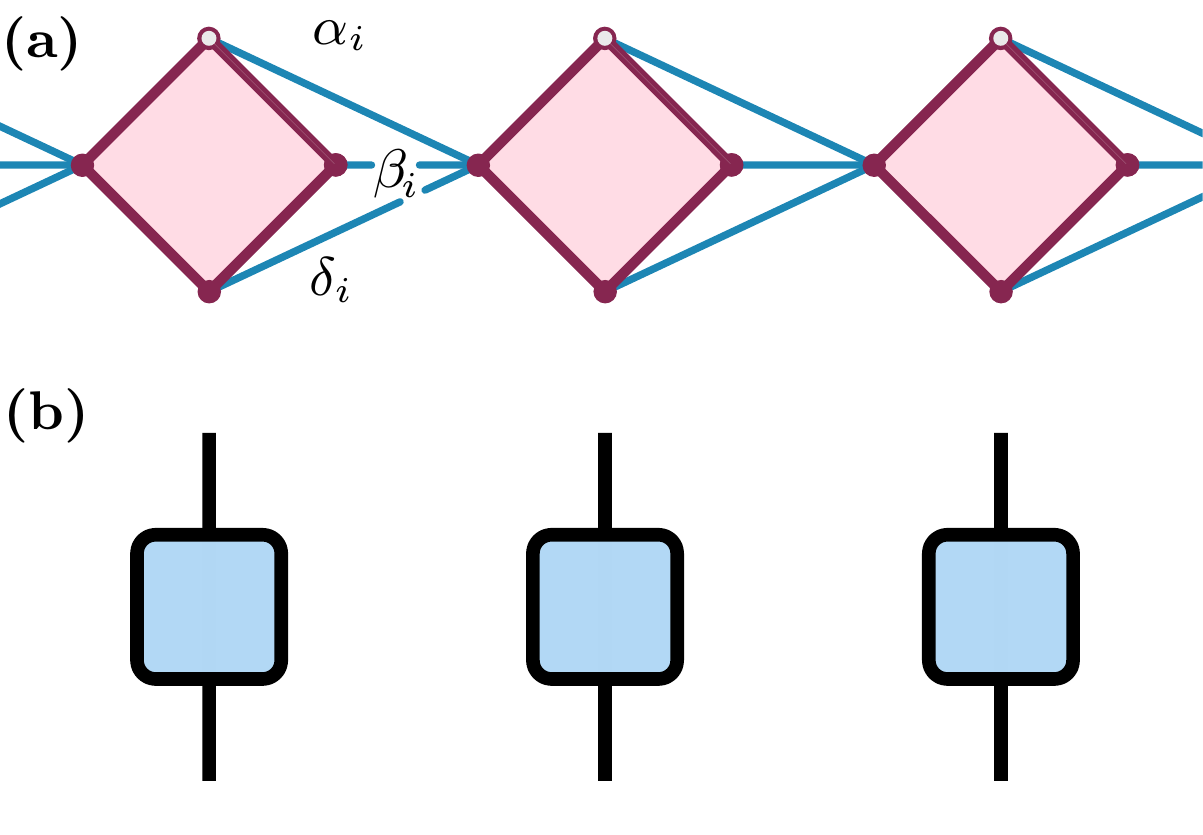}
\caption{(a) MCB ring of tetrons (the chain has periodic boundary conditions), and (b) the leading order of the corresponding effective qubit Hamiltonian (\ref{eq:prod}), a product operator, represented as tensor network.}
\label{fig:tetron_prod}
\end{figure}

Consider a ring of $M$ coupled MCBs, where tunneling bridges only connect neighboring MCBs.  We assume that each MCB contains one MBS at which all tunneling bridges incoming from the left neighbor terminate, as illustrated in Fig.~\ref{fig:tetron_prod}(a) for a tetron ring. In such structures,  loop paths not fully winding around the ring necessarily include one sub-loop of length 2. However, these loops vanish by symmetry, as illustrated in Fig.~\ref{fig:ex2}(c,d). Thus, composite loops containing one or several length-2 sub-loops vanish as well and we only need to consider loop paths with (one or several) full windings around the ring.

We here focus on loops with a single winding as they contribute dominantly to the perturbation expansion \eqref{eq:series}. Noting that hopping processes between adjacent MCBs correspond to a sum of Pauli operators weighted by the respective tunneling amplitudes, $\hat q_i$, we obtain the oriented loop operator 
\begin{eqnarray}
\hat O_+^{(M)} &=& \frac{-\mathrm i^M}{(2E_C)^{M-1}} \hat q_1 \hat q_2 \cdots \hat q_M  \label{eq:prod} \;,\\ 
\hat q_i &=& \delta_i \hat x_i + \beta_i \hat y_i + \alpha_i \hat z_i \;.
\end{eqnarray}
The complex constants $\alpha_i,\beta_i,\delta_i$ describe the three tunneling amplitudes connecting the MCB labeled by $i$ to its left neighbor, see Fig.~\ref{fig:tetron_prod}(a). We next observe that for loops with even (odd) length $M$ and purely real (imaginary) tunneling amplitudes, $\hat O^{(M)}_+$ is Hermitian.  In that case, the low-energy Hamiltonian, $\hat H_{\rm eff}=2\hat O_+^{(M)}$, implements a Hermitian product 
operator of qubits, where
the associated tensor network is shown in Fig.~\ref{fig:tetron_prod}(b).
We also note that by detuning the tunneling phases away from the fine-tuned points above, one may generate operators with stronger entanglement, corresponding to the sum of two product operators. 

\begin{figure}
\includegraphics[width=0.4\textwidth]{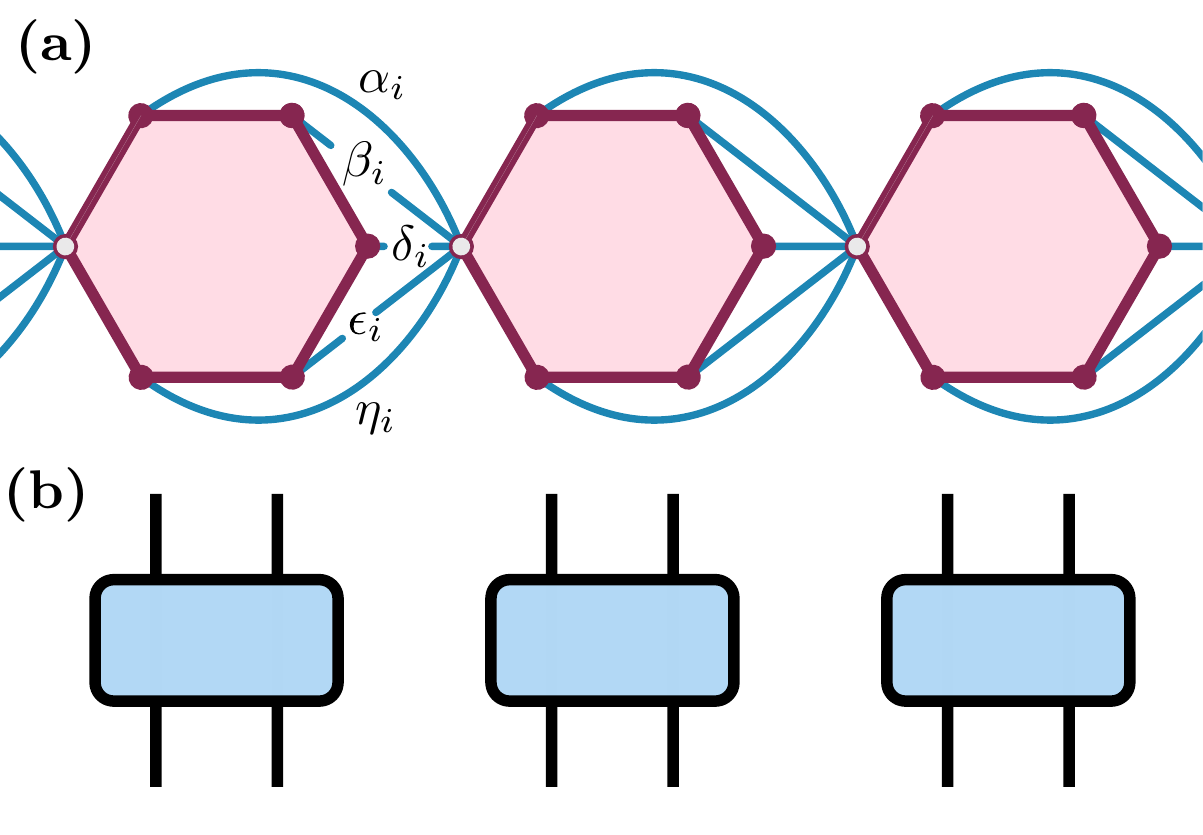}
\caption{(a) MCB ring of hexons, and (b) the leading order of its effective qubit Hamiltonian, a product operator composed from two-qubit operators, depicted as tensor network.}
\label{fig:hexon_prod}
\end{figure}

The above product operator design can also be implemented for hexons or mixed tetron-hexon structures. For hexons, Eq.~(\ref{eq:prod}) is still an appropriate description. However, the individual $\hat q_i$ operators are now replaced by two-qubit operators. For instance, for the example in Fig.~\ref{fig:hexon_prod}(a), we have
\begin{eqnarray}
\hat q_i &=& -\alpha_i (\hat x \otimes \id)_i + 
\beta_i (\hat y \otimes \id)_i \nonumber \\
&+& \delta_i (\hat z \otimes \hat x)_i +\epsilon_i 
(\hat z \otimes \hat y )_i +\eta_i (\hat z \otimes \hat z)_i\;,
\end{eqnarray}
where the constants $\alpha_i,\beta_i,\delta_i,\epsilon_i,\eta_i$ refer to the five tunneling amplitudes connecting the MCB labeled by $i$ to its left neighbor, see Fig.~\ref{fig:hexon_prod}(a).
The corresponding tensor network is shown in Fig.~\ref{fig:hexon_prod}(b). Since the total operator is just a single product over operators, $\hat q_i$ specific to each $i$ (rather than a sum over products), the network does not carry a 'virtual' internal index, as indicated by the absence of  horizontal links. The two-qubit nature of the compound operators $\hat q_i$ is indicated by the presence of two vertical index lines at each block.   More complex structures can be designed by 
increasing the number of tunneling bridges connecting neighboring MCBs. 
This idea naturally leads us to MPOs.

\subsection{Matrix product operators}\label{sec3b}

\begin{figure}
\includegraphics[width=0.62\columnwidth]{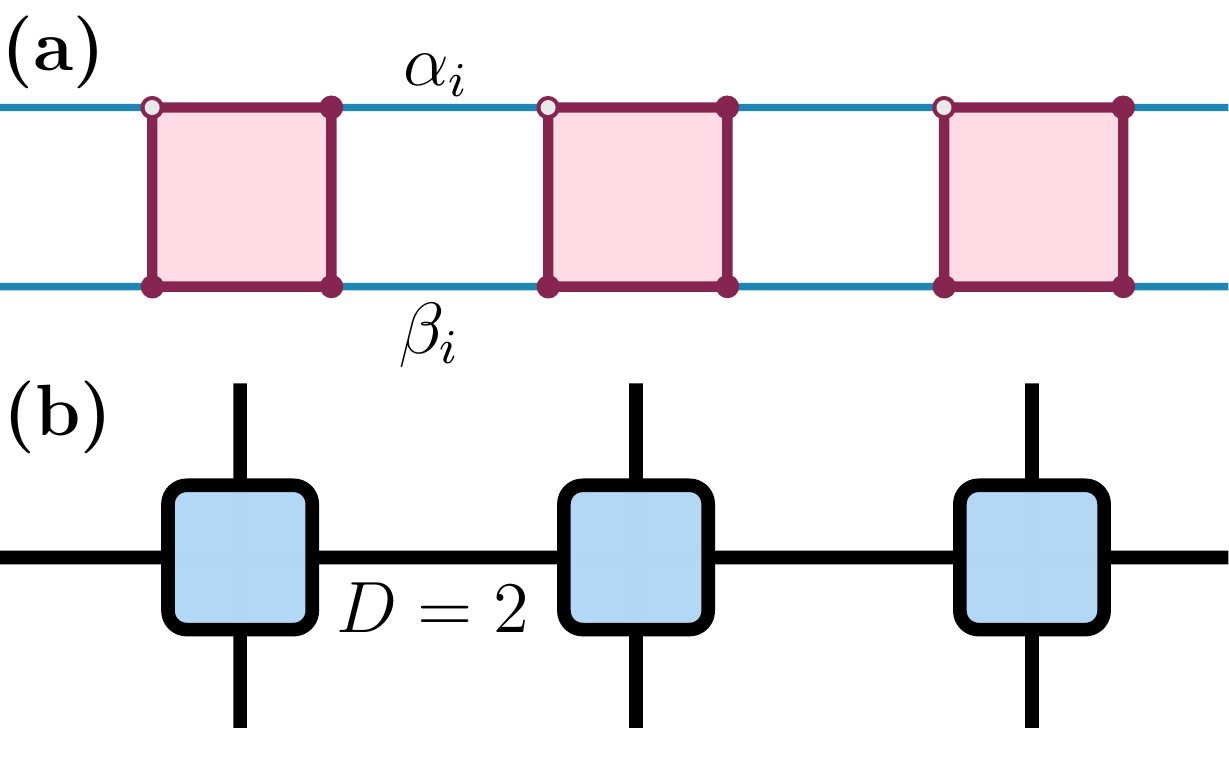}
\caption{Matrix product operators. (a) Tetron ring of length $M=3$.  (b) Tensor network representation for case (a) as a qubit MPO. The bond dimension of this MPO is $D=2$. }
\label{fig:mpo}
\end{figure}

We now consider rings where neighboring MCBs are coupled by tunneling bridges connecting 
at least two MBSs for each MCB. First consider 
 a tetron ring as in Fig.~\ref{fig:mpo}(a). Denoting the tunneling amplitudes of the upper (lower) link by  
 \begin{eqnarray}
 \alpha_i=|\alpha_i|e^{i\phi_{\alpha_i}} \; \left(\beta_i=|\beta_i|e^{i\phi_{\beta_i}}\right)\;, 
 \end{eqnarray}
 we focus on the option to tune the 
loop phases, 
 \begin{eqnarray}
\phi_{\rm loop}^{(i)}=\phi_{\alpha_i}-\phi_{\beta_i}\;.
 \end{eqnarray}
Destructive interference occurs for 
 \begin{eqnarray}
\phi_{\rm loop}^{(i)}=\pm \pi/2 \;,
 \end{eqnarray}
 see Eq.~\eqref{eq:cases}. 
If all $\phi_{\rm loop}^{(i)}$ are tuned to this value, the leading-order contributions to the series expansion \eqref{eq:series} are again loops winding around the ring once. In this case, the effective low energy Hamiltonian representing the structure is given by $\hat H_{\rm eff}=\hat O^{(M)}_+ +$~h.c.,
where the oriented loop operators afford the MPO representation 
\begin{eqnarray}\nonumber
\hat O^{(M)}_+ &=& \frac{-\mathrm i^M}{(2E_C)^{M-1}}\sum_{i_1=0,1} \cdots \sum_{i_M=0,1} \hat A^1_{i_1 i_2} \hat  A^{2}_{i_2 i_3}  \cdots \hat A^{M}_{i_M i_1} \\ 
&=&\frac{-\mathrm i^M}{(2E_C)^{M-1}} {\rm Tr}\left(\hat A^1 \hat A^2 \cdots \hat A^M \right)
\;, \label{mpo}
\end{eqnarray}
and the $\hat A^i$ are Pauli operators acting on MCB $i$ weighted by tunneling amplitudes,
 \begin{eqnarray}   
\hat A^i_{00} &=& -\alpha_i \hat x \;, \quad \hat A^i_{01}= \beta_i \hat y\;, 
\\  \hat A^i_{10}&=&\alpha_i \hat y \;, \quad \hat A^i_{11}=\beta_i \hat x \;. \label{eq:mpo_ex}
\end{eqnarray}
The equivalent qubit tensor network representation is depicted in Fig.~\ref{fig:mpo}(b) where vertical lines indicate the indices $i_k=0,1$ and we used Penrose notation, meaning that index lines without open ends are summed over. Since the index $i_k$ can assume two different values, the bond is said to have bond dimension $D=2$.

The structure of Eq.~(\ref{eq:mpo_ex}) does not suffice to realize arbitrary $D=2$ MPOs. Specifically each of the summands in Eq.~\eqref{mpo} is constrained to contain an even number of  Pauli-$\hat y$ operators. The latter limitations can be addressed by crossing wires. In general one may consider more complex wirings of neighboring tetrons but apart from changes of the local Pauli bases this does not generate additional operator contents. In particular, it is unclear at this stage how MPOs with bond dimension $D>2$ could be realized without generating 2-body interactions from additional length-2 loops. Future research should address how such limitations can be overcome.

\subsection{Repetition code}\label{sec3c}

\begin{figure}
\includegraphics[width=0.9\columnwidth]{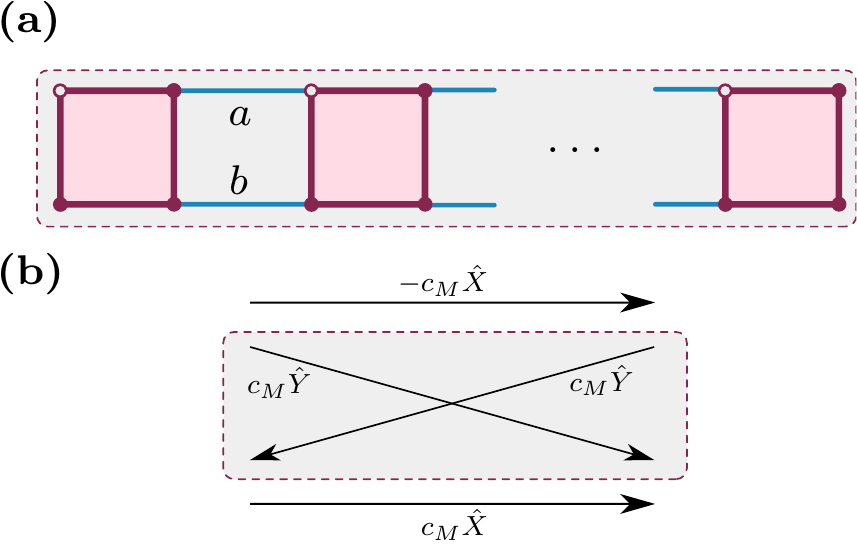}
\caption{Logical qubit in a repetition code. (a) Chain of $M$ tetrons with open boundary conditions. Neighboring tetrons are coupled by tunneling amplitudes $a$ and $b$. (b) Logical Pauli operators $\hat X$ and $\hat Y$ emerge by pumping an electron through the entire chain, see Eq.~(\ref{eq:repoperators}).}
\label{fig:rep}
\end{figure}

We next turn to another useful building block.  Consider an open chain of $M$ tetrons, where neighboring tetrons are coupled by two tunneling bridges with amplitudes $a$ and $b$ (assumed identical for all MCB-MCB contacts), 
see Fig.~\ref{fig:rep}(a). 
To leading order, $\hat H_{\rm eff}$ is determined by summing over all length-2 loops,
$\hat H_{\rm eff}\simeq \hat H^{(2)}$. The result describes an Ising spin chain, 
\begin{equation}
\hat H_{\rm eff}= J \sum_{i=1}^{M-1}  \hat z_i \hat z_{i+1} \;,\quad J=\frac{\textrm{Re}(a b^*)}{2 E_C}\;.\label{eq:repHe}
\end{equation}
 Tuning all elementary loop phases, $\phi_{\rm loop}=\phi_a-\phi_b$, such that 
$\pi/2<\phi_{\rm loop}<3\pi/2$, we have ferromagnetic couplings.  
The ground state space of the $M$-qubit chain is then two-fold degenerate and 
we can encode a logical qubit in this \emph{repetition code} \cite{RevModPhys.87.307}. 
Interestingly, this logical qubit may be operated just like a single tetron-based qubit but with enhanced error resilience. 

To that end, we consider processes where a single electron is pumped through the entire MCB chain. (The practical realization of such a process has been discussed, e.g., in Ref.~\cite{PhysRevB.94.174514}.)  
We assume that the electron enters the left end of the chain by tunneling in 
via the MBS located at the top ($j=0$) or bottom ($j=1$) left corner of the leftmost MCB.
After propagating to the other end, it tunnels out of the chain via 
the MBS corresponding to the top ($j'=0$) or bottom ($j'=1$) right corner 
of the rightmost MCB.  The coherent multi-step tunneling process effectively
applies a `string operator' $\hat S_{jj'}$ to the $M$-qubit state, cf.~Ref.~\cite{PhysRevB.94.174514}, where $\hat S_{jj'}$
is a superposition of Pauli product operators. 
We now show that the string operators $\hat S_{jj'}$, when projected to 
the ground state space of $\hat H_{\rm eff}$, act like 
logical $\hat X$ and $\hat Y$ operators, 
as indicated in Fig.~\ref{fig:rep}(b).

We first consider the case $M=2$ with $j=j'=0$. 
The corresponding string operator is given by
\begin{equation}
\hat S_{00}= \frac{-1}{2E_C} \left( a \hat x_1 \hat x_2 + b \hat y_1 \hat y_2\right)\;.
\end{equation}
Using $\hat y= \mathrm i \hat x \hat z$ and $\hat z_1 
\hat z_2=\id_2$ (which holds within the ground-state sector), we obtain 
\begin{equation}
\hat S_{00}= -\frac{a-b}{2E_C} \hat x_1 \hat x_2,
\end{equation} 
which is proportional to the logical Pauli-$\hat X$ operator. 
Generalizing this argument now to arbitrary $M$, 
logical Pauli operators are defined as
$\hat X = \hat x_1 \hat x_2 \cdots \hat x_{M-1}\hat x_M$ and
$\hat Y = \hat x_1 \cdots \hat x_{M-1} \hat y_M$.
We then find   
\begin{equation}\label{eq:repoperators}
\hat S_{00}=- \hat S_{11} = -c_M \hat X \;, \quad
\hat S_{01}= \hat S_{10}=c_M \hat Y \;.
\end{equation}
Apart from the prefactor, $c_M=(-\mathrm i)^M [(a-b)/(2E_C)]^{M-1}$, this result reproduces the mapping of Majorana bilinears to Pauli operators for a single tetron, cf.~Eq.~\eqref{eq:tetronqubit}. 

\subsection{Bell states}\label{sec3d}

\begin{figure}
\includegraphics[width=0.66\columnwidth]{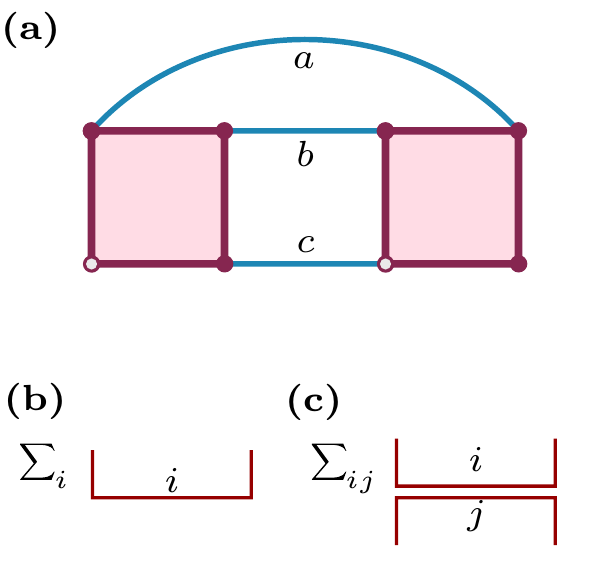}
\caption{Bell pairs. (a) Coupled tetron device. The ground state of the associated effective Hamiltonian \eqref{eq:bell} is a Bell state
for real positive $a=b=c$. (b) Tensor network representation of $\ket{\text{Bell}}=\ket{0,0}+\ket{1,1}$. (c) Tensor network representation of the Bell pair projector $\hat P_{\rm Bell}=\ket{\text{Bell}}\bra{\text{Bell}}$.}
\label{fig:bell}
\end{figure}

We finally show how Bell pair states can be prepared as ground states of tetron structures with tunneling bridges as indicated in Fig.~\ref{fig:bell}(a). 
Bell states are pairs of maximally entangled qubits, e.g., $\ket{\text{Bell}}=\ket{0,0}+\ket{1,1}$. They 
are key to tensor network constructs
like matrix product states (MPS) and their 2D generalizations, PEPS, see Fig.~\ref{fig:bell}(b).
With the projector
$\hat P_{\rm Bell} = \ket{\text{Bell}}\bra{\text{Bell}}$, see Fig.~\ref{fig:bell}(c),
the Hamiltonian 
\begin{equation}\label{eq:bellpair}
\hat H_{\rm Bell}= \varepsilon (1-\hat P_{\rm Bell})\;,
\end{equation}
 $\varepsilon>0$, 
has $\ket{\text{Bell}}$ as its unique ground state.
This Hamiltonian can be realized with two coupled tetrons where a possible setup is shown in Fig.~\ref{fig:bell}(a). We assume that gate calibration has been applied to tune the tunneling amplitudes to real positive values  $a=b=c>0$. In this case, the low-energy Hamiltonian obtained by summation over all length-2 loops reads
\begin{equation}
 \hat H_{\rm eff}= \frac{a^2}{2E_C} 
\left(  - \hat x_1 \hat x_2 - \hat z_1 \hat z_2+ \hat y_1  \hat y_2 \right) \;.  \label{eq:bell}
\end{equation}
This Hamiltonian has $\ket{\text{Bell}}$ as ground state and, for sufficiently strong coupling $\varepsilon=a^2/2E_C$, effectively projects onto this state.

\subsection{Synthesizing design structures}
\label{sec:SynthDesign}

\begin{figure}
\includegraphics[width=0.66\columnwidth]{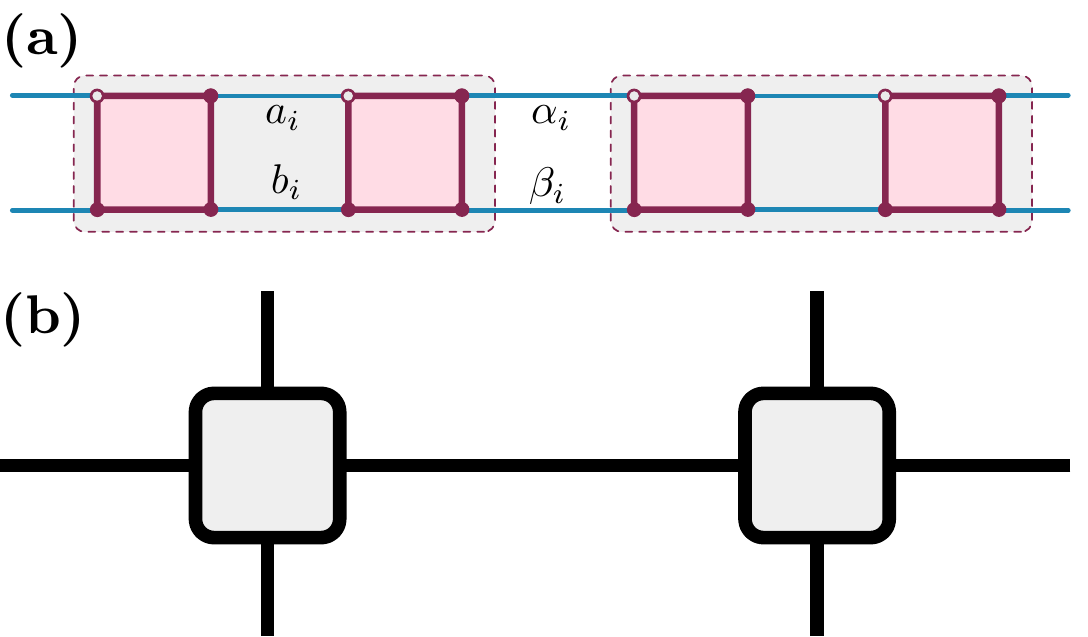}
\caption{(a) A hierarchical structure in which units of two tetrons are first linked by tunneling amplitudes $a_i\simeq -b_i$ to define 
repetition qubits. These blocks are then coupled by amplitudes $\alpha_i, \beta_i$ to an MPO depicted as tensor network in (b).}
\label{fig:RepetitionMPS}
\end{figure}

Since the above structures all make reference to sequential ordering, chains of alternating coupling types may be used to define structures containing several types of functionality. As an example relevant to our construction below we mention MPOs whose base units are repetition code qubits. These are formed  (see Fig.~\ref{fig:RepetitionMPS}) by first  linking blocks of two tetrons each via tunneling amplitudes $\mathrm{Re}(a_ib_i^\ast)<0$. This ferromagnetic coupling defines low energy repetition code qubits on the two tetron compounds. These units may then be coupled by amplitudes $\alpha_i,\beta_i$ to form an MPO structure whose operator units act in the Hilbert spaces of the repetition code qubits.  

\section{Simulating topological tensor networks}\label{sec4}

The MCB networks discussed in Secs.~\ref{sec2} and \ref{sec3}
allow one to realize topological phases with strong entanglement. 
While previous work has shown that Kitaev's toric code can be 
simulated with such constructions \cite{PhysRevLett.116.050501,PhysRevB.94.174514,PhysRevLett.108.260504,PhysRevX.6.031016}, 
it has so far remained open how 
to realize more complicated string-net models.  Using the PEPS tensor network representation for the ground states of string-nets, we 
here discuss how the simplest case beyond Kitaev's toric code,  the \emph{double semion model} \cite{Levin2005,Buerschaper2014,Gu2009}, can be implemented in a 2D network of MCBs.  For pedagogical introductions to tensor networks and PEPS we refer to one of several available reviews, see, 
e.g., Ref.~\cite{Orus-AnnPhys-2014}. While this will provide useful background knowledge, familiarity with these concepts is not essential throughout as all required material is introduced in a self contained manner.
We are confident that by using similar strategies, one could also 
realize more complicated string-nets such as the Fibonacci Levin-Wen model \cite{Levin2005} where, in particular, branching is allowed,
and which leads to schemes of  universal topological quantum computing. 
Our approach relates MPOs to MCB networks where destructive interference mechanisms are exploited to suppress short loop contributions.  The latter, if present, would drive the system into a topologically trivial phase.

To set the stage, we review the basic properties of topological tensor networks from a string-net perspective in Sec.~\ref{sec4a}.  
The Hamiltonian design builds upon seminal work on \emph{Hamiltonian gadgets} \cite{Brell2014PEPS} which 
proposed a perturbative approach to topological tensor networks on the abstract level of qubits. The PEPS tensor network used in such a construction will be discussed in Sec.~\ref{sec4b}. Finally, the MCB network implementation of the PEPS tensor network realizing 
the double semion ground state will be presented in Sec.~\ref{sec4c}.

\subsection{PEPS representation of string-net ground states}\label{sec4a}

String nets have been introduced in Ref.~\cite{Levin2005} as generalizations of Kitaev's toric code and quantum doubles~\cite{Kitaev-AnnPhys-2003}, see also Refs.~\cite{Fidkowski2009,Bonesteel2012,Wen2017,Fendley2013}. While the physical idea behind string-nets is relatively easy to communicate in textual or graphical ways,  quantitative formulations via formulae tend to be cumbersome. The same is true for the representation of string-net ground states as tensor network states. We therefore begin our discussion of the double semion ground states and its simulation in an MCB network with a qualitative discussion of the main principles along the lines of Refs.\ \cite{Buerschaper2009,Gu2009}. 
The language is geared to readers with a background in condensed matter physics for which the tensor network approach to topological phases may be less familiar. 
Throughout, we dispense with technical rigor in exchange for brevity and transparency. For the sake of clarity,  parts of our discussion below are formulated in a general language, before specializing to the double semion case.

\begin{figure}
\includegraphics[width=0.7\columnwidth]{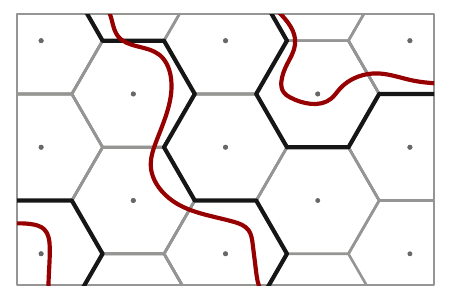}
\caption{Black  lines: configuration of a non-branching colorless string-net on a trivalent lattice. Red lines: equivalent configuration on the corresponding fat lattice. }
\label{fig:TrivalentLattice}
\end{figure}

\emph{String net definition ---} The basis states of a string-net are coverings of a lattice, often chosen as trivalent for convenience, cf.~Fig.~\ref{fig:TrivalentLattice}. Complex string-nets allow for coverings carrying $N$ internal indices (`colors') $i$,  and senses of directions, $i$ vs. $i^\ast$. However, the simple representatives considered here are un-directed and colorless, implying that a covered link may be identified by the label $'1'$, while an empty or vacuum link is identified as $'0'$. We also exclude branching configurations so that the states of the system assume a form as indicated by the pattern of black lines in the figure. For later reference, we denote the linear spaces spanned by $N$ color indices plus vacuum by $\tilde V\simeq \mathbb{C}^{N+1}$.

The physically relevant wave functions, $\Psi$, defined over these sets of basis states are required to satisfy certain equivalence relations. Referring for a full list of five equivalences to Ref.~\cite{Levin2005}, we note that wave functions are to be invariant under topology-preserving deformations (no crossing or tearing) of lines in the net. The inclusion of a closed, simply contractible loop is equivalent to the multiplication of the wave function by a factor $d_i$, which for a general net depends on the color $i$, of the included loop and defines the quantum dimension of the included link species. However, the  most important rule describes what happens under local topology changing re-connections of the net. For a general multi-color string-net with orientation, it states that   
\begin{equation}
\Psi \left(\adjincludegraphics[width=1cm,valign=M]{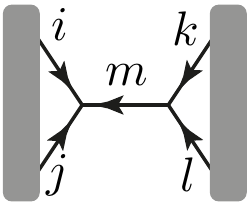}\right)=\sum_n F^{ijm}_{lkn} \Psi  
\left(\adjincludegraphics[width=1cm,valign=M]{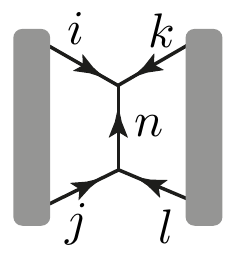}\right)\;,
\end{equation}
where the $F$-\emph{symbols} are scalar coefficients defining the permissible equivalence reconnections of the net~\cite{Levin2005}.
 
 A non-trivial consistent solution $\{F^{ijm}_{lkn},d_i\}$ of all consistency relations defines a topological phase. In the present, colorless, non-oriented, non-branching case, there exist only two solutions, the Kitaev toric code, $d_1=1$, and the double-semion phase, $d_1=-1$. Since the nets are non-branching, the $F$-symbols are defined through just one non-trivial reconnection rule,  
\begin{equation}
\Psi 
\left(\adjincludegraphics[width=0.7cm,valign=M]{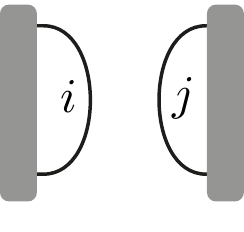}\right)=
F^{110}_{110} \; \Psi  \left(\adjincludegraphics[width=0.7cm,valign=M]
{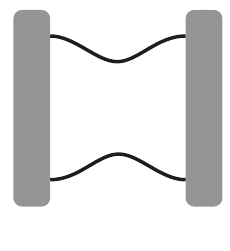}\right)
\end{equation}
with $F^{110}_{110}=\pm 1$, respectively. All other $F$-symbols describing permissible non-branching re-connections (for example, $F^{000}_{111}$) assume the trivial value unity.

From a condensed matter perspective, it may be most natural to describe a string-net in terms of an effective ('fixed point') Hamiltonian 
\begin{equation}
\hat H =-\sum_v \hat Q_v - \sum_p \hat B_p, 
\end{equation}
whose eigenfunctions satisfy the above equivalence relations. Here, $\hat Q_v$ is a projector onto the permissible configurations at each vertex $v$, i.e., a projector enforcing total even spin $0,1$ of the adjacent legs in an identification $(0,1)\leftrightarrow (-1/2,+1/2)$, and $\hat B_p$ is an operator specific to the plaquette, $p$, giving the net dynamics. While the explicit description of $\hat B_p$ in terms of $F$-symbols (the product of six symbols depending on the twelve link states of the plaquette and it adjacent legs) or Pauli spin operators (the tensor product of 12 Pauli operators) is both cumbersome and non-intuitive, a much more intuitive description engages the concept of loop insertion on the 'fat lattice'. In fact, the PEPS construction below and its hardware implementation are closer in spirit to this latter formulation than to the Hamiltonian approach.

\emph{Fat lattice and PEPS string-net representation ---} To obtain the fat lattice, consider the width of the 'physical' honeycomb lattice links in Fig.~\ref{fig:TrivalentLattice} enhanced until the entire plane is covered except for the plaquette center points. Equivalently, it is a planar structure into which holes are drilled at the plaquette centers. A string-net configuration can now be represented in more relaxed ways, as indicated by the red lines. In our tensor network constructions below, state indices $a,b,\dots =0,\dots, N$, carried by lines on the fat lattice, will assume the role of 'virtual indices' and we denote the  space of these indices by $V\simeq \mathbb{C}^{N+1}$. The distinction from the space $\tilde V$ of indices on the physical lattice, $i,j,\dots =0,\dots,N$, is purely syntactic and introduced for conceptual clarity; physically, there is no difference between lines on the physical or fat lattice. 

\begin{figure*}
\includegraphics[width=14cm]{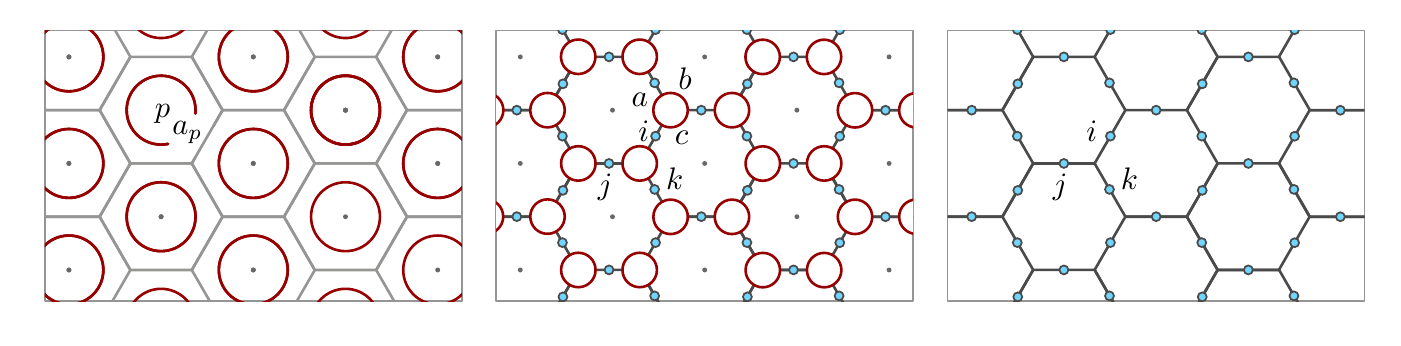}
\caption{Left: Ground state of a string-net Hamiltonian as superposition of fat lattice loop insertions. Middle: A hybrid fat/physical lattice representation obtained after a combination of $F$-moves applied at each lattice center. Right: Full reduction to a physical lattice configuration after reduction of the loops by further $F$-moves.}
\label{fig:PEPSGroundState}
\end{figure*}

The advantage of this reformulation is that it allows for a more flexible representation of configuration rearrangements via $F$-moves. Specifically, the fat lattice affords an intuitive definition of the string-net Hamiltonian. To this end, we note that the insertion of a full set of non-vacuum closed loops $\hat B_p^a$ ($a=1,\dots,N$) around a hole $p$ in the fat lattice, 
\begin{equation}
\hat B_p:= \frac{1}{\mathcal{D}^2} \sum_a d_a \hat B_p^a
\end{equation}
 with $\mathcal{D}^2={\sum_{a=1}^N d_a^2}$, is a projective operation~\cite{Levin2005}. The sum of all these operations \emph{defines} the second term, $-\sum_p \hat B_p$, of the string-net Hamiltonian. An operation of three sequential $F$-moves may then be applied to represent the Hamiltonian entirely through its action on basis states on the physical lattice~\cite{Levin2005}. The latter are  defined through a configuration $i_l$ of states $i=0,\dots, N$ specific to lattice links $l$, and represented in this way the loop insertion assumes the form of a product of six $F$-tensors acting on the link states of the plaquettes surrounding individual loop insertion points. (Following a standard convention we label string types on the fat/physical lattice by $a/i=0,\dots,N$. This distinction will become useful below when these indices correspond to virtual/physical indices of the tensor network framework.)

\begin{figure}
\includegraphics[width=1\columnwidth]{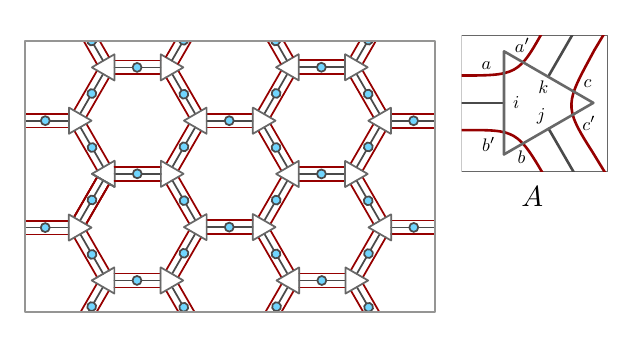}
\caption{Representation of string-net ground state as a tensor network. Vertices of the net carry a tensor $A$. }
\label{fig:ATensorDef}
\end{figure}

In the tensor network description, it is more natural to focus on the Hamiltonian's ground states rather than on the Hamiltonian itself. A ground state must be invariant under the projective action of the Hamiltonian. The projector property implies that, starting from a loopless fat lattice vacuum 
state $|0\rangle$, a ground state is obtained as $|\mathrm{GS}\rangle =\prod_p \hat B_p |0\rangle$, i.e., as an equal weight linear combination of all possible fat lattice elementary loop insertions, see Fig.~\ref{fig:PEPSGroundState}, left. Once again, an operation of sequential $F$-moves may be applied to transform the fat lattice ground state to an equivalent one defined entirely on the physical lattice \cite{Buerschaper2009,Gu2009}: in a first step, three $F$-moves specific to each vertex are applied to turn the configuration to the hybrid shown in Fig.~\ref{fig:PEPSGroundState}, center, where the blue dots on the links indicate that the physical lattice is now carrying index structure. This is followed by two more $F$-moves removing all links in the fat lattice and reducing the state to one on the physical lattice. In this final stage, the state assumes the symbolic form $|\mathrm{GS}\rangle=\sum_{\{i\}} C_{\{i\}}|\{i\}\rangle$, where $\{i\}$ is a basis configuration on the physical lattice specified by a set of indices $i_l$, and the coefficients $C_{\{i\}}$ contain an internal summation over configurations $\{a\}$ originally inserted on the fat lattice. By definition, this makes $C_{\{i\}}$ a tensor network with physical indices $\{i\}$ and virtual indices $\{a\}$. The algebraic representation of $C_{\{i\}}$ for a generic string-net is complicated and contains the  $F$-symbols representing the tensorial structure of individual vertices. Individual of these tensors, represented as triangles in Fig.~\ref{fig:ATensorDef}, are maps $A^{(ijk)}:V\otimes V\otimes V \mapsto V\otimes V\otimes V$ characterized by tensor components $A^{(ijk)}_{abc,a'b'c'}$. (It may be worth repeating that the only distinction between 'physical' ($i,j,k$) and 'virtual' $(a,b,c)$ indices is that the former/latter refer to string states on the physical/fat lattice.) Individual tensor components are defined by $F$-symbols, where  details depend on which sublattice ($1$, corner triangle right, or $2$, corner triangle left) the tensor lives \cite{Buerschaper2009,Gu2009}:

\begin{align*}
1:&\qquad A^{(ijk)}_{abc,a'b'c'}=F^{a^\ast ib}_{j^\ast c k}\delta_{a,a'}
\delta_{b,b'}\delta_{c,c'}\;,\cr
2:&\qquad A^{(ijk)}_{abc,a'b'c'}=\sqrt{d_j}F^{b^\ast jc}_{kai}\delta_{a,a'}\delta_{b,b'}\delta_{c,c'}\;.
\end{align*}
Note the presence of the virtual space Kronecker-$\delta$'s which motivates a representation in which the red virtual lines penetrate the tensorial structure. 

The representation simplifies further in the case of colorless non-branching nets. The absence of branching reconnections implies a locking between virtual and physical indices, and the configuration $(a,b,c)$ determines that of $(i,j,k)$. Specifically, in the double semion system the $A$-tensors of both sublattices are defined as \cite{Gu2009}
\begin{align}
\label{eq:ATensorLWS}
A^{(ijk)}_{abc,a'b'c'}&=A_{abc}\delta_{a,a'}\delta_{b,b'}\delta_{c,c'}\;,\cr
&A_{abc}=
\left\{\begin{array}{ll}
1\;,&\qquad a+b+c=0,3\;,\cr 
\mathrm i \;,&\qquad a+b+c=1\;,\cr 
-\mathrm i\;,&\qquad a+b+c=2\;.
\end{array}\right.
\end{align}
Implicit to this equation is a locking $i=1$ if $(a,b')$ have odd parity [$(0,1)$ or $(1,0)$] and $0$ else. This replacement rule affords an intuitive interpretation \cite{Gu2009}: the ground state of the double semion system is a superposition of all closed loops on the physical lattice, where the coefficient of individual terms is given by the parity $(-)^{\text{no. of loops}}$. The above virtual/physical index locking implies that physical lattice loops are \emph{domain walls} separating hexagons surrounded by virtual loops from those without. The assignment of $\pm \mathrm i$ and $1$ to different virtual index configurations makes sure that each closed domain wall/loop carries the appropriate sign factor.

This concludes our discussion of the PEPS representation of string-nets. In the next subsection we will explore how the above effective mapping of the description from physical to virtual space  provides the key to efficient hardware blueprints simulating string-net ground states.

\subsection{Encoded projected entangled pair states}\label{sec4b}

As outlined in the previous subsection, string-net models are  naturally described using tensor networks. An attractive feature of this representation is that each of the tensors $A$ contains the full information on the system's topological states. At the same time they define a passage from the space of physical indices into the larger space of virtual indices. The advantage gained in exchange for this redundant encoding of information in a larger space is the option of a more local and hardware friendly description of the system. In this subsection, we show how the translation to an encoded description in virtual space is achieved in concrete ways. And in the next, how it is realized as a concrete MCB hardware layout.

\begin{figure*}[t]
\includegraphics[width=2\columnwidth]{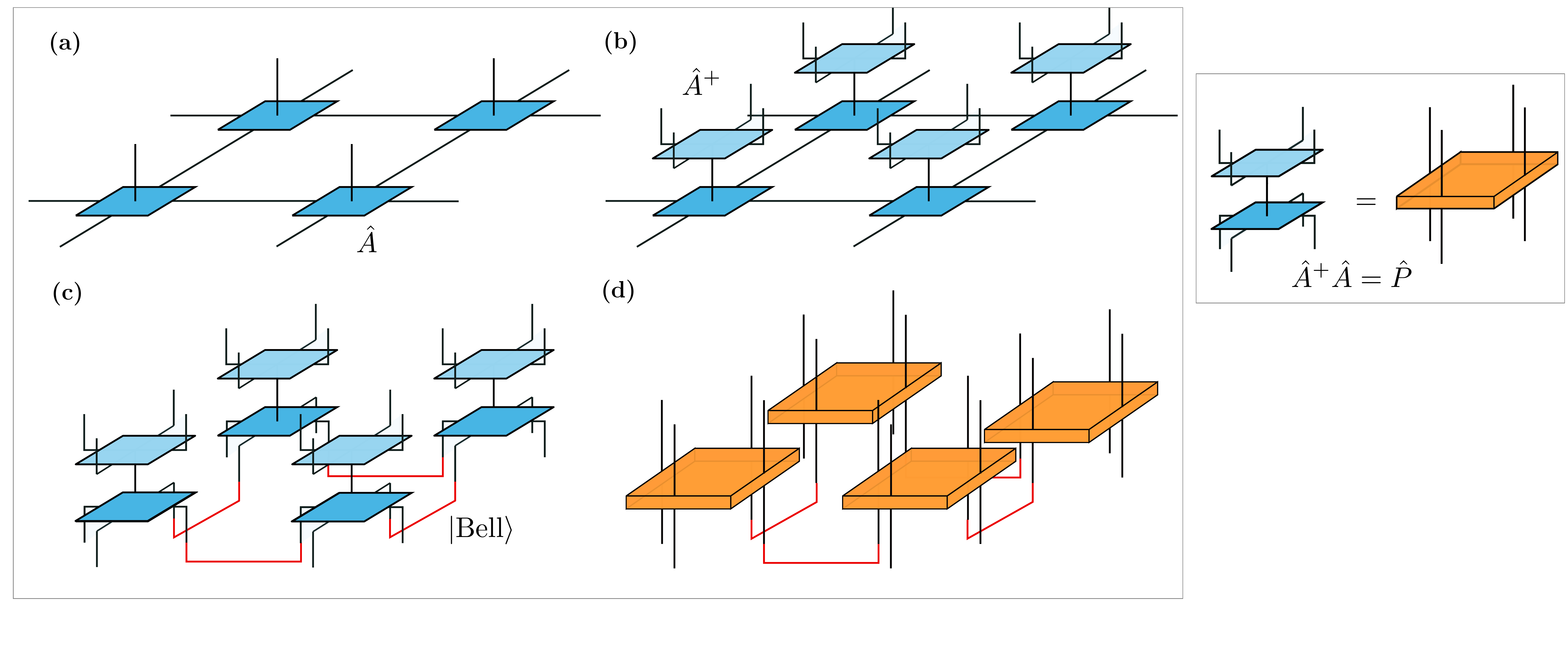}
\caption{
Topological PEPS on a square lattice. (a) PEPS built from local tensors $A$. (b) Encoded PEPS obtained by acting with the pseudo-inverse $\hat A^+$ at every physical site, cf.~Eq.~(\ref{eq:encode}). (c) Encoded PEPS where the deformation of the virtual index lines emphasizes  the local presence of Bell pairs shown in red, cf.~Fig.~\ref{fig:bell}(b). 
(d) Encoded PEPS where $\hat A^+ \hat A$ has been replaced by the projector $\hat P$ (panel to the right) at every site. }
\label{fig:encode}
\end{figure*}

The tensors $A$ define maps, 
\begin{equation}
\hat A = \sum_{a'b'c',abc,ijk} \ket{i,j,k} A^{(ijk)}_{abc,a'b'c'} \bra{a,a',b,b',c,c'}\;,
\end{equation}
between the $2^6$-dimensional virtual space, $\otimes^6 V$, and the $2^3$-dimensional physical space, $\otimes^3 \tilde V$. Due to the different dimensionality of the spaces,  they are neither injective nor invertible. However, to each map $\hat A$ there exists a pseudo-inverse $\hat A^+$ defined by the condition 
\begin{equation}
\hat A^+ \hat A = \hat P \; : \; \otimes^6 V\mapsto \otimes^6 V \;, \label{eq:AAP}
\end{equation}
where $\hat P$ is a projector onto a virtual subspace which is in one-to-one correspondence to the physical space. We will refer to this space as local \emph{code space}. Important properties of this map include $\hat A^+ \hat A \hat A^+ =\hat A^+$, meaning that states in the image of $\hat A^+$ are invariant under application of the projector, and similarly, $\hat A\hat A^+ \hat A=\hat A$. For a more detailed discussion on the properties of \emph{topological PEPS}, see Refs.\ \cite{PEPSTopology,Buerschaper-AnnPhys-2014,1409.2150,Bultinck2017}.

Application of the pseudo-inverse to every physical site of the PEPS ground state yields a state 
\begin{equation}
\ket{\Psi'}=\hat A^+ \otimes \ldots \otimes \hat A^+ \ket{\Psi} \label{eq:encode}
\end{equation}
defined in the larger virtual space. The physical information is now encoded and, following Ref.~\cite{Brell2014PEPS}, we call $\ket{\Psi'}$ the \emph{encoded PEPS}.
The motivation behind Eq.~\eqref{eq:encode}  is that the encoded state will be easier to realize in a physical system. We emphasize again that the un-encoded string-net PEPS, $\ket{\Psi}$, is the ground state of a 12-body Hamiltonian which is extremely difficult to realize in an actual physical system. In contrast the encoded PEPS, $\ket{\Psi'}$, can be obtained perturbatively from a comparatively simple Hamiltonian.

To understand the state $\ket{\Psi'}$ and the alluded Hamiltonian, consider the diagrammatic representation in Fig.~\ref{fig:encode} where we show a square lattice for better visibility. Following standard tensor network notation, horizontal (perpendicular) lines in panel (a) represent contracted virtual (uncontracted physical) indices. Each square indicates the local presence of $\hat A$. Now contract each $\hat A$ with its 
pseudo-inverse, see panel (b). The local building blocks now define 
the projectors $\hat P$ (see panel on the right), and this leads to the representation in  panels (c) and (d). The visualization emphasizes that this state literally is a PEPS, i.e., a state obtained by the local action of projectors $\hat P=\hat A^+ \hat A$ on an assembly of maximally entangled Bell pairs defined on the links of the lattice. 

The entire construction has now been shifted to virtual space. Due to the projective nature of $\hat P=\hat A^+ \hat A$, the encoded state is a ground state of the Hamiltonian
\begin{equation}
\hat H'=\sum_v (1-\hat P) + \varepsilon \sum_e (1-\hat P_\mathrm{Bell}) \label{eq:H} \;,
\end{equation}
where  $v$ runs over the vertices and $e$ over the edges of the underlying lattice, cf.~Eq.~\eqref{eq:bellpair} with $\varepsilon>0$. This Hamiltonian is referred to as the \emph{perturbative parent Hamiltonian}. The first summands ensure that the low energy states lie within the code space and thus can be mapped back to the original physical state $\ket{\Psi}$. \cw{The Bell pair projections act as a perturbation within this (highly degenerate) ground state and effectively reassemble encoded versions of the original string-net Hamiltonian order by order in a 
%(Schrieffer-Wolff type) 
perturbation series expansion.} For further details we refer to Ref.~\cite{Brell2014PEPS}. For an intuitive relation between the original string-net Hamiltonian and the perturbative PEPS parent, we note that the perturbative expansion in the Bell pair projectors, $\hat P_\mathrm{Bell}$, contains  ring exchange processes depicted in Fig.~\ref{fig:encode}(d). Much like the toric code ground state can be obtained via the action of all plaquette operators on a vacuum state, the ground state of $\hat H'$ is obtained by the action of the ring exchange operators on all virtual lattice loops.

Regarding a hardware design realizing a topological ground state our problem is thus reduced to that of understanding the local action of $\hat A^+ \hat A$ and obtaining a good hardware representation of these operators locally. For the concrete case of the double semion model, the $A$ tensor as defined by Eq.~\eqref{eq:ATensorLWS} is given by
\begin{equation}
\hat A = \sum_{a,b,c} \ket{a \oplus  b, b \oplus  c, c \oplus  a} A_{abc} \bra{a,a,b,b,c,c}\;,
\end{equation}
where the addition modulo two, $\oplus$, determines physical indices as required by Eq.~\eqref{eq:ATensorLWS}, $i=a\oplus b$, and so on. 
In this case, the pseudo-inverse has a particularly simple form, $\hat A^+=\frac12 \hat A^\dagger$, where an explicit representation is given by
 \begin{equation}\label{eq:pseudoinv}
\hat A^+= \frac{1}{2}\sum_{i,j,a} \ket{a, a, a \oplus i , a \oplus i, a \oplus j, a \oplus j } A^\ast_{a(a\oplus i)(a\oplus j)} \bra{j,i\oplus j,i} \;.
\end{equation}
Equation  \eqref{eq:pseudoinv} 
 states that the pseudo-inverse will map a general physical state of the system, $\ket{j,i\oplus j,i}$, back to a superposition of virtual states subject to the condition that they (i) have pairwise even parity at the corners and (ii) the parity between states at different corners is determined by the physical state of their edge. This leaves  only one free summation index, $a$, while all others are fixed as indicated.

By a straightforward calculation, we obtain the state projectors as 
\begin{equation}
\hat P =\frac{1}{2} (\hat P_c +\hat X)\;,
\end{equation}
where $\hat P_c=\sum_{a,b,c} \ket{a,a,b,b,c,c} \bra{a,a,b,b,c,c}$
projects from the $2^6$-dimensional space $\otimes^6 V$ of general states, $\ket{a,a',b,b',c,c'}$, onto the $2^3$-dimensional space of 
pairwise even-parity states $\simeq \otimes^3 V$. 
The second operator,
\begin{align}
\hat X=\sum_{a,b,c} &\ket{a\oplus 1,a\oplus 1,b\oplus 1,b\oplus 1,c\oplus 1,c \oplus 1} X_{abc} \nonumber \\ &\bra{a,a,b,b,c,c}
\label{eq:xdef}
\end{align}
likewise acts in $\otimes^3V$ where it flips all states and assigns a sign factor 
\begin{equation}\label{eq:xdef1b}
X_{abc} = \begin{cases} -1\;, \qquad & a+b+c=1,2\;, \\ 1\;, \qquad  &\text{otherwise}\;. 
\end{cases}
\end{equation}
The operator $\hat X$ in Eq.~\eqref{eq:xdef} can alternatively be represented as MPO with bond dimension $D=2$,
\begin{equation}\label{eq:xdef2}
\hat X = - \frac{1}{32} \sum_{i,j,k} \hat A_{ij} \otimes \hat A_{jk} \otimes \hat A_{ki}\;,
\end{equation}
where individual factors, 
\begin{align}\nonumber
\hat A_{00}=& \hat A_{11} = \hat x \otimes \hat x -\hat y \otimes \hat y\;, \\ \label{eq:logpauli}
\hat A_{01}=& \hat A_{10}=\hat x \otimes \hat y + \hat y \otimes \hat x \; ,
\end{align}
act on a qubit pair carried by each of the three corners
of the triangular vertex in Fig.~\ref{fig:ATensorDef}. In passing we note that the representation of the site-local projector $\hat A^+ \hat A$ via a local MPO ring contraction reflects the idea of 'MPO injectivity' introduced in Refs.~\cite{Buerschaper-AnnPhys-2014,1409.2150}.
The advantage of this representation is that the operators $\hat A_{ij}$ act like logical Pauli operators on corner qubit pairs, effectively implementing the logical qubit of a repetition code, see Sec.~\ref{sec3c}.

Since $\hat P_c$ acts as an identity operator within the parity subspace $\otimes^3 V$, the essential information on $\hat P$ is carried by $\hat X$. Specifically, we know that the local ground space of the operators $1-\hat P$ in the parent Hamiltonian in Eq.~\eqref{eq:H} coincides with that of $-\hat X$. Our objective in the next subsection will thus be to engineer an MCB network whose ground space equals that of $-\hat X$. 

\subsection{Double semion MCB network}\label{sec4c}

\begin{figure}
\includegraphics[width=0.5\columnwidth]{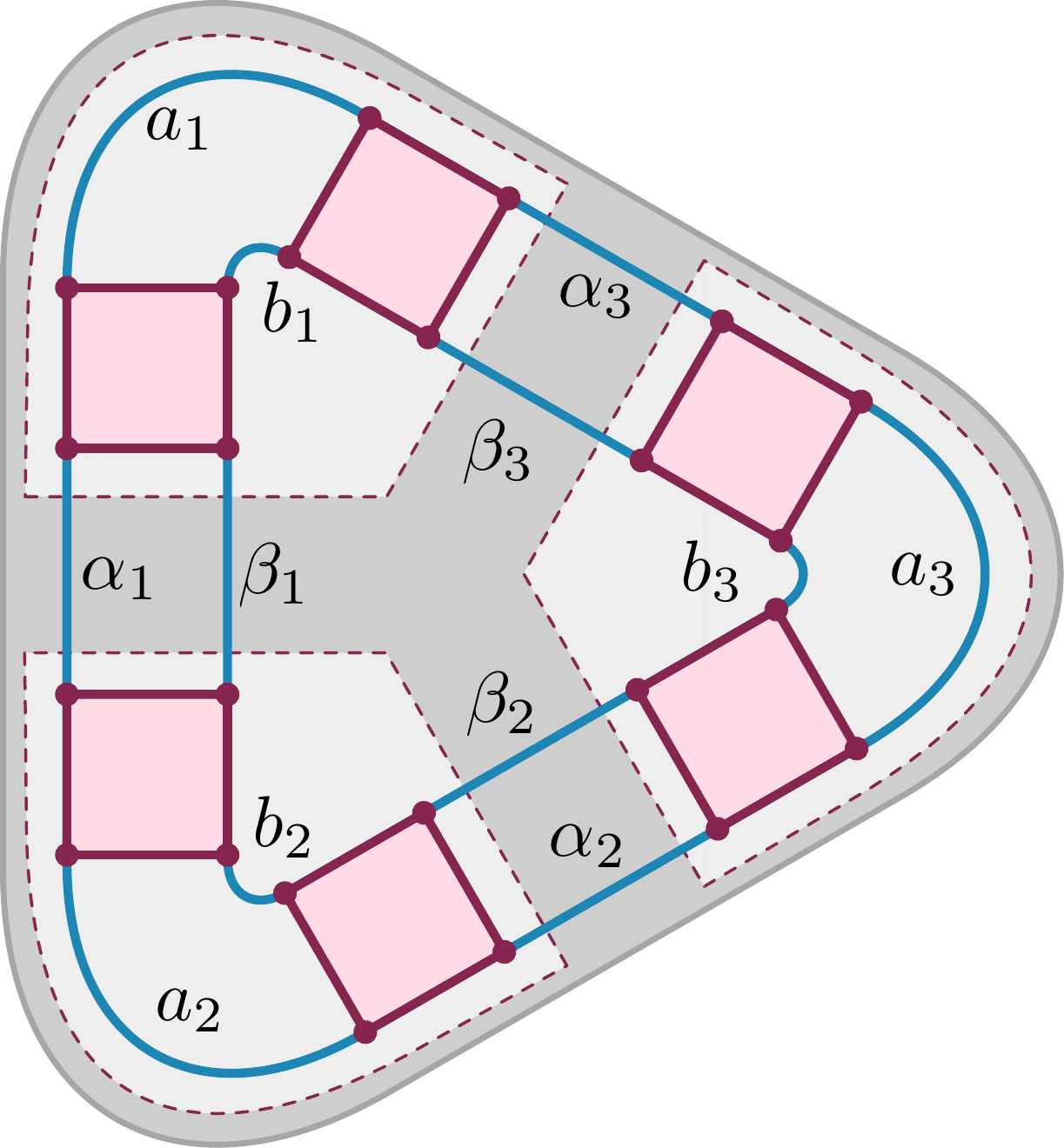}
\caption{Tetron ring (`triangle') used for implementing the effective vertex Hamiltonian, $\hat H_v$.}
\label{fig:DSMPO}
\end{figure}

\begin{figure}
\includegraphics[width=\columnwidth]{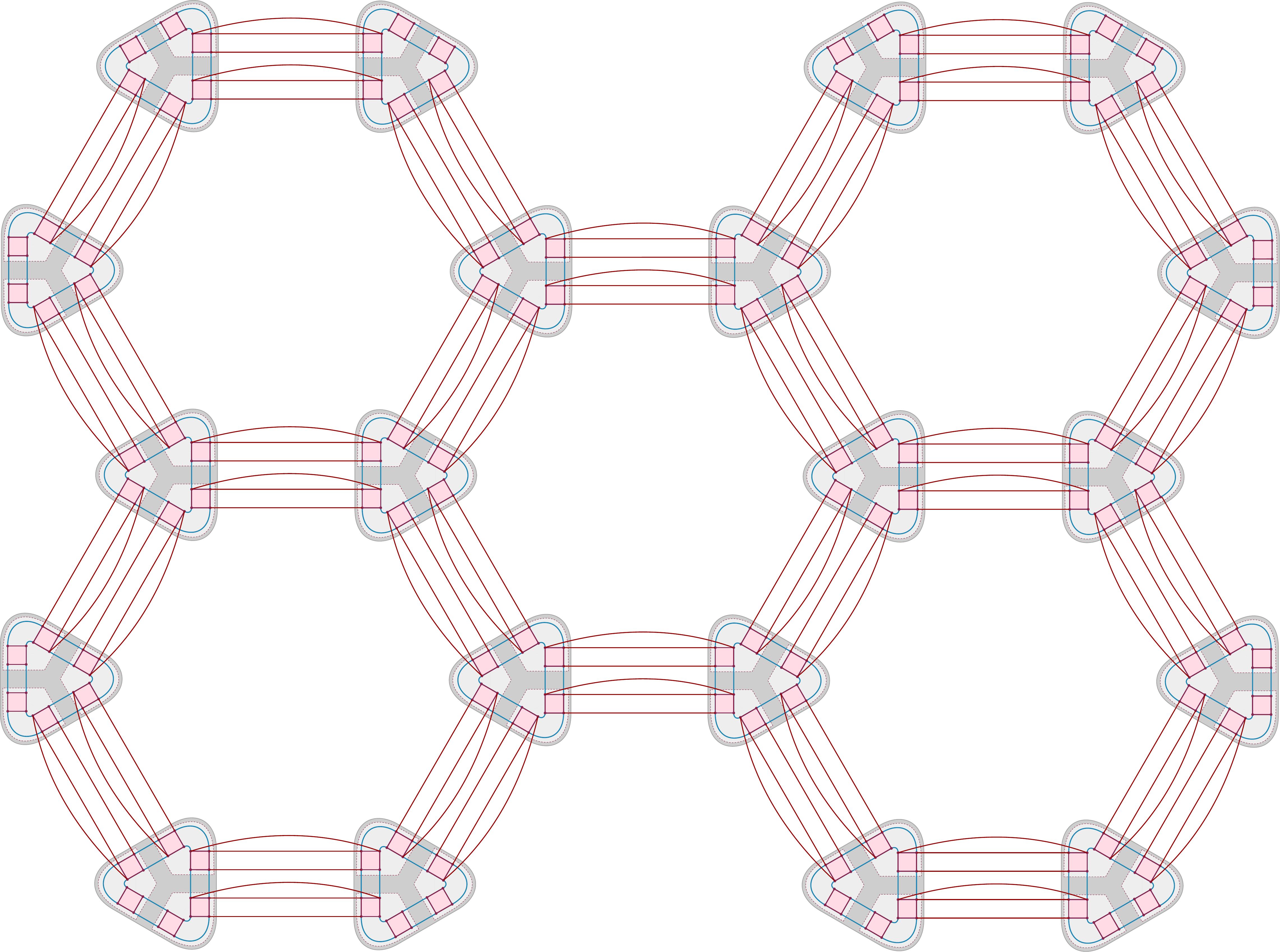}
\caption{Double semion MCB network: Triangular 6-tetron vertices (grey triangles), see Fig.~\ref{fig:DSMPO}, are connected via Bell pair tunneling bridges (red). Note that vertices are arranged in two different sublattices. For a magnified view of the coupling between vertices, see Fig.~\ref{fig:tworings} below.}
\label{fig:DS}
\end{figure}

Above, we have identified the operator $\hat X$, Eq.~\eqref{eq:xdef2}, as key to the description of the PEPS ground state.  The local action of $\hat X$ amounts to an exchange of the three virtual string labels $a,b,c$ in Fig.~\ref{fig:ATensorDef} and the introduction of a sign factor, see Eqs.~\eqref{eq:xdef} and \eqref{eq:xdef1b}. 
This operator can be realized as the tetron ring structure (`triangle') of Fig.~\ref{fig:DSMPO}.
In a second step, adjacent triangles are connected via Bell pair tunneling bridges, resulting in the MCB network depicted in Fig.~\ref{fig:DS}.  This network has the same ground state as the double semion string-net.
 
\emph{Vertex Hamiltonian ---} Let us first discuss how the MCB structure shown in  Fig.~\ref{fig:DSMPO} effectively implements the action of the operator $\hat X$ at the vertices of the lattice. This six tetron structure defines a three unit repetition qubit MPO in the sense of Sec.~\ref{sec:SynthDesign}. The three two-tetron blocks at the corners are linked by tunneling amplitudes (all defined to represent hopping in counter clockwise direction) $a_i\simeq -b_i$. As discussed in Sec.~\ref{sec3c}, this defines three repetition qubits at the corners. Equivalently, we may say that the elementary length two tunneling loops define operators $\frac{\text{Re}\left(a_i  b_i^\ast\right)}{2 E_C} \hat z \otimes  \hat z$, where the two factor Pauli operators are defined to act in the Hilbert spaces of the incoming and outgoing virtual states ($\ket{a}$ and $\ket{a'}$, etc., in Fig.~\ref{fig:ATensorDef}). In this way, the ground state of the system implies the parity projection $\ket{a}=\ket{a'}$ central to the action of $\hat X$. 

The links $\alpha_i, \beta_i$ couple different repetition qubits. The minimal loops formed from these couplings generate an operator $\frac{\text{Re}\left(\alpha_i  \beta_i^\ast\right)}{2 E_C} \hat z \otimes  \hat z$, where the two factor Pauli operators act in the Hilbert spaces of virtual states $\ket{a}$ and $\ket{b}$. Recall that the parity of the latter determines the physical state $\ket{i}=\ket{a \oplus b}$ which is no longer represented by an actual hardware degree of freedom, but can always be deduced from the virtual states. Thus, any preferred alignment of the two spins constitutes an unwanted bias to a product state of $\ket{i}=\ket{0}$ at every site. 
We use the freedom to choose  $\alpha = \mathrm i \beta$ and effectively suppress these terms. 

The most interesting contribution to the tunneling expansion are the  length-6 loops around the ring.  Following the same logics as in Sec.~\ref{sec3b}, the sum of all anti-clockwise oriented tunneling paths defines an MPO 
\begin{equation}
\hat O_+^{(6)}=\frac{63}{8E_C^5} {\rm Tr} \left( \hat B^1  \hat B^2 \hat B^3\right)\;,
\end{equation}
with MPO matrix elements 
\begin{align}
\hat B^i_{00}&=  - \alpha_i ( a_i \hat x \otimes \hat x + b_i  \hat y \otimes \hat y)  \;, \label{eq:B} \\
\hat B^i_{01}&= \beta_i (a_i \hat x \otimes \hat y - b_i \hat y \otimes \hat x)  \;, \nonumber \\
\hat B^i_{10}&=\alpha_i(a_i \hat y \otimes \hat x - b_i \hat x \otimes \hat y)  \;, \nonumber \\
\hat B^i_{11}&=-\beta_i(a_i \hat y \otimes \hat y + b_i \hat x \otimes \hat x) \;,  \nonumber 
\end{align}
acting on the repetition qubits.
Using the above restrictions  $a_i=-b_i$, $\alpha_i = \pm \mathrm i \beta_i$,  
\begin{align}
\hat B^i_{11}&=\pm \mathrm{i}B^i_{00}=  c_i(  \hat x \otimes \hat x -  \hat y \otimes \hat y)  \;, \label{eq:B2} \\
\hat B^i_{01}&=\mp\mathrm{i}B^i_{10}= c_i(\hat x \otimes \hat y + \hat y \otimes \hat x)  \;, \nonumber
\end{align}
with the so far freely adjustable coefficient $c_i:=  \beta_i a_i$. We finally need to choose these parameters such that the sum of the oriented path and its Hermitian conjugate (the reversed path) reproduces the action of $-\hat X$ in~\eqref{eq:xdef2}. A straightforward calculation shows that this is the case for, e.g., $c_i =\mathrm{i}$ and $\alpha_i=\mathrm i \beta_i$. With this choice, the relatively simply MCB network defines an effective  tunneling Hamiltonian, $\hat H_v$ which  encodes the essential structure of the double semion vertex.

\begin{figure}
\includegraphics[width=0.9\columnwidth]{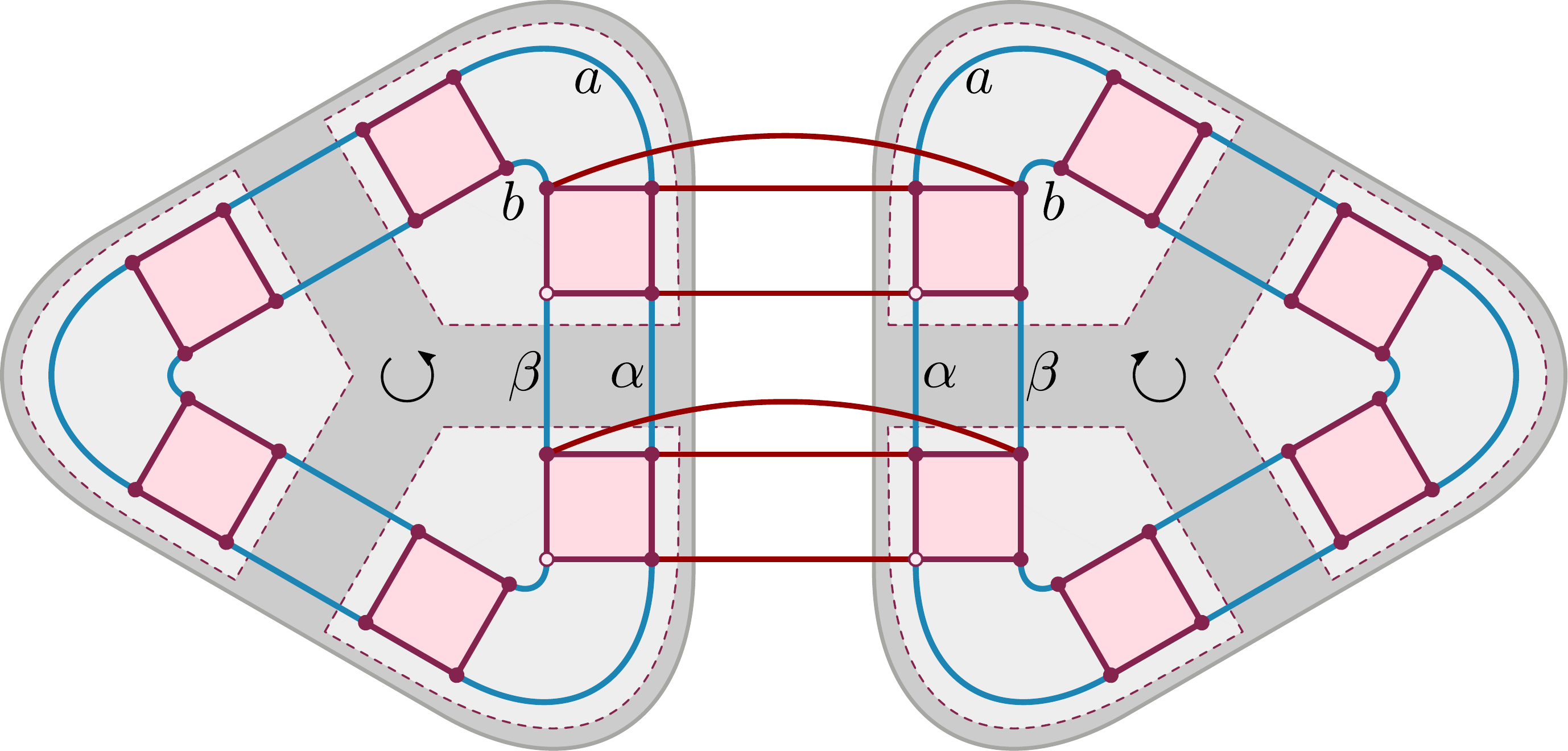}
\caption{Bell pair tunnel bridges 
connecting neighboring vertices in the double semion MCB network of Fig.~\ref{fig:DS}.  }
\label{fig:tworings}
\end{figure}

\emph{Bell pair bridges and double semion network ---}
In a final step, we couple individual vertex structures  via Bell bridges as in Fig.~\ref{fig:bell}. These couplings shown in Fig.~\ref{fig:DS} are to generate a locking of the virtual states of neighboring vertices to a Bell state according to the geometry of the tensor network in Fig. \ref{fig:ATensorDef}. The coupling effectively assigns a Bell pair Hamiltonian, $\hat H_{\rm Bell}$ [Eq.~(\ref{eq:bell})], to all (doubled) edges of the underlying honeycomb lattices connecting two vertices
~\footnote{For completeness, we mention that the coupling always connects vertices of different sublattices. For a connection as indicated in the Fig. \ref{fig:tworings}, the right triangle is identical to the one above. A closer inspection shows that the difference between the operator representing the left triangle differs from the one on the right by interchanging $a$ vs $b$ and $\alpha$ vs $\beta$ which yields a replacement $\hat B_{00}\mapsto \hat B_{11}, \hat B_{11}\mapsto \hat B_{00}, \hat B_{01}\mapsto -\hat B_{10},\hat B_{10}\mapsto - \hat B_{01}$ in Eq. (\ref{eq:B2}). However, since the $B_{01}$, $B_{10}$ always appear in pairs, the MPO $O_6$ is the same.}.

Note that $\hat H_{\rm Bell}$ is generated by length-2 loops while the  tunneling loops within each triangle --- driving the system to the code space --- 
are of length six and thus \emph{a priori} weaker. 
To ensure that the Bell pair Hamiltonians can nonetheless be treated as a perturbation, the respective tunneling strengths $\lambda_{\rm Bell}$ should be sufficiently small ensuring that $\varepsilon:=\lambda^2_{\rm Bell}/E_C$ is small. The effective Hamiltonian of the whole MCB network is given by
\begin{equation}
\hat H_{\rm eff}= \sum_v \hat H_v + \sum_e \hat H_{\rm Bell} + \mathcal O(\varepsilon\lambda^2/E_C^2) \;, \label{eq:DSeff}
\end{equation} 
where $\lambda$ is the absolute value of the tunneling amplitudes $a_i,b_i$ and the leading corrections represent inter-vertex loops of length four, which are parametrically weaker than the leading contributions. In addition they can be controlled to such an extent that they do not influence the result of the perturbative analysis \cite{Brell2014} of the Hamiltonian in Eq.~(\ref{eq:H}). As a conclusion the above effective Hamiltonian has the same ground state space as the perturbative PEPS parent in Eq.~\eqref{eq:H}. Details of the perturbation series analysis are provided in the Appendix.

\emph{Practical implementation ---} Fig.~\ref{fig:DS} shows a schematic of the actual hardware layout implementing the construction. The discussion above reflects the importance of an initial calibration procedure which in turn  necessitates the presence of  local gates near each tunneling link. Specifically,  starting from a configuration in which the ($b_i,\beta_i$) are  turned off, interferometric measurements \cite{Plugge2017,Karzig2017}  between all possible Majorana pair combinations around the outer loop defined by the $(a_i,\alpha_i)$ links, followed by subsequent gate voltage tuning, will be applied to fix a possible choice of relative coupling amplitudes  $a_i=-\alpha_i=\lambda$ with $\lambda$ real. In the next step, interferometric measurements and gate voltage readjustments  are carried out for each length-2 loop defined by $(a_i,b_i)$ and $(\alpha_i,\beta_i$) until one has achieved the values of $b_i=-a_i$ and $\beta_i=\mathrm i a_i$, such that we obtain $c_i =\mathrm i \lambda^2$ and $\lambda^2$ is an inconsequential proportionality factor to the value specified above.  

\alex{
In practice, the above calibration steps would be performed in an automated manner, and 
in view of the presumed robustness of the topological phases, we do not expect that excessive precision need be applied. However, to better understand the consequences of inaccuracies, we note that small errors in the ($\alpha_i,\beta_i$) loop phase,
$\phi_i=\pi/2 +\Delta \phi_i$, 
imply a small but finite contribution to the Hamiltonian, $\hat V_i=(\lambda^2/E_C) \Delta \phi_i \hat z \otimes \hat z$, with likely uncorrelated $\Delta \phi_i$. Depending on the sign of $\Delta \phi_i$, even (odd) parity of the two qubits is now favored and implies  a bias towards the physical $\ket{0}$ ($\ket{1}$) state. This is equivalent to the presence of an effectively random magnetic field in the $z$-direction. The stability of the double semion model against homogeneous magnetic fields has been studied  in Ref.~\cite{Morampudi14}. It turns out that the topological phase persists up to a critical field strength, $h_c$, roughly one order of magnitude smaller than the many-body Hamiltonian gap, $\Delta_g$ (at presumed infinite strength of the vertex operator). The duality of $\mathbb Z_2$ topological ordered Hamiltonians and the random bond Ising model substantiates the intuition that randomly fluctuating fields have a lesser impact and the range of stability is increased~\cite{Tsomokos2011}. 

% Even though the model for randomness in Ref.~\cite{Tsomokos2011} is slightly different than in our problem, 
A  conservative stability estimate follows from the condition that the average absolute value of $\Delta \phi_i$ be less than the critical value for homogeneous magnetic fields, $h_c \approx 0.1 \Delta_g$. In our setup, the double semion model emerges at sixth order of perturbation theory in the hopping amplitudes, $\Delta_g\sim\varepsilon^6/\Delta_v^5$, where $\Delta_v \sim \lambda^6/E_C^5$ is the energy scale of $H_v$ and $\varepsilon \sim \lambda_\text{Bell}^2/E_C$ the energy scale of the Bell pair Hamiltonian. Assuming a  separation of 
energy scales $\varepsilon/\Delta_v$ and $\lambda/ E_C$ by at least one order of magnitude we are led to the conclusion that  that phase variances $|\Delta\phi_i|<10^{-10}$ will certainly be tolerable.
However, this estimate is indeed very conservative. The perturbative constructions of  effective qubit Hamiltonians from Majorana networks tolerate lower energy scale separations than those assumed above \cite{PhysRevLett.108.260504}. Since these ratios enter our construction at $\ge 5$th order it is likely that phase variance exceeding the above estimate by  several orders of magnitudes, and hence within experimental reach, will not jeopardize the integrity of the double semion phase.   }

\section{Conclusions and Outlook}\label{sec5}

In this work, tensor network approaches have been introduced for quantum simulations of complex phases of matter in networks of Majorana Cooper boxes.  Such networks may be experimentally realizable in the near future. We have formulated several design principles generating desired  many-qubit interactions, and suppressing unwanted lower-order interactions via mechanisms of symmetry or engineered destructive interference. 
Specifically, we have studied how tensor networks may serve to simulate topological Levin-Wen string-net models \cite{Levin2005,Wen2017} beyond Kitaev's toric code, a class of systems so far elusive.   As a concrete  example, we have detailed how to simulate the ground state of the double semion model \cite{Levin2005}, the  simplest string-net beyond the toric code.  While the quantum simulation of an exactly solvable model in its pristine form may  provide only limited insights,  our constructions  allow to perturb around the solvable limit in controlled ways and probe the stability of the phase. Similarly, the realization of the net will be a first and necessary step to the creation of excitations and the study of their dynamics.

The present work illustrates the potential of the linkage condensed-matter/tensor network/device implementation, and may actually define an
entire research program. Concrete  directions of research within the general framework include the realization of a   large class of many-qubit interactions from correlated Majorana Cooper box networks, or novel quantum simulation schemes \cite{CiracZollerSimulation,Roadmap} with read-out possibilities \cite{Plugge2017,Karzig2017} unavailable in other architectures.  It should be clear that the local multi-qubit interactions generated by  networks of Majorana Cooper boxes as such already hold the promise of establishing new quantum simulation schemes, regardless of
notions of topological order.

In this work, we have gone further to show how to engineer a restricted class of matrix product operators with bond dimension $D=2$, where network structures emerge by building on Hamiltonian gadget techniques. However, one may go beyond the level of tetrons to design hexon or polygon networks of advanced flexibility and versatility.  At any rate, it will be important to extend the scope and to understand in generality  which types of matrix product operators can be designed in such architectures. The flexibility of the Majorana platform may in fact allow for large classes of matrix product operators while  avoiding undesired few-qubit interactions.  However, further research is required to substantiate this expectation. 

Turning to applications, it will be interesting to further explore the quantum error correcting capabilities of the double semion model
beyond a CSS picture \cite{SemionErrorCorrection}. Given our approach towards realizing this phase of matter,
a natural next step is to quantitatively assess the advantages arising from such a picture of quantum error correction. 
It will be equally exciting  to explore other phases of matter that can be simulated within the 
present framework. For example, the Majorana dimer models \cite{PhysRevB.94.115115,PhysRevB.94.115127} are a class of systems which appear to be within direct reach.
From the perspective of quantum information, realizing
Fibonacci anyon models \cite{Levin2005}  and exploring  implications 
for topological quantum computing is an obvious stepping stone. Given the huge overhead in surface code based topological quantum computing using Clifford operators and magic state distillation \cite{PhysRevA.71.022316}, 
a comprehensive analysis of such alternative approaches seems highly desirable. 
We are confident that our work will also stimulate research along this direction.

\begin{acknowledgements}
We warmly thank A.\ Bauer for various helpful discussions and acknowledge funding by the DFG (CRC TR 183 within project C4 and B1, 
EI 519/7-1, and EI 519/15-1),
the Studienstiftung des Deutschen Volkes, and the ERC (TAQ). This work has also received funding from the European Union's Horizon 2020 research and innovation 
programme under grant agreement No 817482 (\je{PASQuanS}). 

\emph{Note added:}  During the writing of this manuscript, we became aware of two  interesting preprints
\cite{Sagi2018,Thomson2018}  suggesting the application of  MCB networks  to the quantum simulation 
of spin liquid phases. These works, too, exploit the freedom of  engineered  
interactions in Majorana networks. However, the focus is more on generating tailored spin correlations,  and the design principles of tensor networks  or string-net phases are not considered.  
\end{acknowledgements}

\appendix*
\section{Perturbation analysis} \label{secA}

We consider the MCB network in Fig.~\ref{fig:DS} with the Hamiltonian in Eq.~(\ref{eq:DSeff}), where we assume that the energy scale $\lambda^6/E_C^5$ characterizing the vertex part, $\hat H_v$, is much larger than the scale $\varepsilon=\lambda_{\rm Bell}^2/E_C$ corresponding to Bell pair tunneling links between neighboring vertices. Under this assumption, the analysis of Ref.~\cite{Brell2014PEPS} applies where one treats the Bell pair Hamiltonians $\hat H_{\rm Bell}$ as perturbation to $\sum_v \hat H_v$.
In addition, in our case, terms of order $\mathcal O(\varepsilon\lambda^2/E_c^2)$ arise due to loops of length 4 (and beyond) which 
involve MCBs on different vertices. Such contributions need to be
included in a perturbative analysis aimed at checking that the low-energy Hamiltonian of the perturbative PEPS parent theory is indeed the parent Hamiltonian of the encoded PEPS, Eq.~(\ref{eq:encode}). While the full analysis combining global and local Schrieffer-Wolff transformations \cite{Bravyi2011} is cumbersome, a simplified calculation \cite{Brell2014} for the double semion model considers a self-energy expansion series order by order and establishes the physical meaning of each term. As in our perturbative expansion for the low-energy physics of MCB networks in Sec.~\ref{sec2}, 
only operators acting within the ground state space of the dominant Hamiltonian term, here the code space, are considered. 

The first order gives no contribution since  $\hat P \hat H_{\rm Bell}\hat P=0$, where $\hat P$ is the projector to the ground state of $\sum_v \hat H_v$. In fact,
all odd orders do not contribute apart from trivial energy shifts.
At second order, the simultaneous action of Bell pair operators on two neighboring vertices yields a contribution favoring even parity states for the corresponding four qubits. This term has a clear physical meaning. 
Indeed, mapping the encoded state back to the physical state by applying the PEPS map $A$, one finds that the physical state (which used to live on the edge of the honeycomb lattice) is overdetermined. Thanks to the even parity constraint for the four qubits, this overdetermination is resolved and one recovers a consistent mapping from code space to physical space.  At fourth order, products of second-order terms but no  qualitatively new contributions emerge. Such contributions arise, however, 
in sixth order from six Bell pairs Hamiltonians acting on 12 qubits around one plaquette. These terms correspond to encoded versions of the string-net plaquette operator and are essential to drive the system into a topological phase. Note that up to the fifth order, the effective Hamiltonian only contains (encoded versions of) vertex projections and terms ensuring consistency of different local decodings, where one does not expect a topologically ordered ground state.
It is thus crucial that the effect of the plaquette operator is relevant for the system. In particular, if we repeat the perturbative analysis for the MCB network Hamiltonian, taking into account perturbations due to loops of length 4 and beyond, we have to check that no terms overwhelming the plaquette operator are present which can potentially destroy the topological order of the ground state.   

The design of the MCB network in Sec.~\ref{sec4c} overcomes this obstacle in a natural way by making use of destructive interference mechanisms. Consider a junction of two vertices connected by two Bell pair bridges in Fig.~\ref{fig:tworings}. There are 36 different length-4 loops that emerge from tunneling from one MCB at one vertex to the neighboring MCB on the same vertex (via $\alpha$ or $\beta$) then to the other vertex via one of the three Bell bridges, then (again via $\alpha$ or $\beta$) to the neighboring MCB and finally back to the starting point via one of the three Bell bridges. It is useful to partition these 36 loops into four different groups labeled by the loop phase $ \phi_{\alpha/
\beta}-\phi_{\alpha/ \beta}$. Let us analyse the different $\phi_\alpha - \phi_\beta$ loops. There is one loop that is threatening the emergence of topological order. It corresponds to the $\hat z \hat z$ interactions on neighboring qubits, interactions that we needed to suppress already in the design of the vertex Hamiltonian. Because the parity of two different corner qubits determines the physical state uniquely, a system that favors a specific parity of those two qubits corresponds to a trivial product state. Since this state is part of the code space there is no chance to argue that it will not survive the code space projection $P$. This is why we have chosen the tunneling links on different sublattices according to Fig. \ref{fig:tworings}. The loop phase $\phi_\alpha -\phi_\beta=\pi/2$ and likewise $\phi_\beta-\phi_\alpha$ leads to a complete cancelation of all those loops (note that the Bell pair wires don't add phase shifts) including the one that would otherwise lead to the adverse $\hat z \hat z$ interaction. The other half of the length-4 loops does not vanish but also poses no threat to the formation of topological order.
They are  
of the form  $\hat O^{(4)}\propto\hat  p \hat p \otimes \hat q \hat q $
where the tensor product indicates the separation into two Bell pairs and $\hat p$, $\hat q$ denote different Pauli operators.  Contributions of this form also arise in second order perturbation theory in $\varepsilon$ from the Bell pair 
Hamiltonian and therefore do not yield qualitatively new contributions. In addition they are completely outside the code space and do not contribute at all when treated as a first order correction to $\sum_v H_v$. Similar arguments apply to higher order loop terms. Prominently among them are detour-loops that wind once around a vertex ring but with a detour via two qubits of a neighboring vertex. Those detour terms either leave the code space or yield terms that correspond to a product of $O^{(6)}$ and $\hat z \hat z \otimes \hat z \hat z$. These terms cause a splitting of the ground state space degeneracy that is in agreement with the aforementioned consistency of physical qubit states and do not cause a phase transition. Detour terms involving more than one detour can be treated similarly.

To conclude, we have verified that the analysis of Ref.~\cite{Brell2014}, and therefore our encoded PEPS approach to the ground state of the double semion model, also applies in the presence of higher-order terms of ${\cal O}(\varepsilon\lambda^2/E_C^2)$ in Eq.~(\ref{eq:DSeff}). 

%\bibliography{MCBReferences,Reinholdref}

\begin{thebibliography}{88}%
\makeatletter
\providecommand \@ifxundefined [1]{%
 \@ifx{#1\undefined}
}%
\providecommand \@ifnum [1]{%
 \ifnum #1\expandafter \@firstoftwo
 \else \expandafter \@secondoftwo
 \fi
}%
\providecommand \@ifx [1]{%
 \ifx #1\expandafter \@firstoftwo
 \else \expandafter \@secondoftwo
 \fi
}%
\providecommand \natexlab [1]{#1}%
\providecommand \enquote  [1]{``#1''}%
\providecommand \bibnamefont  [1]{#1}%
\providecommand \bibfnamefont [1]{#1}%
\providecommand \citenamefont [1]{#1}%
\providecommand \href@noop [0]{\@secondoftwo}%
\providecommand \href [0]{\begingroup \@sanitize@url \@href}%
\providecommand \@href[1]{\@@startlink{#1}\@@href}%
\providecommand \@@href[1]{\endgroup#1\@@endlink}%
\providecommand \@sanitize@url [0]{\catcode `\\12\catcode `\$12\catcode
  `\&12\catcode `\#12\catcode `\^12\catcode `\_12\catcode `\%12\relax}%
\providecommand \@@startlink[1]{}%
\providecommand \@@endlink[0]{}%
\providecommand \url  [0]{\begingroup\@sanitize@url \@url }%
\providecommand \@url [1]{\endgroup\@href {#1}{\urlprefix }}%
\providecommand \urlprefix  [0]{URL }%
\providecommand \Eprint [0]{\href }%
\providecommand \doibase [0]{http://dx.doi.org/}%
\providecommand \selectlanguage [0]{\@gobble}%
\providecommand \bibinfo  [0]{\@secondoftwo}%
\providecommand \bibfield  [0]{\@secondoftwo}%
\providecommand \translation [1]{[#1]}%
\providecommand \BibitemOpen [0]{}%
\providecommand \bibitemStop [0]{}%
\providecommand \bibitemNoStop [0]{.\EOS\space}%
\providecommand \EOS [0]{\spacefactor3000\relax}%
\providecommand \BibitemShut  [1]{\csname bibitem#1\endcsname}%
\let\auto@bib@innerbib\@empty
%</preamble>
\bibitem [{\citenamefont {Nayak}\ \emph {et~al.}(2008)\citenamefont {Nayak},
  \citenamefont {Simon}, \citenamefont {Stern}, \citenamefont {Freedman},\ and\
  \citenamefont {Das~Sarma}}]{qc3}%
  \BibitemOpen
  \bibfield  {author} {\bibinfo {author} {\bibfnamefont {C.}~\bibnamefont
  {Nayak}}, \bibinfo {author} {\bibfnamefont {S.~H.}\ \bibnamefont {Simon}},
  \bibinfo {author} {\bibfnamefont {A.}~\bibnamefont {Stern}}, \bibinfo
  {author} {\bibfnamefont {M.}~\bibnamefont {Freedman}}, \ and\ \bibinfo
  {author} {\bibfnamefont {S.}~\bibnamefont {Das~Sarma}},\ }\href {\doibase
  10.1103/RevModPhys.80.1083} {\bibfield  {journal} {\bibinfo  {journal} {Rev.
  Mod. Phys.}\ }\textbf {\bibinfo {volume} {80}},\ \bibinfo {pages} {1083}
  (\bibinfo {year} {2008})}\BibitemShut {NoStop}%
\bibitem [{\citenamefont {Wen}(2017)}]{Wen2017}%
  \BibitemOpen
  \bibfield  {author} {\bibinfo {author} {\bibfnamefont {X.~G.}\ \bibnamefont
  {Wen}},\ }\href
  {https://journals.aps.org/rmp/abstract/10.1103/RevModPhys.89.041004}
  {\bibfield  {journal} {\bibinfo  {journal} {Rev. Mod. Phys.}\ }\textbf
  {\bibinfo {volume} {89}},\ \bibinfo {pages} {041004} (\bibinfo {year}
  {2017})}\BibitemShut {NoStop}%
\bibitem [{\citenamefont {Dennis}\ \emph {et~al.}(2002)\citenamefont {Dennis},
  \citenamefont {Kitaev}, \citenamefont {Landahl},\ and\ \citenamefont
  {Preskill}}]{TopologicalQuantumMemory}%
  \BibitemOpen
  \bibfield  {author} {\bibinfo {author} {\bibfnamefont {E.}~\bibnamefont
  {Dennis}}, \bibinfo {author} {\bibfnamefont {A.}~\bibnamefont {Kitaev}},
  \bibinfo {author} {\bibfnamefont {A.}~\bibnamefont {Landahl}}, \ and\
  \bibinfo {author} {\bibfnamefont {J.}~\bibnamefont {Preskill}},\ }\href
  {\doibase 10.1063/1.1499754} {\bibfield  {journal} {\bibinfo  {journal} {J.
  Math. Phys.}\ }\textbf {\bibinfo {volume} {43}},\ \bibinfo {pages} {4452}
  (\bibinfo {year} {2002})}\BibitemShut {NoStop}%
\bibitem [{\citenamefont {Bravyi}\ and\ \citenamefont
  {Kitaev}(2005)}]{PhysRevA.71.022316}%
  \BibitemOpen
  \bibfield  {author} {\bibinfo {author} {\bibfnamefont {S.}~\bibnamefont
  {Bravyi}}\ and\ \bibinfo {author} {\bibfnamefont {A.}~\bibnamefont
  {Kitaev}},\ }\href {\doibase 10.1103/PhysRevA.71.022316} {\bibfield
  {journal} {\bibinfo  {journal} {Phys. Rev. A}\ }\textbf {\bibinfo {volume}
  {71}},\ \bibinfo {pages} {022316} (\bibinfo {year} {2005})}\BibitemShut
  {NoStop}%
\bibitem [{\citenamefont {Kitaev}(2003)}]{Kitaev-AnnPhys-2003}%
  \BibitemOpen
  \bibfield  {author} {\bibinfo {author} {\bibfnamefont {A.~Y.}\ \bibnamefont
  {Kitaev}},\ }\href {\doibase 10.1016/S0003-4916(02)00018-0} {\bibfield
  {journal} {\bibinfo  {journal} {Ann. Phys.}\ }\textbf {\bibinfo {volume}
  {303}},\ \bibinfo {pages} {2 } (\bibinfo {year} {2003})}\BibitemShut
  {NoStop}%
\bibitem [{\citenamefont {Terhal}(2015)}]{RevModPhys.87.307}%
  \BibitemOpen
  \bibfield  {author} {\bibinfo {author} {\bibfnamefont {B.~M.}\ \bibnamefont
  {Terhal}},\ }\href {\doibase 10.1103/RevModPhys.87.307} {\bibfield  {journal}
  {\bibinfo  {journal} {Rev. Mod. Phys.}\ }\textbf {\bibinfo {volume} {87}},\
  \bibinfo {pages} {307} (\bibinfo {year} {2015})}\BibitemShut {NoStop}%
\bibitem [{\citenamefont {Levin}\ and\ \citenamefont {Wen}(2005)}]{Levin2005}%
  \BibitemOpen
  \bibfield  {author} {\bibinfo {author} {\bibfnamefont {M.}~\bibnamefont
  {Levin}}\ and\ \bibinfo {author} {\bibfnamefont {X.~G.}\ \bibnamefont
  {Wen}},\ }\href
  {https://journals.aps.org/prb/abstract/10.1103/PhysRevB.71.045110} {\bibfield
   {journal} {\bibinfo  {journal} {Phys. Rev. B}\ }\textbf {\bibinfo {volume}
  {71}},\ \bibinfo {pages} {045110} (\bibinfo {year} {2005})}\BibitemShut
  {NoStop}%
\bibitem [{\citenamefont {Levin}\ and\ \citenamefont {Wen}(2006)}]{LevinWen}%
  \BibitemOpen
  \bibfield  {author} {\bibinfo {author} {\bibfnamefont {M.}~\bibnamefont
  {Levin}}\ and\ \bibinfo {author} {\bibfnamefont {X.-G.}\ \bibnamefont
  {Wen}},\ }\href {\doibase 10.1103/PhysRevLett.96.110405} {\bibfield
  {journal} {\bibinfo  {journal} {Phys. Rev. Lett.}\ }\textbf {\bibinfo
  {volume} {96}},\ \bibinfo {pages} {110405} (\bibinfo {year}
  {2006})}\BibitemShut {NoStop}%
\bibitem [{\citenamefont {Cirac}\ and\ \citenamefont
  {Zoller}(2012)}]{CiracZollerSimulation}%
  \BibitemOpen
  \bibfield  {author} {\bibinfo {author} {\bibfnamefont {J.~I.}\ \bibnamefont
  {Cirac}}\ and\ \bibinfo {author} {\bibfnamefont {P.}~\bibnamefont {Zoller}},\
  }\href {\doibase 10.1038/nphys2275} {\bibfield  {journal} {\bibinfo
  {journal} {Nature Phys.}\ }\textbf {\bibinfo {volume} {8}},\ \bibinfo {pages}
  {264} (\bibinfo {year} {2012})}\BibitemShut {NoStop}%
\bibitem [{\citenamefont {Acin}\ \emph {et~al.}(2018)\citenamefont {Acin},
  \citenamefont {Bloch}, \citenamefont {Buhrman}, \citenamefont {Calarco},
  \citenamefont {Eichler}, \citenamefont {Eisert}, \citenamefont {Esteve},
  \citenamefont {Gisin}, \citenamefont {Glaser}, \citenamefont {Jelezko},
  \citenamefont {Kuhr}, \citenamefont {Lewenstein}, \citenamefont {Riedel},
  \citenamefont {Schmidt}, \citenamefont {Thew}, \citenamefont {Wallraff},
  \citenamefont {Walmsley},\ and\ \citenamefont {Wilhelm}}]{Roadmap}%
  \BibitemOpen
  \bibfield  {author} {\bibinfo {author} {\bibfnamefont {A.}~\bibnamefont
  {Acin}}, \bibinfo {author} {\bibfnamefont {I.}~\bibnamefont {Bloch}},
  \bibinfo {author} {\bibfnamefont {H.}~\bibnamefont {Buhrman}}, \bibinfo
  {author} {\bibfnamefont {T.}~\bibnamefont {Calarco}}, \bibinfo {author}
  {\bibfnamefont {C.}~\bibnamefont {Eichler}}, \bibinfo {author} {\bibfnamefont
  {J.}~\bibnamefont {Eisert}}, \bibinfo {author} {\bibfnamefont
  {D.}~\bibnamefont {Esteve}}, \bibinfo {author} {\bibfnamefont
  {N.}~\bibnamefont {Gisin}}, \bibinfo {author} {\bibfnamefont {S.~J.}\
  \bibnamefont {Glaser}}, \bibinfo {author} {\bibfnamefont {F.}~\bibnamefont
  {Jelezko}}, \bibinfo {author} {\bibfnamefont {S.}~\bibnamefont {Kuhr}},
  \bibinfo {author} {\bibfnamefont {M.}~\bibnamefont {Lewenstein}}, \bibinfo
  {author} {\bibfnamefont {M.~F.}\ \bibnamefont {Riedel}}, \bibinfo {author}
  {\bibfnamefont {P.~O.}\ \bibnamefont {Schmidt}}, \bibinfo {author}
  {\bibfnamefont {R.}~\bibnamefont {Thew}}, \bibinfo {author} {\bibfnamefont
  {A.}~\bibnamefont {Wallraff}}, \bibinfo {author} {\bibfnamefont
  {I.}~\bibnamefont {Walmsley}}, \ and\ \bibinfo {author} {\bibfnamefont
  {F.~K.}\ \bibnamefont {Wilhelm}},\ }\href {\doibase 10.1088/1367-2630/aad1ea}
  {\bibfield  {journal} {\bibinfo  {journal} {New J. Phys.}\ }\textbf {\bibinfo
  {volume} {20}},\ \bibinfo {pages} {080201} (\bibinfo {year}
  {2018})}\BibitemShut {NoStop}%
\bibitem [{\citenamefont {Bloch}\ \emph {et~al.}(2012)\citenamefont {Bloch},
  \citenamefont {Dalibard},\ and\ \citenamefont
  {Nascimbene}}]{BlochSimulation}%
  \BibitemOpen
  \bibfield  {author} {\bibinfo {author} {\bibfnamefont {I.}~\bibnamefont
  {Bloch}}, \bibinfo {author} {\bibfnamefont {J.}~\bibnamefont {Dalibard}}, \
  and\ \bibinfo {author} {\bibfnamefont {S.}~\bibnamefont {Nascimbene}},\
  }\href {\doibase doi.org/10.1038/nphys2259} {\bibfield  {journal} {\bibinfo
  {journal} {Nature Phys.}\ }\textbf {\bibinfo {volume} {8}},\ \bibinfo {pages}
  {267} (\bibinfo {year} {2012})}\BibitemShut {NoStop}%
\bibitem [{\citenamefont {Vijay}\ \emph {et~al.}(2015)\citenamefont {Vijay},
  \citenamefont {Hsieh},\ and\ \citenamefont {Fu}}]{Vijay2015}%
  \BibitemOpen
  \bibfield  {author} {\bibinfo {author} {\bibfnamefont {S.}~\bibnamefont
  {Vijay}}, \bibinfo {author} {\bibfnamefont {T.~H.}\ \bibnamefont {Hsieh}}, \
  and\ \bibinfo {author} {\bibfnamefont {L.}~\bibnamefont {Fu}},\ }\href
  {https://journals.aps.org/prx/abstract/10.1103/PhysRevX.5.041038} {\bibfield
  {journal} {\bibinfo  {journal} {Phys. Rev. X}\ }\textbf {\bibinfo {volume}
  {5}},\ \bibinfo {pages} {041038} (\bibinfo {year} {2015})}\BibitemShut
  {NoStop}%
\bibitem [{\citenamefont {Plugge}\ \emph {et~al.}(2016)\citenamefont {Plugge},
  \citenamefont {Landau}, \citenamefont {Sela}, \citenamefont {Altland},
  \citenamefont {Flensberg},\ and\ \citenamefont {Egger}}]{PhysRevB.94.174514}%
  \BibitemOpen
  \bibfield  {author} {\bibinfo {author} {\bibfnamefont {S.}~\bibnamefont
  {Plugge}}, \bibinfo {author} {\bibfnamefont {L.~A.}\ \bibnamefont {Landau}},
  \bibinfo {author} {\bibfnamefont {E.}~\bibnamefont {Sela}}, \bibinfo {author}
  {\bibfnamefont {A.}~\bibnamefont {Altland}}, \bibinfo {author} {\bibfnamefont
  {K.}~\bibnamefont {Flensberg}}, \ and\ \bibinfo {author} {\bibfnamefont
  {R.}~\bibnamefont {Egger}},\ }\href {\doibase 10.1103/PhysRevB.94.174514}
  {\bibfield  {journal} {\bibinfo  {journal} {Phys. Rev. B}\ }\textbf {\bibinfo
  {volume} {94}},\ \bibinfo {pages} {174514} (\bibinfo {year}
  {2016})}\BibitemShut {NoStop}%
\bibitem [{\citenamefont {Landau}\ \emph {et~al.}(2016)\citenamefont {Landau},
  \citenamefont {Plugge}, \citenamefont {Sela}, \citenamefont {Altland},
  \citenamefont {Albrecht},\ and\ \citenamefont
  {Egger}}]{PhysRevLett.116.050501}%
  \BibitemOpen
  \bibfield  {author} {\bibinfo {author} {\bibfnamefont {L.~A.}\ \bibnamefont
  {Landau}}, \bibinfo {author} {\bibfnamefont {S.}~\bibnamefont {Plugge}},
  \bibinfo {author} {\bibfnamefont {E.}~\bibnamefont {Sela}}, \bibinfo {author}
  {\bibfnamefont {A.}~\bibnamefont {Altland}}, \bibinfo {author} {\bibfnamefont
  {S.~M.}\ \bibnamefont {Albrecht}}, \ and\ \bibinfo {author} {\bibfnamefont
  {R.}~\bibnamefont {Egger}},\ }\href {\doibase 10.1103/PhysRevLett.116.050501}
  {\bibfield  {journal} {\bibinfo  {journal} {Phys. Rev. Lett.}\ }\textbf
  {\bibinfo {volume} {116}},\ \bibinfo {pages} {050501} (\bibinfo {year}
  {2016})}\BibitemShut {NoStop}%
\bibitem [{\citenamefont {Terhal}\ \emph {et~al.}(2012)\citenamefont {Terhal},
  \citenamefont {Hassler},\ and\ \citenamefont
  {DiVincenzo}}]{PhysRevLett.108.260504}%
  \BibitemOpen
  \bibfield  {author} {\bibinfo {author} {\bibfnamefont {B.~M.}\ \bibnamefont
  {Terhal}}, \bibinfo {author} {\bibfnamefont {F.}~\bibnamefont {Hassler}}, \
  and\ \bibinfo {author} {\bibfnamefont {D.~P.}\ \bibnamefont {DiVincenzo}},\
  }\href {\doibase 10.1103/PhysRevLett.108.260504} {\bibfield  {journal}
  {\bibinfo  {journal} {Phys. Rev. Lett.}\ }\textbf {\bibinfo {volume} {108}},\
  \bibinfo {pages} {260504} (\bibinfo {year} {2012})}\BibitemShut {NoStop}%
\bibitem [{\citenamefont {Hoffman}\ \emph {et~al.}(2016)\citenamefont
  {Hoffman}, \citenamefont {Schrade}, \citenamefont {Klinovaja},\ and\
  \citenamefont {Loss}}]{Hoffman2016}%
  \BibitemOpen
  \bibfield  {author} {\bibinfo {author} {\bibfnamefont {S.}~\bibnamefont
  {Hoffman}}, \bibinfo {author} {\bibfnamefont {C.}~\bibnamefont {Schrade}},
  \bibinfo {author} {\bibfnamefont {J.}~\bibnamefont {Klinovaja}}, \ and\
  \bibinfo {author} {\bibfnamefont {D.}~\bibnamefont {Loss}},\ }\href {\doibase
  10.1103/PhysRevB.94.045316} {\bibfield  {journal} {\bibinfo  {journal} {Phys.
  Rev. B}\ }\textbf {\bibinfo {volume} {94}},\ \bibinfo {pages} {045316}
  (\bibinfo {year} {2016})}\BibitemShut {NoStop}%
\bibitem [{\citenamefont {Dauphinais}\ and\ \citenamefont
  {Poulin}(2016)}]{NonAbelianDecoders}%
  \BibitemOpen
  \bibfield  {author} {\bibinfo {author} {\bibfnamefont {G.}~\bibnamefont
  {Dauphinais}}\ and\ \bibinfo {author} {\bibfnamefont {D.}~\bibnamefont
  {Poulin}},\ }\href {\doibase 10.1007/s00220-017-2923-9} {\bibfield  {journal}
  {\bibinfo  {journal} {Commun. Math. Phys.}\ }\textbf {\bibinfo {volume}
  {355}},\ \bibinfo {pages} {519} (\bibinfo {year} {2016})}\BibitemShut
  {NoStop}%
\bibitem [{\citenamefont {Burton}\ \emph {et~al.}(2017)\citenamefont {Burton},
  \citenamefont {Brell},\ and\ \citenamefont {Flammia}}]{PhysRevA.95.022309}%
  \BibitemOpen
  \bibfield  {author} {\bibinfo {author} {\bibfnamefont {S.}~\bibnamefont
  {Burton}}, \bibinfo {author} {\bibfnamefont {C.~G.}\ \bibnamefont {Brell}}, \
  and\ \bibinfo {author} {\bibfnamefont {S.~T.}\ \bibnamefont {Flammia}},\
  }\href {\doibase 10.1103/PhysRevA.95.022309} {\bibfield  {journal} {\bibinfo
  {journal} {Phys. Rev. A}\ }\textbf {\bibinfo {volume} {95}},\ \bibinfo
  {pages} {022309} (\bibinfo {year} {2017})}\BibitemShut {NoStop}%
\bibitem [{\citenamefont {Dauphinais}\ \emph {et~al.}(2018)\citenamefont
  {Dauphinais}, \citenamefont {Ortiz}, \citenamefont {Varona},\ and\
  \citenamefont {Martin-Delgado}}]{SemionErrorCorrection}%
  \BibitemOpen
  \bibfield  {author} {\bibinfo {author} {\bibfnamefont {G.}~\bibnamefont
  {Dauphinais}}, \bibinfo {author} {\bibfnamefont {L.}~\bibnamefont {Ortiz}},
  \bibinfo {author} {\bibfnamefont {S.}~\bibnamefont {Varona}}, \ and\ \bibinfo
  {author} {\bibfnamefont {M.~A.}\ \bibnamefont {Martin-Delgado}},\ }\href@noop
  {} {\  (\bibinfo {year} {2018})},\ \Eprint {http://arxiv.org/abs/1810.08204}
  {arXiv:1810.08204} \BibitemShut {NoStop}%
\bibitem [{\citenamefont {Or\'us}(2014)}]{Orus-AnnPhys-2014}%
  \BibitemOpen
  \bibfield  {author} {\bibinfo {author} {\bibfnamefont {R.}~\bibnamefont
  {Or\'us}},\ }\href {\doibase 10.1016/j.aop.2014.06.013} {\bibfield  {journal}
  {\bibinfo  {journal} {Ann. Phys.}\ }\textbf {\bibinfo {volume} {349}},\
  \bibinfo {pages} {117} (\bibinfo {year} {2014})}\BibitemShut {NoStop}%
\bibitem [{\citenamefont {Eisert}\ \emph {et~al.}(2010)\citenamefont {Eisert},
  \citenamefont {Cramer},\ and\ \citenamefont {Plenio}}]{AreaReview}%
  \BibitemOpen
  \bibfield  {author} {\bibinfo {author} {\bibfnamefont {J.}~\bibnamefont
  {Eisert}}, \bibinfo {author} {\bibfnamefont {M.}~\bibnamefont {Cramer}}, \
  and\ \bibinfo {author} {\bibfnamefont {M.~B.}\ \bibnamefont {Plenio}},\
  }\href {\doibase 10.1103/RevModPhys.82.277} {\bibfield  {journal} {\bibinfo
  {journal} {Rev. Mod. Phys.}\ }\textbf {\bibinfo {volume} {82}},\ \bibinfo
  {pages} {277} (\bibinfo {year} {2010})}\BibitemShut {NoStop}%
\bibitem [{\citenamefont {Verstraete}\ \emph {et~al.}(2008)\citenamefont
  {Verstraete}, \citenamefont {Cirac},\ and\ \citenamefont
  {Murg}}]{VerstraeteBig}%
  \BibitemOpen
  \bibfield  {author} {\bibinfo {author} {\bibfnamefont {F.}~\bibnamefont
  {Verstraete}}, \bibinfo {author} {\bibfnamefont {J.~I.}\ \bibnamefont
  {Cirac}}, \ and\ \bibinfo {author} {\bibfnamefont {V.}~\bibnamefont {Murg}},\
  }\href {\doibase 10.1080/14789940801912366} {\bibfield  {journal} {\bibinfo
  {journal} {Adv. Phys.}\ }\textbf {\bibinfo {volume} {57}},\ \bibinfo {pages}
  {143} (\bibinfo {year} {2008})}\BibitemShut {NoStop}%
\bibitem [{\citenamefont {Schuch}(2013)}]{SchuchReview}%
  \BibitemOpen
  \bibfield  {author} {\bibinfo {author} {\bibfnamefont {N.}~\bibnamefont
  {Schuch}},\ }\href@noop {} {\bibfield  {journal} {\bibinfo  {journal}
  {Lecture notes for the 44th IFF Spring School ``Quantum Information
  Processing'' in Juelich}\ } (\bibinfo {year} {2013})}\BibitemShut {NoStop}%
\bibitem [{\citenamefont {Alicea}(2012)}]{Alicea2012}%
  \BibitemOpen
  \bibfield  {author} {\bibinfo {author} {\bibfnamefont {J.}~\bibnamefont
  {Alicea}},\ }\href {http://stacks.iop.org/0034-4885/75/i=7/a=076501}
  {\bibfield  {journal} {\bibinfo  {journal} {Rep. Prog. Phys.}\ }\textbf
  {\bibinfo {volume} {75}},\ \bibinfo {pages} {076501} (\bibinfo {year}
  {2012})}\BibitemShut {NoStop}%
\bibitem [{\citenamefont {Leijnse}\ and\ \citenamefont
  {Flensberg}(2012)}]{Leijnse2012}%
  \BibitemOpen
  \bibfield  {author} {\bibinfo {author} {\bibfnamefont {M.}~\bibnamefont
  {Leijnse}}\ and\ \bibinfo {author} {\bibfnamefont {K.}~\bibnamefont
  {Flensberg}},\ }\href {https://doi.org/10.1088/0268-1242/27/12/124003}
  {\bibfield  {journal} {\bibinfo  {journal} {Semicond. Sci. Techn.}\ }\textbf
  {\bibinfo {volume} {27}},\ \bibinfo {pages} {124003} (\bibinfo {year}
  {2012})}\BibitemShut {NoStop}%
\bibitem [{\citenamefont {Beenakker}(2013)}]{Beenakker2013}%
  \BibitemOpen
  \bibfield  {author} {\bibinfo {author} {\bibfnamefont {C.~W.~J.}\
  \bibnamefont {Beenakker}},\ }\href
  {https://doi.org/10.1146/annurev-conmatphys-030212-184337} {\bibfield
  {journal} {\bibinfo  {journal} {Annu. Rev. Con. Mat. Phys.}\ }\textbf
  {\bibinfo {volume} {4}},\ \bibinfo {pages} {113} (\bibinfo {year}
  {2013})}\BibitemShut {NoStop}%
\bibitem [{\citenamefont {Sarma}\ \emph {et~al.}(2015)\citenamefont {Sarma},
  \citenamefont {Freedman},\ and\ \citenamefont {Nayak}}]{Sarma2015}%
  \BibitemOpen
  \bibfield  {author} {\bibinfo {author} {\bibfnamefont {S.~D.}\ \bibnamefont
  {Sarma}}, \bibinfo {author} {\bibfnamefont {M.}~\bibnamefont {Freedman}}, \
  and\ \bibinfo {author} {\bibfnamefont {C.}~\bibnamefont {Nayak}},\ }\href
  {http://dx.doi.org/10.1038/npjqi.2015.1} {\bibfield  {journal} {\bibinfo
  {journal} {njp Quant. Inf.}\ }\textbf {\bibinfo {volume} {1}},\ \bibinfo
  {pages} {15001} (\bibinfo {year} {2015})}\BibitemShut {NoStop}%
\bibitem [{\citenamefont {Aguado}(2017)}]{Aguado2017}%
  \BibitemOpen
  \bibfield  {author} {\bibinfo {author} {\bibfnamefont {R.}~\bibnamefont
  {Aguado}},\ }\href {http://dx.doi.org/10.1393/ncr/i2017-10141-9} {\bibfield
  {journal} {\bibinfo  {journal} {Rivista del Nuovo Cimento}\ }\textbf
  {\bibinfo {volume} {40}},\ \bibinfo {pages} {523} (\bibinfo {year}
  {2017})}\BibitemShut {NoStop}%
\bibitem [{\citenamefont {Lutchyn}\ \emph {et~al.}(2017)\citenamefont
  {Lutchyn}, \citenamefont {Bakkers}, \citenamefont {Kouwenhoven},
  \citenamefont {Krogstrup}, \citenamefont {Marcus},\ and\ \citenamefont
  {Oreg}}]{Lutchyn2018}%
  \BibitemOpen
  \bibfield  {author} {\bibinfo {author} {\bibfnamefont {R.~M.}\ \bibnamefont
  {Lutchyn}}, \bibinfo {author} {\bibfnamefont {E.~P. A.~M.}\ \bibnamefont
  {Bakkers}}, \bibinfo {author} {\bibfnamefont {L.~P.}\ \bibnamefont
  {Kouwenhoven}}, \bibinfo {author} {\bibfnamefont {P.}~\bibnamefont
  {Krogstrup}}, \bibinfo {author} {\bibfnamefont {C.~M.}\ \bibnamefont
  {Marcus}}, \ and\ \bibinfo {author} {\bibfnamefont {Y.}~\bibnamefont
  {Oreg}},\ }\href {https://doi.org/10.1038/s41578-018-0003-1} {\bibfield
  {journal} {\bibinfo  {journal} {Nat. Rev. Mat.}\ }\textbf {\bibinfo {volume}
  {3}},\ \bibinfo {pages} {52} (\bibinfo {year} {2017})}\BibitemShut {NoStop}%
\bibitem [{\citenamefont {Mourik}\ \emph {et~al.}(2012)\citenamefont {Mourik},
  \citenamefont {Zuo}, \citenamefont {Frolov}, \citenamefont {Plissard},
  \citenamefont {Bakkers},\ and\ \citenamefont {Kouwenhoven}}]{Mourik2012}%
  \BibitemOpen
  \bibfield  {author} {\bibinfo {author} {\bibfnamefont {V.}~\bibnamefont
  {Mourik}}, \bibinfo {author} {\bibfnamefont {K.}~\bibnamefont {Zuo}},
  \bibinfo {author} {\bibfnamefont {S.~M.}\ \bibnamefont {Frolov}}, \bibinfo
  {author} {\bibfnamefont {S.~R.}\ \bibnamefont {Plissard}}, \bibinfo {author}
  {\bibfnamefont {E.~P. A.~M.}\ \bibnamefont {Bakkers}}, \ and\ \bibinfo
  {author} {\bibfnamefont {L.~P.}\ \bibnamefont {Kouwenhoven}},\ }\href
  {\doibase 10.1126/science.1222360} {\bibfield  {journal} {\bibinfo  {journal}
  {Science}\ }\textbf {\bibinfo {volume} {336}},\ \bibinfo {pages} {1003}
  (\bibinfo {year} {2012})}\BibitemShut {NoStop}%
\bibitem [{\citenamefont {Albrecht}\ \emph {et~al.}(2016)\citenamefont
  {Albrecht}, \citenamefont {Higginbotham}, \citenamefont {Madsen},
  \citenamefont {Kuemmeth}, \citenamefont {Jespersen}, \citenamefont
  {Nyg{\aa}rd}, \citenamefont {Krogstrup},\ and\ \citenamefont
  {Marcus}}]{Albrecht2016}%
  \BibitemOpen
  \bibfield  {author} {\bibinfo {author} {\bibfnamefont {S.~M.}\ \bibnamefont
  {Albrecht}}, \bibinfo {author} {\bibfnamefont {A.~P.}\ \bibnamefont
  {Higginbotham}}, \bibinfo {author} {\bibfnamefont {M.}~\bibnamefont
  {Madsen}}, \bibinfo {author} {\bibfnamefont {F.}~\bibnamefont {Kuemmeth}},
  \bibinfo {author} {\bibfnamefont {T.~S.}\ \bibnamefont {Jespersen}}, \bibinfo
  {author} {\bibfnamefont {J.}~\bibnamefont {Nyg{\aa}rd}}, \bibinfo {author}
  {\bibfnamefont {P.}~\bibnamefont {Krogstrup}}, \ and\ \bibinfo {author}
  {\bibfnamefont {C.~M.}\ \bibnamefont {Marcus}},\ }\href
  {http://dx.doi.org/10.1038/nature17162} {\bibfield  {journal} {\bibinfo
  {journal} {Nature}\ }\textbf {\bibinfo {volume} {531}},\ \bibinfo {pages}
  {206} (\bibinfo {year} {2016})}\BibitemShut {NoStop}%
\bibitem [{\citenamefont {Deng}\ \emph {et~al.}(2016)\citenamefont {Deng},
  \citenamefont {Vaitiekenas}, \citenamefont {Hansen}, \citenamefont {Danon},
  \citenamefont {Leijnse}, \citenamefont {Flensberg}, \citenamefont
  {Nyg{\aa}rd}, \citenamefont {Krogstrup},\ and\ \citenamefont
  {Marcus}}]{Deng2016}%
  \BibitemOpen
  \bibfield  {author} {\bibinfo {author} {\bibfnamefont {M.~T.}\ \bibnamefont
  {Deng}}, \bibinfo {author} {\bibfnamefont {S.}~\bibnamefont {Vaitiekenas}},
  \bibinfo {author} {\bibfnamefont {E.~B.}\ \bibnamefont {Hansen}}, \bibinfo
  {author} {\bibfnamefont {J.}~\bibnamefont {Danon}}, \bibinfo {author}
  {\bibfnamefont {M.}~\bibnamefont {Leijnse}}, \bibinfo {author} {\bibfnamefont
  {K.}~\bibnamefont {Flensberg}}, \bibinfo {author} {\bibfnamefont
  {J.}~\bibnamefont {Nyg{\aa}rd}}, \bibinfo {author} {\bibfnamefont
  {P.}~\bibnamefont {Krogstrup}}, \ and\ \bibinfo {author} {\bibfnamefont
  {C.~M.}\ \bibnamefont {Marcus}},\ }\href
  {https://doi.org/10.1126/science.aaf3961} {\bibfield  {journal} {\bibinfo
  {journal} {Science}\ }\textbf {\bibinfo {volume} {354}},\ \bibinfo {pages}
  {1557} (\bibinfo {year} {2016})}\BibitemShut {NoStop}%
\bibitem [{\citenamefont {Nichele}\ \emph {et~al.}(2017)\citenamefont
  {Nichele}, \citenamefont {Drachmann}, \citenamefont {Whiticar}, \citenamefont
  {O'Farrell}, \citenamefont {Suominen}, \citenamefont {Fornieri},
  \citenamefont {Wang}, \citenamefont {Gardner}, \citenamefont {Thomas},
  \citenamefont {Hatke}, \citenamefont {Krogstrup}, \citenamefont {Manfra},
  \citenamefont {Flensberg},\ and\ \citenamefont {Marcus}}]{Nichele2017}%
  \BibitemOpen
  \bibfield  {author} {\bibinfo {author} {\bibfnamefont {F.}~\bibnamefont
  {Nichele}}, \bibinfo {author} {\bibfnamefont {A.~C.~C.}\ \bibnamefont
  {Drachmann}}, \bibinfo {author} {\bibfnamefont {A.~M.}\ \bibnamefont
  {Whiticar}}, \bibinfo {author} {\bibfnamefont {E.~C.~T.}\ \bibnamefont
  {O'Farrell}}, \bibinfo {author} {\bibfnamefont {H.~J.}\ \bibnamefont
  {Suominen}}, \bibinfo {author} {\bibfnamefont {A.}~\bibnamefont {Fornieri}},
  \bibinfo {author} {\bibfnamefont {T.}~\bibnamefont {Wang}}, \bibinfo {author}
  {\bibfnamefont {G.~C.}\ \bibnamefont {Gardner}}, \bibinfo {author}
  {\bibfnamefont {C.}~\bibnamefont {Thomas}}, \bibinfo {author} {\bibfnamefont
  {A.~T.}\ \bibnamefont {Hatke}}, \bibinfo {author} {\bibfnamefont
  {P.}~\bibnamefont {Krogstrup}}, \bibinfo {author} {\bibfnamefont {M.~J.}\
  \bibnamefont {Manfra}}, \bibinfo {author} {\bibfnamefont {K.}~\bibnamefont
  {Flensberg}}, \ and\ \bibinfo {author} {\bibfnamefont {C.~M.}\ \bibnamefont
  {Marcus}},\ }\href
  {https://journals.aps.org/prl/abstract/10.1103/PhysRevLett.119.136803}
  {\bibfield  {journal} {\bibinfo  {journal} {Phys. Rev. Lett.}\ }\textbf
  {\bibinfo {volume} {119}},\ \bibinfo {pages} {136803} (\bibinfo {year}
  {2017})}\BibitemShut {NoStop}%
\bibitem [{\citenamefont {Gazibegovich}\ \emph {et~al.}(2017)\citenamefont
  {Gazibegovich}, \citenamefont {Car}, \citenamefont {Zhang}, \citenamefont
  {Balk}, \citenamefont {Logan}, \citenamefont {de~Moor}, \citenamefont
  {Cassidy}, \citenamefont {Schmits}, \citenamefont {Xu}, \citenamefont {Wang},
  \citenamefont {Krogstrup}, \citenamefont {Op~het Veld}, \citenamefont {Shen},
  \citenamefont {Bouman}, \citenamefont {Shojaei}, \citenamefont {Pennachio},
  \citenamefont {Lee}, \citenamefont {van Veldhoven}, \citenamefont {Koelling},
  \citenamefont {Verheijen}, \citenamefont {Kouwenhoven}, \citenamefont
  {Palmstr{\o}m},\ and\ \citenamefont {Bakkers}}]{Gazi2017}%
  \BibitemOpen
  \bibfield  {author} {\bibinfo {author} {\bibfnamefont {S.}~\bibnamefont
  {Gazibegovich}}, \bibinfo {author} {\bibfnamefont {D.}~\bibnamefont {Car}},
  \bibinfo {author} {\bibfnamefont {H.}~\bibnamefont {Zhang}}, \bibinfo
  {author} {\bibfnamefont {S.~C.}\ \bibnamefont {Balk}}, \bibinfo {author}
  {\bibfnamefont {J.~A.}\ \bibnamefont {Logan}}, \bibinfo {author}
  {\bibfnamefont {M.~W.~A.}\ \bibnamefont {de~Moor}}, \bibinfo {author}
  {\bibfnamefont {M.~C.}\ \bibnamefont {Cassidy}}, \bibinfo {author}
  {\bibfnamefont {R.}~\bibnamefont {Schmits}}, \bibinfo {author} {\bibfnamefont
  {D.}~\bibnamefont {Xu}}, \bibinfo {author} {\bibfnamefont {G.}~\bibnamefont
  {Wang}}, \bibinfo {author} {\bibfnamefont {P.}~\bibnamefont {Krogstrup}},
  \bibinfo {author} {\bibfnamefont {R.~L.~M.}\ \bibnamefont {Op~het Veld}},
  \bibinfo {author} {\bibfnamefont {J.}~\bibnamefont {Shen}}, \bibinfo {author}
  {\bibfnamefont {D.}~\bibnamefont {Bouman}}, \bibinfo {author} {\bibfnamefont
  {B.}~\bibnamefont {Shojaei}}, \bibinfo {author} {\bibfnamefont
  {D.}~\bibnamefont {Pennachio}}, \bibinfo {author} {\bibfnamefont {J.~S.}\
  \bibnamefont {Lee}}, \bibinfo {author} {\bibfnamefont {P.~J.}\ \bibnamefont
  {van Veldhoven}}, \bibinfo {author} {\bibfnamefont {S.}~\bibnamefont
  {Koelling}}, \bibinfo {author} {\bibfnamefont {M.~A.}\ \bibnamefont
  {Verheijen}}, \bibinfo {author} {\bibfnamefont {L.~P.}\ \bibnamefont
  {Kouwenhoven}}, \bibinfo {author} {\bibfnamefont {C.~J.}\ \bibnamefont
  {Palmstr{\o}m}}, \ and\ \bibinfo {author} {\bibfnamefont {E.~P. A.~M.}\
  \bibnamefont {Bakkers}},\ }\href
  {https://www.nature.com/articles/nature23468} {\bibfield  {journal} {\bibinfo
   {journal} {Nature}\ }\textbf {\bibinfo {volume} {548}},\ \bibinfo {pages}
  {434} (\bibinfo {year} {2017})}\BibitemShut {NoStop}%
\bibitem [{\citenamefont {Zhang}\ \emph {et~al.}(2018)\citenamefont {Zhang},
  \citenamefont {Liu}, \citenamefont {Gazibegovich}, \citenamefont {Xu},
  \citenamefont {Logan}, \citenamefont {Wang}, \citenamefont {van Loo},
  \citenamefont {Bommer}, \citenamefont {de~Moor}, \citenamefont {Car},
  \citenamefont {Op~het Veld}, \citenamefont {Veldhoven}, \citenamefont
  {Koelling}, \citenamefont {Verheijen}, \citenamefont {Palmstr{\o}m},
  \citenamefont {Pendharkar}, \citenamefont {Pennachio}, \citenamefont
  {Shojaei}, \citenamefont {Lee}, \citenamefont {Bakkers},\ and\ \citenamefont
  {Kouwenhoven}}]{Zhang2018}%
  \BibitemOpen
  \bibfield  {author} {\bibinfo {author} {\bibfnamefont {H.}~\bibnamefont
  {Zhang}}, \bibinfo {author} {\bibfnamefont {C.~X.}\ \bibnamefont {Liu}},
  \bibinfo {author} {\bibfnamefont {S.}~\bibnamefont {Gazibegovich}}, \bibinfo
  {author} {\bibfnamefont {D.}~\bibnamefont {Xu}}, \bibinfo {author}
  {\bibfnamefont {J.}~\bibnamefont {Logan}}, \bibinfo {author} {\bibfnamefont
  {G.}~\bibnamefont {Wang}}, \bibinfo {author} {\bibfnamefont {N.}~\bibnamefont
  {van Loo}}, \bibinfo {author} {\bibfnamefont {J.~D.~S.}\ \bibnamefont
  {Bommer}}, \bibinfo {author} {\bibfnamefont {M.~W.~A.}\ \bibnamefont
  {de~Moor}}, \bibinfo {author} {\bibfnamefont {D.}~\bibnamefont {Car}},
  \bibinfo {author} {\bibfnamefont {R.~L.~M.}\ \bibnamefont {Op~het Veld}},
  \bibinfo {author} {\bibfnamefont {P.~J.}\ \bibnamefont {Veldhoven}}, \bibinfo
  {author} {\bibfnamefont {S.}~\bibnamefont {Koelling}}, \bibinfo {author}
  {\bibfnamefont {M.~A.}\ \bibnamefont {Verheijen}}, \bibinfo {author}
  {\bibfnamefont {C.~J.}\ \bibnamefont {Palmstr{\o}m}}, \bibinfo {author}
  {\bibfnamefont {M.}~\bibnamefont {Pendharkar}}, \bibinfo {author}
  {\bibfnamefont {D.~J.}\ \bibnamefont {Pennachio}}, \bibinfo {author}
  {\bibfnamefont {B.}~\bibnamefont {Shojaei}}, \bibinfo {author} {\bibfnamefont
  {J.~S.}\ \bibnamefont {Lee}}, \bibinfo {author} {\bibfnamefont {E.~P. A.~M.}\
  \bibnamefont {Bakkers}}, \ and\ \bibinfo {author} {\bibfnamefont {L.~P.}\
  \bibnamefont {Kouwenhoven}},\ }\href
  {https://www.nature.com/articles/nature26142} {\bibfield  {journal} {\bibinfo
   {journal} {Nature}\ }\textbf {\bibinfo {volume} {556}},\ \bibinfo {pages}
  {74} (\bibinfo {year} {2018})}\BibitemShut {NoStop}%
\bibitem [{\citenamefont {B{\'e}ri}\ and\ \citenamefont
  {Cooper}(2012)}]{Beri2012}%
  \BibitemOpen
  \bibfield  {author} {\bibinfo {author} {\bibfnamefont {B.}~\bibnamefont
  {B{\'e}ri}}\ and\ \bibinfo {author} {\bibfnamefont {N.~R.}\ \bibnamefont
  {Cooper}},\ }\href
  {https://journals.aps.org/prl/abstract/10.1103/PhysRevLett.109.156803}
  {\bibfield  {journal} {\bibinfo  {journal} {Phys. Rev. Lett.}\ }\textbf
  {\bibinfo {volume} {109}},\ \bibinfo {pages} {156803} (\bibinfo {year}
  {2012})}\BibitemShut {NoStop}%
\bibitem [{\citenamefont {B{\'e}ri}(2013)}]{Beri2013}%
  \BibitemOpen
  \bibfield  {author} {\bibinfo {author} {\bibfnamefont {B.}~\bibnamefont
  {B{\'e}ri}},\ }\href
  {https://journals.aps.org/prl/abstract/10.1103/PhysRevLett.110.216803}
  {\bibfield  {journal} {\bibinfo  {journal} {Phys. Rev. Lett.}\ }\textbf
  {\bibinfo {volume} {110}},\ \bibinfo {pages} {216803} (\bibinfo {year}
  {2013})}\BibitemShut {NoStop}%
\bibitem [{\citenamefont {Altland}\ and\ \citenamefont
  {Egger}(2013)}]{Altland2013}%
  \BibitemOpen
  \bibfield  {author} {\bibinfo {author} {\bibfnamefont {A.}~\bibnamefont
  {Altland}}\ and\ \bibinfo {author} {\bibfnamefont {R.}~\bibnamefont
  {Egger}},\ }\href
  {https://journals.aps.org/prl/abstract/10.1103/PhysRevLett.110.196401}
  {\bibfield  {journal} {\bibinfo  {journal} {Phys. Rev. Lett.}\ }\textbf
  {\bibinfo {volume} {110}},\ \bibinfo {pages} {196401} (\bibinfo {year}
  {2013})}\BibitemShut {NoStop}%
\bibitem [{\citenamefont {Plugge}\ \emph {et~al.}(2017)\citenamefont {Plugge},
  \citenamefont {Rasmussen}, \citenamefont {Egger},\ and\ \citenamefont
  {Flensberg}}]{Plugge2017}%
  \BibitemOpen
  \bibfield  {author} {\bibinfo {author} {\bibfnamefont {S.}~\bibnamefont
  {Plugge}}, \bibinfo {author} {\bibfnamefont {A.}~\bibnamefont {Rasmussen}},
  \bibinfo {author} {\bibfnamefont {R.}~\bibnamefont {Egger}}, \ and\ \bibinfo
  {author} {\bibfnamefont {K.}~\bibnamefont {Flensberg}},\ }\href
  {http://iopscience.iop.org/article/10.1088/1367-2630/aa54e1} {\bibfield
  {journal} {\bibinfo  {journal} {New J. Phys.}\ }\textbf {\bibinfo {volume}
  {19}},\ \bibinfo {pages} {012001} (\bibinfo {year} {2017})}\BibitemShut
  {NoStop}%
\bibitem [{\citenamefont {Karzig}\ \emph {et~al.}(2017)\citenamefont {Karzig},
  \citenamefont {Knapp}, \citenamefont {Lutchyn}, \citenamefont {Bonderson},
  \citenamefont {Hastings}, \citenamefont {Nayak}, \citenamefont {Alicea},
  \citenamefont {Flensberg}, \citenamefont {Plugge}, \citenamefont {Oreg},
  \citenamefont {Marcus},\ and\ \citenamefont {Freedman}}]{Karzig2017}%
  \BibitemOpen
  \bibfield  {author} {\bibinfo {author} {\bibfnamefont {T.}~\bibnamefont
  {Karzig}}, \bibinfo {author} {\bibfnamefont {C.}~\bibnamefont {Knapp}},
  \bibinfo {author} {\bibfnamefont {R.~M.}\ \bibnamefont {Lutchyn}}, \bibinfo
  {author} {\bibfnamefont {P.}~\bibnamefont {Bonderson}}, \bibinfo {author}
  {\bibfnamefont {M.~B.}\ \bibnamefont {Hastings}}, \bibinfo {author}
  {\bibfnamefont {C.}~\bibnamefont {Nayak}}, \bibinfo {author} {\bibfnamefont
  {J.}~\bibnamefont {Alicea}}, \bibinfo {author} {\bibfnamefont
  {K.}~\bibnamefont {Flensberg}}, \bibinfo {author} {\bibfnamefont
  {S.}~\bibnamefont {Plugge}}, \bibinfo {author} {\bibfnamefont
  {Y.}~\bibnamefont {Oreg}}, \bibinfo {author} {\bibfnamefont {C.~M.}\
  \bibnamefont {Marcus}}, \ and\ \bibinfo {author} {\bibfnamefont {M.~H.}\
  \bibnamefont {Freedman}},\ }\href
  {https://journals.aps.org/prb/abstract/10.1103/PhysRevB.95.235305} {\bibfield
   {journal} {\bibinfo  {journal} {Phys. Rev. B}\ }\textbf {\bibinfo {volume}
  {95}},\ \bibinfo {pages} {235305} (\bibinfo {year} {2017})}\BibitemShut
  {NoStop}%
\bibitem [{\citenamefont {Gu}\ \emph {et~al.}(2009)\citenamefont {Gu},
  \citenamefont {Levin}, \citenamefont {Swingle},\ and\ \citenamefont
  {Wen}}]{Gu2009}%
  \BibitemOpen
  \bibfield  {author} {\bibinfo {author} {\bibfnamefont {Z.-C.}\ \bibnamefont
  {Gu}}, \bibinfo {author} {\bibfnamefont {M.}~\bibnamefont {Levin}}, \bibinfo
  {author} {\bibfnamefont {B.}~\bibnamefont {Swingle}}, \ and\ \bibinfo
  {author} {\bibfnamefont {X.-G.}\ \bibnamefont {Wen}},\ }\href {\doibase
  10.1103/PhysRevB.79.085118} {\bibfield  {journal} {\bibinfo  {journal} {Phys.
  Rev. B}\ }\textbf {\bibinfo {volume} {79}},\ \bibinfo {pages} {085118}
  (\bibinfo {year} {2009})}\BibitemShut {NoStop}%
\bibitem [{\citenamefont {Buerschaper}\ \emph {et~al.}(2009)\citenamefont
  {Buerschaper}, \citenamefont {Aguado},\ and\ \citenamefont
  {Vidal}}]{Buerschaper2009}%
  \BibitemOpen
  \bibfield  {author} {\bibinfo {author} {\bibfnamefont {O.}~\bibnamefont
  {Buerschaper}}, \bibinfo {author} {\bibfnamefont {M.}~\bibnamefont {Aguado}},
  \ and\ \bibinfo {author} {\bibfnamefont {G.}~\bibnamefont {Vidal}},\ }\href
  {\doibase 10.1103/PhysRevB.79.085119} {\bibfield  {journal} {\bibinfo
  {journal} {Phys. Rev. B}\ }\textbf {\bibinfo {volume} {79}},\ \bibinfo
  {pages} {085119} (\bibinfo {year} {2009})}\BibitemShut {NoStop}%
\bibitem [{\citenamefont {Schuch}\ \emph {et~al.}(2008)\citenamefont {Schuch},
  \citenamefont {Wolf}, \citenamefont {Verstraete},\ and\ \citenamefont
  {Cirac}}]{Schuch_MPS}%
  \BibitemOpen
  \bibfield  {author} {\bibinfo {author} {\bibfnamefont {N.}~\bibnamefont
  {Schuch}}, \bibinfo {author} {\bibfnamefont {M.~M.}\ \bibnamefont {Wolf}},
  \bibinfo {author} {\bibfnamefont {F.}~\bibnamefont {Verstraete}}, \ and\
  \bibinfo {author} {\bibfnamefont {J.~I.}\ \bibnamefont {Cirac}},\ }\href
  {\doibase 10.1103/PhysRevLett.100.030504} {\bibfield  {journal} {\bibinfo
  {journal} {Phys. Rev. Lett.}\ }\textbf {\bibinfo {volume} {100}},\ \bibinfo
  {pages} {030504} (\bibinfo {year} {2008})}\BibitemShut {NoStop}%
\bibitem [{\citenamefont {Sahinoglu}\ \emph {et~al.}()\citenamefont
  {Sahinoglu}, \citenamefont {Williamson}, \citenamefont {Bultinck},
  \citenamefont {Marien}, \citenamefont {Haegeman}, \citenamefont {Schuch},\
  and\ \citenamefont {Verstraete}}]{1409.2150}%
  \BibitemOpen
  \bibfield  {author} {\bibinfo {author} {\bibfnamefont {M.~B.}\ \bibnamefont
  {Sahinoglu}}, \bibinfo {author} {\bibfnamefont {D.}~\bibnamefont
  {Williamson}}, \bibinfo {author} {\bibfnamefont {N.}~\bibnamefont
  {Bultinck}}, \bibinfo {author} {\bibfnamefont {M.}~\bibnamefont {Marien}},
  \bibinfo {author} {\bibfnamefont {J.}~\bibnamefont {Haegeman}}, \bibinfo
  {author} {\bibfnamefont {N.}~\bibnamefont {Schuch}}, \ and\ \bibinfo {author}
  {\bibfnamefont {F.}~\bibnamefont {Verstraete}},\ }\href@noop {} {\ }\Eprint
  {http://arxiv.org/abs/1409.2150} {arxiv:1409.2150} \BibitemShut {NoStop}%
\bibitem [{\citenamefont {Williamson}\ \emph {et~al.}()\citenamefont
  {Williamson}, \citenamefont {Bultinck}, \citenamefont {Haegeman},\ and\
  \citenamefont {Verstraete}}]{FermionicMPO}%
  \BibitemOpen
  \bibfield  {author} {\bibinfo {author} {\bibfnamefont {D.~J.}\ \bibnamefont
  {Williamson}}, \bibinfo {author} {\bibfnamefont {N.}~\bibnamefont
  {Bultinck}}, \bibinfo {author} {\bibfnamefont {J.}~\bibnamefont {Haegeman}},
  \ and\ \bibinfo {author} {\bibfnamefont {F.}~\bibnamefont {Verstraete}},\
  }\href@noop {} {\ }\Eprint {http://arxiv.org/abs/1609.02897}
  {arXiv:1609.02897} \BibitemShut {NoStop}%
\bibitem [{\citenamefont {Wille}\ \emph {et~al.}(2017)\citenamefont {Wille},
  \citenamefont {Buerschaper},\ and\ \citenamefont
  {Eisert}}]{PhysRevB.95.245127}%
  \BibitemOpen
  \bibfield  {author} {\bibinfo {author} {\bibfnamefont {C.}~\bibnamefont
  {Wille}}, \bibinfo {author} {\bibfnamefont {O.}~\bibnamefont {Buerschaper}},
  \ and\ \bibinfo {author} {\bibfnamefont {J.}~\bibnamefont {Eisert}},\ }\href
  {\doibase 10.1103/PhysRevB.95.245127} {\bibfield  {journal} {\bibinfo
  {journal} {Phys. Rev. B}\ }\textbf {\bibinfo {volume} {95}},\ \bibinfo
  {pages} {245127} (\bibinfo {year} {2017})}\BibitemShut {NoStop}%
\bibitem [{\citenamefont {Brell}\ \emph
  {et~al.}(2014{\natexlab{a}})\citenamefont {Brell}, \citenamefont {Bartlett},\
  and\ \citenamefont {Doherty}}]{Brell2014PEPS}%
  \BibitemOpen
  \bibfield  {author} {\bibinfo {author} {\bibfnamefont {C.~G.}\ \bibnamefont
  {Brell}}, \bibinfo {author} {\bibfnamefont {S.~D.}\ \bibnamefont {Bartlett}},
  \ and\ \bibinfo {author} {\bibfnamefont {A.~C.}\ \bibnamefont {Doherty}},\
  }\href {http://stacks.iop.org/1367-2630/16/i=12/a=123056} {\bibfield
  {journal} {\bibinfo  {journal} {New J. Phys.}\ }\textbf {\bibinfo {volume}
  {16}},\ \bibinfo {pages} {123056} (\bibinfo {year}
  {2014}{\natexlab{a}})}\BibitemShut {NoStop}%
\bibitem [{\citenamefont {Kempe}\ \emph {et~al.}(2006)\citenamefont {Kempe},
  \citenamefont {Kitaev},\ and\ \citenamefont {Regev}}]{Kempe-SIAM-2006}%
  \BibitemOpen
  \bibfield  {author} {\bibinfo {author} {\bibfnamefont {J.}~\bibnamefont
  {Kempe}}, \bibinfo {author} {\bibfnamefont {A.}~\bibnamefont {Kitaev}}, \
  and\ \bibinfo {author} {\bibfnamefont {O.}~\bibnamefont {Regev}},\ }\href
  {\doibase 10.1137/s0097539704445226} {\bibfield  {journal} {\bibinfo
  {journal} {SIAM J. Comp.}\ }\textbf {\bibinfo {volume} {35}},\ \bibinfo
  {pages} {1070} (\bibinfo {year} {2006})}\BibitemShut {NoStop}%
\bibitem [{\citenamefont {Jordan}\ and\ \citenamefont
  {Farhi}(2008)}]{PhysRevA.77.062329}%
  \BibitemOpen
  \bibfield  {author} {\bibinfo {author} {\bibfnamefont {S.~P.}\ \bibnamefont
  {Jordan}}\ and\ \bibinfo {author} {\bibfnamefont {E.}~\bibnamefont {Farhi}},\
  }\href {\doibase 10.1103/PhysRevA.77.062329} {\bibfield  {journal} {\bibinfo
  {journal} {Phys. Rev. A}\ }\textbf {\bibinfo {volume} {77}},\ \bibinfo
  {pages} {062329} (\bibinfo {year} {2008})}\BibitemShut {NoStop}%
\bibitem [{Note1()}]{Note1}%
  \BibitemOpen
  \bibinfo {note} {{\protect \color {alex} For the double semion model
  restricted to the ground state space of all vertex operators, there is a
  mapping from a Hamiltonian with 12-local to 6-local plaquette operators \cite
  {SemionErrorCorrection}. However, the resulting 6-local operators have a more
  complicated structure and, in particular, they are no longer given by product
  operators. As a consequence, there is no straightforward way to implement the
  corresponding Hamiltonian.}}\BibitemShut {Stop}%
\bibitem [{\citenamefont {Aasen}\ \emph {et~al.}(2016)\citenamefont {Aasen},
  \citenamefont {Hell}, \citenamefont {Mishmash}, \citenamefont {Higginbotham},
  \citenamefont {Danon}, \citenamefont {Leijnse}, \citenamefont {Jespersen},
  \citenamefont {Folk}, \citenamefont {Marcus}, \citenamefont {Flensberg},\
  and\ \citenamefont {Alicea}}]{PhysRevX.6.031016}%
  \BibitemOpen
  \bibfield  {author} {\bibinfo {author} {\bibfnamefont {D.}~\bibnamefont
  {Aasen}}, \bibinfo {author} {\bibfnamefont {M.}~\bibnamefont {Hell}},
  \bibinfo {author} {\bibfnamefont {R.~V.}\ \bibnamefont {Mishmash}}, \bibinfo
  {author} {\bibfnamefont {A.}~\bibnamefont {Higginbotham}}, \bibinfo {author}
  {\bibfnamefont {J.}~\bibnamefont {Danon}}, \bibinfo {author} {\bibfnamefont
  {M.}~\bibnamefont {Leijnse}}, \bibinfo {author} {\bibfnamefont {T.~S.}\
  \bibnamefont {Jespersen}}, \bibinfo {author} {\bibfnamefont {J.~A.}\
  \bibnamefont {Folk}}, \bibinfo {author} {\bibfnamefont {C.~M.}\ \bibnamefont
  {Marcus}}, \bibinfo {author} {\bibfnamefont {K.}~\bibnamefont {Flensberg}}, \
  and\ \bibinfo {author} {\bibfnamefont {J.}~\bibnamefont {Alicea}},\ }\href
  {\doibase 10.1103/PhysRevX.6.031016} {\bibfield  {journal} {\bibinfo
  {journal} {Phys. Rev. X}\ }\textbf {\bibinfo {volume} {6}},\ \bibinfo {pages}
  {031016} (\bibinfo {year} {2016})}\BibitemShut {NoStop}%
\bibitem [{\citenamefont {Roy}\ \emph {et~al.}(2017)\citenamefont {Roy},
  \citenamefont {Terhal},\ and\ \citenamefont {Hassler}}]{Roy2017}%
  \BibitemOpen
  \bibfield  {author} {\bibinfo {author} {\bibfnamefont {A.}~\bibnamefont
  {Roy}}, \bibinfo {author} {\bibfnamefont {B.~M.}\ \bibnamefont {Terhal}}, \
  and\ \bibinfo {author} {\bibfnamefont {F.}~\bibnamefont {Hassler}},\ }\href
  {\doibase 10.1103/PhysRevLett.119.180508} {\bibfield  {journal} {\bibinfo
  {journal} {Phys. Rev. Lett.}\ }\textbf {\bibinfo {volume} {119}},\ \bibinfo
  {pages} {180508} (\bibinfo {year} {2017})}\BibitemShut {NoStop}%
\bibitem [{\citenamefont {Verstraete}\ and\ \citenamefont {Cirac}()}]{PEPSOld}%
  \BibitemOpen
  \bibfield  {author} {\bibinfo {author} {\bibfnamefont {F.}~\bibnamefont
  {Verstraete}}\ and\ \bibinfo {author} {\bibfnamefont {J.~I.}\ \bibnamefont
  {Cirac}},\ }\href@noop {} {}\Eprint {http://arxiv.org/abs/0407066}
  {cond-mat:0407066} \BibitemShut {NoStop}%
\bibitem [{\citenamefont {Phien}\ \emph {et~al.}(2015)\citenamefont {Phien},
  \citenamefont {Bengua}, \citenamefont {Tuan}, \citenamefont {Corboz},\ and\
  \citenamefont {Orus}}]{iPEPS}%
  \BibitemOpen
  \bibfield  {author} {\bibinfo {author} {\bibfnamefont {H.~N.}\ \bibnamefont
  {Phien}}, \bibinfo {author} {\bibfnamefont {J.~A.}\ \bibnamefont {Bengua}},
  \bibinfo {author} {\bibfnamefont {H.~D.}\ \bibnamefont {Tuan}}, \bibinfo
  {author} {\bibfnamefont {P.}~\bibnamefont {Corboz}}, \ and\ \bibinfo {author}
  {\bibfnamefont {R.}~\bibnamefont {Orus}},\ }\href {\doibase
  10.1103/PhysRevB.92.035142} {\bibfield  {journal} {\bibinfo  {journal} {Phys.
  Rev. B}\ }\textbf {\bibinfo {volume} {92}},\ \bibinfo {pages} {035142}
  (\bibinfo {year} {2015})}\BibitemShut {NoStop}%
\bibitem [{\citenamefont {White}(1992)}]{DMRGWhite92}%
  \BibitemOpen
  \bibfield  {author} {\bibinfo {author} {\bibfnamefont {S.~R.}\ \bibnamefont
  {White}},\ }\href {\doibase 10.1103/PhysRevLett.69.2863} {\bibfield
  {journal} {\bibinfo  {journal} {Phys. Rev. Lett.}\ }\textbf {\bibinfo
  {volume} {69}},\ \bibinfo {pages} {2863} (\bibinfo {year}
  {1992})}\BibitemShut {NoStop}%
\bibitem [{\citenamefont {Schuch}\ \emph {et~al.}(2013)\citenamefont {Schuch},
  \citenamefont {Poilblanc}, \citenamefont {Cirac},\ and\ \citenamefont
  {Perez-Garcia}}]{TopologicalOrderInPEPS}%
  \BibitemOpen
  \bibfield  {author} {\bibinfo {author} {\bibfnamefont {N.}~\bibnamefont
  {Schuch}}, \bibinfo {author} {\bibfnamefont {D.}~\bibnamefont {Poilblanc}},
  \bibinfo {author} {\bibfnamefont {J.~I.}\ \bibnamefont {Cirac}}, \ and\
  \bibinfo {author} {\bibfnamefont {D.}~\bibnamefont {Perez-Garcia}},\ }\href
  {\doibase 10.1103/PhysRevLett.111.090501} {\bibfield  {journal} {\bibinfo
  {journal} {Phys. Rev. Lett.}\ }\textbf {\bibinfo {volume} {111}},\ \bibinfo
  {pages} {090501} (\bibinfo {year} {2013})}\BibitemShut {NoStop}%
\bibitem [{\citenamefont {Schuch}\ \emph {et~al.}(2010)\citenamefont {Schuch},
  \citenamefont {Cirac},\ and\ \citenamefont {Perez-Garcia}}]{PEPSTopology}%
  \BibitemOpen
  \bibfield  {author} {\bibinfo {author} {\bibfnamefont {N.}~\bibnamefont
  {Schuch}}, \bibinfo {author} {\bibfnamefont {J.~I.}\ \bibnamefont {Cirac}}, \
  and\ \bibinfo {author} {\bibfnamefont {D.}~\bibnamefont {Perez-Garcia}},\
  }\href {\doibase 10.1016/j.aop.2010.05.008} {\bibfield  {journal} {\bibinfo
  {journal} {Ann. Phys.}\ }\textbf {\bibinfo {volume} {325}},\ \bibinfo {pages}
  {2153} (\bibinfo {year} {2010})}\BibitemShut {NoStop}%
\bibitem [{\citenamefont {Schuch}\ \emph {et~al.}(2011)\citenamefont {Schuch},
  \citenamefont {Perez-Garcia},\ and\ \citenamefont
  {Cirac}}]{ClassificationPhases}%
  \BibitemOpen
  \bibfield  {author} {\bibinfo {author} {\bibfnamefont {N.}~\bibnamefont
  {Schuch}}, \bibinfo {author} {\bibfnamefont {D.}~\bibnamefont
  {Perez-Garcia}}, \ and\ \bibinfo {author} {\bibfnamefont {I.}~\bibnamefont
  {Cirac}},\ }\href {\doibase 10.1103/PhysRevB.84.165139} {\bibfield  {journal}
  {\bibinfo  {journal} {Phys. Rev. B}\ }\textbf {\bibinfo {volume} {84}},\
  \bibinfo {pages} {165139} (\bibinfo {year} {2011})}\BibitemShut {NoStop}%
\bibitem [{\citenamefont {Pollmann}\ \emph {et~al.}(2010)\citenamefont
  {Pollmann}, \citenamefont {Turner}, \citenamefont {Berg},\ and\ \citenamefont
  {Oshikawa}}]{PhysRevB.81.064439}%
  \BibitemOpen
  \bibfield  {author} {\bibinfo {author} {\bibfnamefont {F.}~\bibnamefont
  {Pollmann}}, \bibinfo {author} {\bibfnamefont {A.~M.}\ \bibnamefont
  {Turner}}, \bibinfo {author} {\bibfnamefont {E.}~\bibnamefont {Berg}}, \ and\
  \bibinfo {author} {\bibfnamefont {M.}~\bibnamefont {Oshikawa}},\ }\href
  {\doibase 10.1103/PhysRevB.81.064439} {\bibfield  {journal} {\bibinfo
  {journal} {Phys. Rev. B}\ }\textbf {\bibinfo {volume} {81}},\ \bibinfo
  {pages} {064439} (\bibinfo {year} {2010})}\BibitemShut {NoStop}%
\bibitem [{\citenamefont {Xu}\ and\ \citenamefont {Fu}(2010)}]{Xu2010}%
  \BibitemOpen
  \bibfield  {author} {\bibinfo {author} {\bibfnamefont {C.}~\bibnamefont
  {Xu}}\ and\ \bibinfo {author} {\bibfnamefont {L.}~\bibnamefont {Fu}},\ }\href
  {https://journals.aps.org/prb/abstract/10.1103/PhysRevB.81.134435} {\bibfield
   {journal} {\bibinfo  {journal} {Phys. Rev. B}\ }\textbf {\bibinfo {volume}
  {81}},\ \bibinfo {pages} {134435} (\bibinfo {year} {2010})}\BibitemShut
  {NoStop}%
\bibitem [{\citenamefont {Nussinov}\ \emph {et~al.}(2012)\citenamefont
  {Nussinov}, \citenamefont {Ortiz},\ and\ \citenamefont
  {Cobanera}}]{Nussinov2012}%
  \BibitemOpen
  \bibfield  {author} {\bibinfo {author} {\bibfnamefont {Z.}~\bibnamefont
  {Nussinov}}, \bibinfo {author} {\bibfnamefont {G.}~\bibnamefont {Ortiz}}, \
  and\ \bibinfo {author} {\bibfnamefont {E.}~\bibnamefont {Cobanera}},\ }\href
  {https://journals.aps.org/prb/abstract/10.1103/PhysRevB.86.085415} {\bibfield
   {journal} {\bibinfo  {journal} {Phys. Rev. B}\ }\textbf {\bibinfo {volume}
  {86}},\ \bibinfo {pages} {085415} (\bibinfo {year} {2012})}\BibitemShut
  {NoStop}%
\bibitem [{\citenamefont {Vijay}\ and\ \citenamefont {Fu}(2016)}]{Vijay2016}%
  \BibitemOpen
  \bibfield  {author} {\bibinfo {author} {\bibfnamefont {S.}~\bibnamefont
  {Vijay}}\ and\ \bibinfo {author} {\bibfnamefont {L.}~\bibnamefont {Fu}},\
  }\href {https://journals.aps.org/prb/abstract/10.1103/PhysRevB.94.235446}
  {\bibfield  {journal} {\bibinfo  {journal} {Phys. Rev. B}\ }\textbf {\bibinfo
  {volume} {94}},\ \bibinfo {pages} {235446} (\bibinfo {year}
  {2016})}\BibitemShut {NoStop}%
\bibitem [{\citenamefont {Litinski}\ \emph {et~al.}(2017)\citenamefont
  {Litinski}, \citenamefont {Kesselring}, \citenamefont {Eisert},\ and\
  \citenamefont {Oppen}}]{Litinski2017}%
  \BibitemOpen
  \bibfield  {author} {\bibinfo {author} {\bibfnamefont {D.}~\bibnamefont
  {Litinski}}, \bibinfo {author} {\bibfnamefont {M.~S.}\ \bibnamefont
  {Kesselring}}, \bibinfo {author} {\bibfnamefont {J.}~\bibnamefont {Eisert}},
  \ and\ \bibinfo {author} {\bibfnamefont {F.~v.}\ \bibnamefont {Oppen}},\
  }\href {https://journals.aps.org/prx/abstract/10.1103/PhysRevX.7.031048}
  {\bibfield  {journal} {\bibinfo  {journal} {Phys. Rev. X}\ }\textbf {\bibinfo
  {volume} {7}},\ \bibinfo {pages} {031048} (\bibinfo {year}
  {2017})}\BibitemShut {NoStop}%
\bibitem [{\citenamefont {Fu}(2010)}]{Fu2010}%
  \BibitemOpen
  \bibfield  {author} {\bibinfo {author} {\bibfnamefont {L.}~\bibnamefont
  {Fu}},\ }\href
  {https://journals.aps.org/prl/abstract/10.1103/PhysRevLett.104.056402}
  {\bibfield  {journal} {\bibinfo  {journal} {Phys. Rev. Lett.}\ }\textbf
  {\bibinfo {volume} {104}},\ \bibinfo {pages} {056402} (\bibinfo {year}
  {2010})}\BibitemShut {NoStop}%
\bibitem [{\citenamefont {Hyart}\ \emph {et~al.}(2013)\citenamefont {Hyart},
  \citenamefont {van Heck}, \citenamefont {Fulga}, \citenamefont {Burrello},
  \citenamefont {Akhmerov},\ and\ \citenamefont {Beenakker}}]{Hyart2013}%
  \BibitemOpen
  \bibfield  {author} {\bibinfo {author} {\bibfnamefont {T.}~\bibnamefont
  {Hyart}}, \bibinfo {author} {\bibfnamefont {B.}~\bibnamefont {van Heck}},
  \bibinfo {author} {\bibfnamefont {I.~C.}\ \bibnamefont {Fulga}}, \bibinfo
  {author} {\bibfnamefont {M.}~\bibnamefont {Burrello}}, \bibinfo {author}
  {\bibfnamefont {A.~R.}\ \bibnamefont {Akhmerov}}, \ and\ \bibinfo {author}
  {\bibfnamefont {C.~W.~J.}\ \bibnamefont {Beenakker}},\ }\href
  {https://journals.aps.org/prb/abstract/10.1103/PhysRevB.88.035121} {\bibfield
   {journal} {\bibinfo  {journal} {Phys. Rev. B}\ }\textbf {\bibinfo {volume}
  {88}},\ \bibinfo {pages} {035121} (\bibinfo {year} {2013})}\BibitemShut
  {NoStop}%
\bibitem [{\citenamefont {Kitaev}(2006)}]{Kitaev06}%
  \BibitemOpen
  \bibfield  {author} {\bibinfo {author} {\bibfnamefont {A.}~\bibnamefont
  {Kitaev}},\ }\href {\doibase https://doi.org/10.1016/j.aop.2005.10.005}
  {\bibfield  {journal} {\bibinfo  {journal} {Ann. Phys.}\ }\textbf {\bibinfo
  {volume} {321}},\ \bibinfo {pages} {2 } (\bibinfo {year} {2006})}\BibitemShut
  {NoStop}%
\bibitem [{\citenamefont {Brell}\ \emph {et~al.}(2011)\citenamefont {Brell},
  \citenamefont {Flammia}, \citenamefont {Bartlett},\ and\ \citenamefont
  {Doherty}}]{1367-2630-13-5-053039}%
  \BibitemOpen
  \bibfield  {author} {\bibinfo {author} {\bibfnamefont {C.~G.}\ \bibnamefont
  {Brell}}, \bibinfo {author} {\bibfnamefont {S.~T.}\ \bibnamefont {Flammia}},
  \bibinfo {author} {\bibfnamefont {S.~D.}\ \bibnamefont {Bartlett}}, \ and\
  \bibinfo {author} {\bibfnamefont {A.~C.}\ \bibnamefont {Doherty}},\ }\href
  {http://stacks.iop.org/1367-2630/13/i=5/a=053039} {\bibfield  {journal}
  {\bibinfo  {journal} {New J. Phys.}\ }\textbf {\bibinfo {volume} {13}},\
  \bibinfo {pages} {053039} (\bibinfo {year} {2011})}\BibitemShut {NoStop}%
\bibitem [{\citenamefont {Fidkowski}\ \emph {et~al.}(2009)\citenamefont
  {Fidkowski}, \citenamefont {Freedman}, \citenamefont {Nayak}, \citenamefont
  {Walker},\ and\ \citenamefont {Wang}}]{Fidkowski2009}%
  \BibitemOpen
  \bibfield  {author} {\bibinfo {author} {\bibfnamefont {L.}~\bibnamefont
  {Fidkowski}}, \bibinfo {author} {\bibfnamefont {M.}~\bibnamefont {Freedman}},
  \bibinfo {author} {\bibfnamefont {C.}~\bibnamefont {Nayak}}, \bibinfo
  {author} {\bibfnamefont {K.}~\bibnamefont {Walker}}, \ and\ \bibinfo {author}
  {\bibfnamefont {Z.}~\bibnamefont {Wang}},\ }\href
  {https://doi.org/10.1007/s00220-009-0757-9} {\bibfield  {journal} {\bibinfo
  {journal} {Comm. Math. Phys.}\ }\textbf {\bibinfo {volume} {287}},\ \bibinfo
  {pages} {805} (\bibinfo {year} {2009})}\BibitemShut {NoStop}%
\bibitem [{\citenamefont {Bravyi}\ \emph {et~al.}(2010)\citenamefont {Bravyi},
  \citenamefont {Hastings},\ and\ \citenamefont {Michalakis}}]{1001.0344}%
  \BibitemOpen
  \bibfield  {author} {\bibinfo {author} {\bibfnamefont {S.}~\bibnamefont
  {Bravyi}}, \bibinfo {author} {\bibfnamefont {M.~B.}\ \bibnamefont
  {Hastings}}, \ and\ \bibinfo {author} {\bibfnamefont {S.}~\bibnamefont
  {Michalakis}},\ }\href {\doibase 10.1063/1.3490195} {\bibfield  {journal}
  {\bibinfo  {journal} {J. Math. Phys.}\ }\textbf {\bibinfo {volume} {51}},\
  \bibinfo {pages} {093512} (\bibinfo {year} {2010})}\BibitemShut {NoStop}%
\bibitem [{\citenamefont {Bartlett}\ and\ \citenamefont
  {Rudolph}(2006)}]{Bartlett06}%
  \BibitemOpen
  \bibfield  {author} {\bibinfo {author} {\bibfnamefont {S.~D.}\ \bibnamefont
  {Bartlett}}\ and\ \bibinfo {author} {\bibfnamefont {T.}~\bibnamefont
  {Rudolph}},\ }\href {\doibase 10.1103/PhysRevA.74.040302} {\bibfield
  {journal} {\bibinfo  {journal} {Phys. Rev. A}\ }\textbf {\bibinfo {volume}
  {74}},\ \bibinfo {pages} {040302} (\bibinfo {year} {2006})}\BibitemShut
  {NoStop}%
\bibitem [{\citenamefont {Verstraete}\ \emph {et~al.}(2004)\citenamefont
  {Verstraete}, \citenamefont {Garcia-Ripoll},\ and\ \citenamefont
  {Cirac}}]{Mixed}%
  \BibitemOpen
  \bibfield  {author} {\bibinfo {author} {\bibfnamefont {F.}~\bibnamefont
  {Verstraete}}, \bibinfo {author} {\bibfnamefont {J.~J.}\ \bibnamefont
  {Garcia-Ripoll}}, \ and\ \bibinfo {author} {\bibfnamefont {J.~I.}\
  \bibnamefont {Cirac}},\ }\href {\doibase 10.1103/PhysRevLett.93.207204}
  {\bibfield  {journal} {\bibinfo  {journal} {Phys. Rev. Lett.}\ }\textbf
  {\bibinfo {volume} {93}},\ \bibinfo {pages} {207204} (\bibinfo {year}
  {2004})}\BibitemShut {NoStop}%
\bibitem [{\citenamefont {Fannes}\ \emph {et~al.}(1992)\citenamefont {Fannes},
  \citenamefont {Nachtergaele},\ and\ \citenamefont {Werner}}]{raey}%
  \BibitemOpen
  \bibfield  {author} {\bibinfo {author} {\bibfnamefont {M.}~\bibnamefont
  {Fannes}}, \bibinfo {author} {\bibfnamefont {B.}~\bibnamefont
  {Nachtergaele}}, \ and\ \bibinfo {author} {\bibfnamefont {R.}~\bibnamefont
  {Werner}},\ }\href {\doibase 10.1007/BF02099178} {\bibfield  {journal}
  {\bibinfo  {journal} {Commun. Math. Phys.}\ }\textbf {\bibinfo {volume}
  {144}},\ \bibinfo {pages} {443} (\bibinfo {year} {1992})}\BibitemShut
  {NoStop}%
\bibitem [{\citenamefont {Pirvu}\ \emph {et~al.}(2010)\citenamefont {Pirvu},
  \citenamefont {Murg}, \citenamefont {Cirac},\ and\ \citenamefont
  {Verstraete}}]{MPO_Representations}%
  \BibitemOpen
  \bibfield  {author} {\bibinfo {author} {\bibfnamefont {B.}~\bibnamefont
  {Pirvu}}, \bibinfo {author} {\bibfnamefont {V.}~\bibnamefont {Murg}},
  \bibinfo {author} {\bibfnamefont {J.~I.}\ \bibnamefont {Cirac}}, \ and\
  \bibinfo {author} {\bibfnamefont {F.}~\bibnamefont {Verstraete}},\ }\href
  {\doibase 10.1088/1367-2630/12/2/025012} {\bibfield  {journal} {\bibinfo
  {journal} {New J. Phys.}\ }\textbf {\bibinfo {volume} {12}},\ \bibinfo
  {pages} {025012} (\bibinfo {year} {2010})}\BibitemShut {NoStop}%
\bibitem [{\citenamefont {Kliesch}\ \emph {et~al.}(2014)\citenamefont
  {Kliesch}, \citenamefont {Gross},\ and\ \citenamefont
  {Eisert}}]{UndecidableMPO}%
  \BibitemOpen
  \bibfield  {author} {\bibinfo {author} {\bibfnamefont {M.}~\bibnamefont
  {Kliesch}}, \bibinfo {author} {\bibfnamefont {D.}~\bibnamefont {Gross}}, \
  and\ \bibinfo {author} {\bibfnamefont {J.}~\bibnamefont {Eisert}},\ }\href
  {\doibase 10.1103/PhysRevLett.113.160503} {\bibfield  {journal} {\bibinfo
  {journal} {Phys. Rev. Lett.}\ }\textbf {\bibinfo {volume} {113}},\ \bibinfo
  {pages} {160503} (\bibinfo {year} {2014})}\BibitemShut {NoStop}%
\bibitem [{\citenamefont {Bultinck}\ \emph {et~al.}(2017)\citenamefont
  {Bultinck}, \citenamefont {Marien}, \citenamefont {Williamson}, \citenamefont
  {Sahinoglu}, \citenamefont {Haegeman},\ and\ \citenamefont
  {Verstraete}}]{Bultinck2017}%
  \BibitemOpen
  \bibfield  {author} {\bibinfo {author} {\bibfnamefont {N.}~\bibnamefont
  {Bultinck}}, \bibinfo {author} {\bibfnamefont {M.}~\bibnamefont {Marien}},
  \bibinfo {author} {\bibfnamefont {D.}~\bibnamefont {Williamson}}, \bibinfo
  {author} {\bibfnamefont {M.~B.}\ \bibnamefont {Sahinoglu}}, \bibinfo {author}
  {\bibfnamefont {J.}~\bibnamefont {Haegeman}}, \ and\ \bibinfo {author}
  {\bibfnamefont {F.}~\bibnamefont {Verstraete}},\ }\href {\doibase
  https://doi.org/10.1016/j.aop.2017.01.004} {\bibfield  {journal} {\bibinfo
  {journal} {Ann. Phys.}\ }\textbf {\bibinfo {volume} {378}},\ \bibinfo {pages}
  {183 } (\bibinfo {year} {2017})}\BibitemShut {NoStop}%
\bibitem [{\citenamefont {Buerschaper}\ \emph {et~al.}(2014)\citenamefont
  {Buerschaper}, \citenamefont {Morampudi},\ and\ \citenamefont
  {Pollmann}}]{Buerschaper2014}%
  \BibitemOpen
  \bibfield  {author} {\bibinfo {author} {\bibfnamefont {O.}~\bibnamefont
  {Buerschaper}}, \bibinfo {author} {\bibfnamefont {S.~C.}\ \bibnamefont
  {Morampudi}}, \ and\ \bibinfo {author} {\bibfnamefont {F.}~\bibnamefont
  {Pollmann}},\ }\href {\doibase 10.1103/PhysRevB.90.195148} {\bibfield
  {journal} {\bibinfo  {journal} {Phys. Rev. B}\ }\textbf {\bibinfo {volume}
  {90}},\ \bibinfo {pages} {195148} (\bibinfo {year} {2014})}\BibitemShut
  {NoStop}%
\bibitem [{\citenamefont {Bonesteel}\ and\ \citenamefont
  {DiVincenzo}(2012)}]{Bonesteel2012}%
  \BibitemOpen
  \bibfield  {author} {\bibinfo {author} {\bibfnamefont {N.~E.}\ \bibnamefont
  {Bonesteel}}\ and\ \bibinfo {author} {\bibfnamefont {D.~P.}\ \bibnamefont
  {DiVincenzo}},\ }\href
  {https://journals.aps.org/prb/abstract/10.1103/PhysRevB.86.165113} {\bibfield
   {journal} {\bibinfo  {journal} {Phys. Rev. B}\ }\textbf {\bibinfo {volume}
  {86}},\ \bibinfo {pages} {165113} (\bibinfo {year} {2012})}\BibitemShut
  {NoStop}%
\bibitem [{\citenamefont {Fendley}\ \emph {et~al.}(2013)\citenamefont
  {Fendley}, \citenamefont {Isakov},\ and\ \citenamefont
  {Troyer}}]{Fendley2013}%
  \BibitemOpen
  \bibfield  {author} {\bibinfo {author} {\bibfnamefont {P.}~\bibnamefont
  {Fendley}}, \bibinfo {author} {\bibfnamefont {S.~V.}\ \bibnamefont {Isakov}},
  \ and\ \bibinfo {author} {\bibfnamefont {M.}~\bibnamefont {Troyer}},\ }\href
  {https://journals.aps.org/prl/abstract/10.1103/PhysRevLett.110.260408}
  {\bibfield  {journal} {\bibinfo  {journal} {Phys. Rev. Lett.}\ }\textbf
  {\bibinfo {volume} {110}},\ \bibinfo {pages} {260408} (\bibinfo {year}
  {2013})}\BibitemShut {NoStop}%
\bibitem [{\citenamefont {Buerschaper}(2014)}]{Buerschaper-AnnPhys-2014}%
  \BibitemOpen
  \bibfield  {author} {\bibinfo {author} {\bibfnamefont {O.}~\bibnamefont
  {Buerschaper}},\ }\href {\doibase 10.1016/j.aop.2014.09.007} {\bibfield
  {journal} {\bibinfo  {journal} {Ann. Phys.}\ }\textbf {\bibinfo {volume}
  {351}},\ \bibinfo {pages} {447} (\bibinfo {year} {2014})}\BibitemShut
  {NoStop}%
\bibitem [{Note2()}]{Note2}%
  \BibitemOpen
  \bibinfo {note} {For completeness, we mention that the coupling always
  connects vertices of different sublattices. For a connection as indicated in
  the Fig. \ref {fig:tworings}, the right triangle is identical to the one
  above. A closer inspection shows that the difference between the operator
  representing the left triangle differs from the one on the right by
  interchanging $a$ vs $b$ and $\alpha $ vs $\beta $ which yields a replacement
  $\protect \mathaccentV {hat}05EB_{00}\DOTSB \mapstochar \rightarrow \protect
  \mathaccentV {hat}05EB_{11}, \protect \mathaccentV {hat}05EB_{11}\DOTSB
  \mapstochar \rightarrow \protect \mathaccentV {hat}05EB_{00}, \protect
  \mathaccentV {hat}05EB_{01}\DOTSB \mapstochar \rightarrow -\protect
  \mathaccentV {hat}05EB_{10},\protect \mathaccentV {hat}05EB_{10}\DOTSB
  \mapstochar \rightarrow - \protect \mathaccentV {hat}05EB_{01}$ in Eq. (\ref
  {eq:B2}). However, since the $B_{01}$, $B_{10}$ always appear in pairs, the
  MPO $O_6$ is the same.}\BibitemShut {Stop}%
\bibitem [{\citenamefont {Brell}\ \emph
  {et~al.}(2014{\natexlab{b}})\citenamefont {Brell}, \citenamefont {Burton},
  \citenamefont {Dauphinais}, \citenamefont {Flammia},\ and\ \citenamefont
  {Poulin}}]{Brell2014}%
  \BibitemOpen
  \bibfield  {author} {\bibinfo {author} {\bibfnamefont {C.~G.}\ \bibnamefont
  {Brell}}, \bibinfo {author} {\bibfnamefont {S.}~\bibnamefont {Burton}},
  \bibinfo {author} {\bibfnamefont {G.}~\bibnamefont {Dauphinais}}, \bibinfo
  {author} {\bibfnamefont {S.~T.}\ \bibnamefont {Flammia}}, \ and\ \bibinfo
  {author} {\bibfnamefont {D.}~\bibnamefont {Poulin}},\ }\href {\doibase
  10.1103/PhysRevX.4.031058} {\bibfield  {journal} {\bibinfo  {journal} {Phys.
  Rev. X}\ }\textbf {\bibinfo {volume} {4}},\ \bibinfo {pages} {031058}
  (\bibinfo {year} {2014}{\natexlab{b}})}\BibitemShut {NoStop}%
\bibitem [{\citenamefont {Morampudi}\ \emph {et~al.}(2014)\citenamefont
  {Morampudi}, \citenamefont {von Keyserlingk},\ and\ \citenamefont
  {Pollmann}}]{Morampudi14}%
  \BibitemOpen
  \bibfield  {author} {\bibinfo {author} {\bibfnamefont {S.~C.}\ \bibnamefont
  {Morampudi}}, \bibinfo {author} {\bibfnamefont {C.}~\bibnamefont {von
  Keyserlingk}}, \ and\ \bibinfo {author} {\bibfnamefont {F.}~\bibnamefont
  {Pollmann}},\ }\href {\doibase 10.1103/PhysRevB.90.035117} {\bibfield
  {journal} {\bibinfo  {journal} {Phys. Rev. B}\ }\textbf {\bibinfo {volume}
  {90}},\ \bibinfo {pages} {035117} (\bibinfo {year} {2014})}\BibitemShut
  {NoStop}%
\bibitem [{\citenamefont {Tsomokos}\ \emph {et~al.}(2011)\citenamefont
  {Tsomokos}, \citenamefont {Osborne},\ and\ \citenamefont
  {Castelnovo}}]{Tsomokos2011}%
  \BibitemOpen
  \bibfield  {author} {\bibinfo {author} {\bibfnamefont {D.~I.}\ \bibnamefont
  {Tsomokos}}, \bibinfo {author} {\bibfnamefont {T.~J.}\ \bibnamefont
  {Osborne}}, \ and\ \bibinfo {author} {\bibfnamefont {C.}~\bibnamefont
  {Castelnovo}},\ }\href {\doibase 10.1103/PhysRevB.83.075124} {\bibfield
  {journal} {\bibinfo  {journal} {Phys. Rev. B}\ }\textbf {\bibinfo {volume}
  {83}},\ \bibinfo {pages} {075124} (\bibinfo {year} {2011})}\BibitemShut
  {NoStop}%
\bibitem [{\citenamefont {Tarantino}\ and\ \citenamefont
  {Fidkowski}(2016)}]{PhysRevB.94.115115}%
  \BibitemOpen
  \bibfield  {author} {\bibinfo {author} {\bibfnamefont {N.}~\bibnamefont
  {Tarantino}}\ and\ \bibinfo {author} {\bibfnamefont {L.}~\bibnamefont
  {Fidkowski}},\ }\href {\doibase 10.1103/PhysRevB.94.115115} {\bibfield
  {journal} {\bibinfo  {journal} {Phys. Rev. B}\ }\textbf {\bibinfo {volume}
  {94}},\ \bibinfo {pages} {115115} (\bibinfo {year} {2016})}\BibitemShut
  {NoStop}%
\bibitem [{\citenamefont {Ware}\ \emph {et~al.}(2016)\citenamefont {Ware},
  \citenamefont {Son}, \citenamefont {Cheng}, \citenamefont {Mishmash},
  \citenamefont {Alicea},\ and\ \citenamefont {Bauer}}]{PhysRevB.94.115127}%
  \BibitemOpen
  \bibfield  {author} {\bibinfo {author} {\bibfnamefont {B.}~\bibnamefont
  {Ware}}, \bibinfo {author} {\bibfnamefont {J.~H.}\ \bibnamefont {Son}},
  \bibinfo {author} {\bibfnamefont {M.}~\bibnamefont {Cheng}}, \bibinfo
  {author} {\bibfnamefont {R.~V.}\ \bibnamefont {Mishmash}}, \bibinfo {author}
  {\bibfnamefont {J.}~\bibnamefont {Alicea}}, \ and\ \bibinfo {author}
  {\bibfnamefont {B.}~\bibnamefont {Bauer}},\ }\href {\doibase
  10.1103/PhysRevB.94.115127} {\bibfield  {journal} {\bibinfo  {journal} {Phys.
  Rev. B}\ }\textbf {\bibinfo {volume} {94}},\ \bibinfo {pages} {115127}
  (\bibinfo {year} {2016})}\BibitemShut {NoStop}%
\bibitem [{\citenamefont {Sagi}\ \emph {et~al.}(2019)\citenamefont {Sagi},
  \citenamefont {Ebisu}, \citenamefont {Tanaka}, \citenamefont {Stern},\ and\
  \citenamefont {Oreg}}]{Sagi2018}%
  \BibitemOpen
  \bibfield  {author} {\bibinfo {author} {\bibfnamefont {E.}~\bibnamefont
  {Sagi}}, \bibinfo {author} {\bibfnamefont {H.}~\bibnamefont {Ebisu}},
  \bibinfo {author} {\bibfnamefont {Y.}~\bibnamefont {Tanaka}}, \bibinfo
  {author} {\bibfnamefont {A.}~\bibnamefont {Stern}}, \ and\ \bibinfo {author}
  {\bibfnamefont {Y.}~\bibnamefont {Oreg}},\ }\href {\doibase
  10.1103/PhysRevB.99.075107} {\bibfield  {journal} {\bibinfo  {journal} {Phys.
  Rev. B}\ }\textbf {\bibinfo {volume} {99}},\ \bibinfo {pages} {075107}
  (\bibinfo {year} {2019})}\BibitemShut {NoStop}%
\bibitem [{\citenamefont {Thomson}\ and\ \citenamefont
  {Pientka}(2018)}]{Thomson2018}%
  \BibitemOpen
  \bibfield  {author} {\bibinfo {author} {\bibfnamefont {A.}~\bibnamefont
  {Thomson}}\ and\ \bibinfo {author} {\bibfnamefont {F.}~\bibnamefont
  {Pientka}},\ }\href@noop {} {\  (\bibinfo {year} {2018})},\ \Eprint
  {http://arxiv.org/abs/arXiv:1807.09291v1} {arXiv:1807.09291v1} \BibitemShut
  {NoStop}%
\bibitem [{\citenamefont {Bravyi}\ \emph {et~al.}(2011)\citenamefont {Bravyi},
  \citenamefont {DiVincenzo},\ and\ \citenamefont {Loss}}]{Bravyi2011}%
  \BibitemOpen
  \bibfield  {author} {\bibinfo {author} {\bibfnamefont {S.}~\bibnamefont
  {Bravyi}}, \bibinfo {author} {\bibfnamefont {D.~P.}\ \bibnamefont
  {DiVincenzo}}, \ and\ \bibinfo {author} {\bibfnamefont {D.}~\bibnamefont
  {Loss}},\ }\href {https://doi.org/10.1016/j.aop.2011.06.004} {\bibfield
  {journal} {\bibinfo  {journal} {Ann. Phys.}\ }\textbf {\bibinfo {volume}
  {326}},\ \bibinfo {pages} {2793} (\bibinfo {year} {2011})}\BibitemShut
  {NoStop}%
\end{thebibliography}

\end{document}